\pdfoutput=1
\documentclass{aastex631}
\usepackage{amsmath}
\usepackage{appendix}
\usepackage{longtable}
\usepackage{url}
\shorttitle{Variability of magnetic hot stars in TESS}
\shortauthors{D.-X. Shen et al.}
\graphicspath{{./}{figures/}}
\begin{document}

\title{Variability of magnetic hot stars from the TESS observations}

\author[0000-0002-1926-8248]{Shen, Dong-Xiang}
\affiliation{School of Physical Science and Technology, Xinjiang University, Urumqi 830046, People's Republic of China; {\it chunhuazhu@sina.cn}}

\author[0000-0001-9313-251X]{Li, Gang}
\affiliation{Institute of Astronomy, KU Leuven, Celestijnenlaan 200D, B-3001 Leuven, Belgium;}

\author{Abdusamatjan Iskandar}
\affiliation{Xinjiang Astronomical Observatory, National Astronomical Observatories, Chinese Academy of Sciences, Urumqi 830000, China; {\it liujinzh@xao.ac.cn}}
\affiliation{School of Astronomy and Space Science, University of Chinese Academy of Sciences, Beijing 100049, People's Republic of China}

\author[0000-0001-8241-1740]{Fu, Jian-Ning}
\affiliation{Department of Astronomy, Beijing Normal University, Beijing, 100871, People's Republic of China}

\author{Zhu, Chun-Hua}
\affiliation{School of Physical Science and Technology, Xinjiang University, Urumqi 830046, People's Republic of China; {\it chunhuazhu@sina.cn}}

\author{Yu, Jin-Long}
\affiliation{College of Mechanical and Electronic Engineering, Tarim University. 843300, Alar, P.R.China}

\author{Zhang, Yu}
\affiliation{Xinjiang Astronomical Observatory, National Astronomical Observatories, Chinese Academy of Sciences, Urumqi 830000, China; {\it liujinzh@xao.ac.cn}}

\author{L$\ddot{\rm u}$ , Guo-Liang}
\affiliation{Xinjiang Astronomical Observatory, National Astronomical Observatories, Chinese Academy of Sciences, Urumqi 830000, China; {\it liujinzh@xao.ac.cn}}

\author[0009-0008-3490-7934]{Zhai, Nan-Nan}
\affiliation{Xinjiang Astronomical Observatory, National Astronomical Observatories, Chinese Academy of Sciences, Urumqi 830000, China; {\it liujinzh@xao.ac.cn}}

\author{Liu, Jin-Zhong}
\affiliation{Xinjiang Astronomical Observatory, National Astronomical Observatories, Chinese Academy of Sciences, Urumqi 830000, China; {\it liujinzh@xao.ac.cn}}



\begin{abstract}

    Magnetic hot stars refer to the stars, which effective temperatures approximately in the range from 7,000 to 50,000 K, and with large-scale globally organized magnetic fields. These magnetic fields exhibit strengths ranging from tens of Gauss to tens of kilo-Gauss. They are key in understanding the effects caused by magnetic fields in the stellar evolution. However, there are only three magnetic hot stars studied via a combination of spectropolarimetric and asteroseismic modeling. Combined with $Transiting\;Exoplanet\;Survey\;Satellite\;(TESS)$ 1-56 sectors data sets, we provided a photometric variability and stochastic low frequency (SLF) variability study of 118 magnetic hot stars. 9 new rotating variable stars are identified. Using the Bayesian Markov Chain Monte Carlo (MCMC) framework, we fitted the morphologies of SLF variability for magnetic hot stars. Our analysis reveals that the magnetic hot stars in our sample have $\gamma < 5.5$ with the vast majority having $1 \leq \gamma \leq 3$. The $\nu_{\rm char}$ is primarily in the ranges of $0\;\text{d}^{-1} < \nu_{\rm char} < 6.3\;\text{d}^{-1}$. The amplitude of SLF variability, log$\alpha_{\rm 0}$, shows a dominant distribution ranging from 0.8 to 3. No significant correlations are observed between the luminosity and fitting parameters, suggesting no clear dependence of SLF variability on stellar mass for our sample of magnetic hot stars with masses between approximately $1.5 M_{\odot}< M < 20 M_{\odot}$. We found a significant negative correlation between the $B_{\rm p}$ and $\nu_{char}$. This suppression effect of magnetic fields on $\nu_{\rm char}$ may be a result of their inhibition of macroturbulence.

%Asteroseismology is a powerful tool for calibrating the free parameters in structure and evolution theory of magnetic hot stars.
\end{abstract}



\section{Introduction}
Magnetic hot stars have been revealed by recent large surveys \citep[e.g.][]{2014A&A...562A.143F,2015A&A...583A.115B,2017MNRAS.465.2432G}. Their effective temperatures range from roughly 7,000 to 50,000 K \citep{2021mobs.confE..17N}. The majority of the detected large-scale magnetic fields of the magnetic hot stars are stable over timespans of up to decades \citep[e.g.][]{2012MNRAS.419..959O,2019MNRAS.483.2300S}, and have a predominantly dipole topological structure inclined to the rotational axis, with magnetic field strengths ranging from tens of Gauss to tens of kilo-Gauss \citep[e.g.][]{2018MNRAS.475.5144S,2021MNRAS.501.4534J}. Strong magnetic fields have been considered to explain the particular properties of these hot stars, such as photometric variations \citep[e.g.][]{2015MNRAS.451.2015O,2022arXiv221100271S}, H$\alpha$ line variations \citep[e.g.][]{2015ApJ...811L..26W,2017MNRAS.467L..81H,2020MNRAS.499.5379S}, the UV resonance line variations \citep[e.g.][]{1990ApJ...365..665S,2022arXiv220612838U}, the X-ray properties \citep[e.g.][]{2012MNRAS.425.1278W,2015MNRAS.453.3288P}, and the radio emissions \citep[e.g.][]{2017MNRAS.467.2820L,2022MNRAS.513.1429S}.

The strong, large-scale dipole magnetic field in a hot star could be a fossil field from the pre-main sequence (pre-MS), or the result of a violent binary interaction such as a merger \citep{2003ApJ...586..480M, 2004Natur.431..819B, 2005ApJ...629..461B, 2006A&A...450.1077B, 2010A&A...523A..41P, 2011A&A...532L..13P, 2011A&A...534A.140C, 2013MNRAS.429.1001A, 2015A&A...575A.106J}. Theories and numerical simulations considered various effects caused by strong magnetic fields in hot stars, i.e.: magneto-convection \citep[e.g.][]{2020ApJ...900..113J}, mass-loss quenching \citep[e.g.][]{2019PhDT.......122K, 2019MNRAS.485.5843K}, internal angular momentum transport \citep[e.g.][]{2013ApJ...772...21R, 2023arXiv230207811M}, magnetic braking \citep[e.g.][]{2020MNRAS.493..518K}, and mass leakage \citep[e.g.][]{2018MNRAS.474.3090O}. On the other hand, observations contributed more and more pieces of evidence of these effects \citep[e.g.][]{2011MNRAS.416.1456O, 2012MNRAS.419.1610G, 2012A&A...545A.119H, 2014A&A...562A.143F, 2015A&A...574A..20F, 2020MNRAS.499.5379S}.

The number of known magnetic stars is steadily increased with the dedicated surveys of spectropolarimetry and high-resolution spectrometry. FOcal Reducer and low dispersion Spectrograph (FORS1) catalog include about 1500 observing series, for a total of more than about 12 000 scientific frames, revealing dozens of magnetic hot stars \citep{2015A&A...583A.115B}. The B fields in OB stars (BOB) Collaboration observed hundreds of OB-type stars to characterize the incidence of large-scale magnetic fields in slowly rotating main-sequence massive stars \citep{2014Msngr.157...27M,2015IAUS..307..342M}. Magnetism in Massive Stars (MiMeS) Survey also observed numbers of massive stars to search magnetic hot stars \citep{2017MNRAS.465.2432G}. These surveys indicate about $\sim 10 \%$ of the B stars host strong and large-scale magnetic fields, while only $\sim 6 \%$ of O-type \citep{2012ASPC..465...42G, 2017MNRAS.465.2432G}. Up to now, more and more magnetic hot stars are still detected by the spectrometric or photometric surveys \citep[e.g.][]{2014ApJ...784L..30E,2019ApJ...873L...5C,2022ApJ...924L..10J,2022MNRAS.516.2812C}.

Despite tremendous contributions made by spectropolarimetric and high-resolution spectrometric measurements, it is difficult to calibrate the physical properties of stellar interiors. Asteroseismology is a powerful tool that allows us to directly access the interiors of early-type stars and improve the physics of stellar interiors \citep{2020FrASS...7...70B, 2021RvMP...93a5001A}. Analysis of stellar pulsations allows deriving direct inferences of the internal rotation and chemical stratification from the pulsation frequencies or period spacing patterns in stars born with a convective core \citep{2015ApJS..218...27V,2019A&A...626A.121O,2020MNRAS.491.3586L}. Furthermore, forward seismic modeling of the frequencies restricts the masses and ages of stars \citep{2016A&A...592A.116S,2018ApJS..237...15A,2019A&A...624A..75A}. The high-duty cycles, long-term, near-continuous, and high-precision data sets are necessary for asteroseismology. The fragmented time series obtained with ground-based telescopes are notoriously limited by their poor duty cycles which complicates mode identification \citep[e.g.][]{2005MNRAS.362..612H,2009MNRAS.396.1460D,2013MNRAS.431.3396D}. In contrast to ground-based telescopes, the $Transiting\;Exoplanet\;Survey\;Satellite$ \citep[$TESS$;][]{2015JATIS...1a4003R} is designed to observe the whole sky, and provide high-precision photometric data with a wide field of view. \textbf{The asteroseismic potential of TESS is evidenced by many studies \citep[e.g.][]{2019ApJ...872L...9P, 2020A&A...640A..36B, 2022ApJ...924L..10J, 2020MNRAS.493.5871B, 2023NatAs.tmp..161B}.} 

There have two main coherent pulsations in magnetic hot stars: one is the gravity modes ($g$ modes) where the restoring force is buoyancy, and another one is pressure modes ($p$ modes) where the restoring force is pressure \citep{2010aste.book.....A}. They are standing waves motivated by heat engine driving mechanism \citep{2010aste.book.....A}. Gravity modes normally associated with the Slowly Pulsating B-type stars \citep{2002A&A...393..965D,2013ASSP...31..191W, 2015A&A...580A..27M}, while $p$ modes more occurred in $\beta$ Cephei variables \citep{2005ApJS..158..193S}. The theoretical instability strips can be calculated using non-adiabatic calculations performed with Aarhus Pulsation code \citep[ADIPLS;][]{2008Ap&SS.316..113C} or GYRE  \citep[e.g.][]{2013MNRAS.435.3406T}. The stellar interior rotation profiles, temperature gradient in the core boundary layer, and envelope mixing profiles are revealed by analyzing the coherent pulsations \citep[e.g.][]{2009A&A...506..111D, 2010Natur.464..259D, 2013MNRAS.431.3396D, 2018MNRAS.478.2243S, 2021A&A...650A.175M, 2022MNRAS.511.1529S}. \textbf{The core masses and ages for a number of OB-type stars are also constrained by the method of asteroseismology \citep[e.g.][]{2008MNRAS.385.2061D, 2019ApJ...886...35D, 2021NatAs...5..715P, 2022A&A...664L...1S, 2023NatAs.tmp..161B}.} 

However, there are few inferences as to the magnetic field strength and geometry within the deep interiors of stars. Only $\beta\;Cep$ \citep[e.g.][]{2000ApJ...531L.143S, 2013A&A...555A..46H}, V2052 Oph \citep[e.g.][]{2012A&A...537A.148N, 2012MNRAS.424.2380H, 2012MNRAS.427..483B}, and HD\;43317 \citep[][]{2017A&A...605A.104B, 2018A&A...616A.148B, 2022MNRAS.512L..16L} have been studied using combined asteroseismic and magnetometric analyses. Despite the interaction of IGWs with a magnetic field is discussed extensively using the numerical simulations \citep[e.g.][]{2017MNRAS.466.2181L, 2023MNRAS.521.5372J, 2023arXiv230308147R}, there only have rare observational evidences \citep[][]{2015Sci...350..423F, 2022MNRAS.512L..16L}. The lack of large samples of the pulsating magnetic stars has impeded the exploration of pulsating variability, magnetic fields, and stellar evolution in the magnetic hot star regime.

Apart from coherent pulsations, internal gravity waves (IGWs) are predicted to exist in stars with convective regions, and their properties vary based on the radial rotation profile \citep[e.g.][]{2007A&A...474..155P, 2013ApJ...772...21R, 2014A&A...565A..47M, 2015ApJ...815L..30R} or the strength of the internal magnetic field \citep[e.g.][]{2010MNRAS.401..191R, 2011MNRAS.410..946R, 2011A&A...526A..65M, 2012A&A...540A..37M, 2016ApJ...829...92A, 2017MNRAS.466.2181L}. The amplitudes of IGWs are expected to scale with stellar mass, while their frequencies are predicted to inversely scale with stellar mass on the main sequence \citep[e.g.][]{2010Ap&SS.328..253S, 2013MNRAS.430.1736S, 2015ApJ...815L..30R}. Although two- and three-dimensional hydrodynamical simulations of IGWs for massive stars with $M > 8\;M_{\odot}$ have also been done \citep[e.g.][]{2023arXiv230506379R,2023arXiv230306125T, 2023arXiv230305495H}, the influence of magnetic fields on these stars is still not widely studied. Stochastic low frequency (SLF) variability is observed in upper main sequence stars \citep[e.g.][]{2019A&A...621A.135B, 2019NatAs...3..760B, 2020A&A...640A..36B, 2022A&A...668A.134B}. The SLF variability is seen across a wide range in mass, age, and metallicity in massive stars \citep{2020A&A...640A..36B, 2019NatAs...3..760B}. The origin of SLF variability is currently debated. It could be caused by the IGWs \citep[e.g.][]{2019ApJ...876....4E, 2019A&A...621A.135B, 2020A&A...640A..36B, 2022A&A...668A.134B}, or by subsurface convection zones \citep[e.g.][]{2011A&A...533A...4B, 2019ApJ...883..106C, 2019ApJ...886L..15L, 2021ApJ...915..112C}. \cite{2019NatAs...3..760B} investigated the 167 OB-type stars within Milky Way and Large Magellanic Cloud (LMC) galaxy, and reported that SLF variability is insensitive to the metallicity of the star. While how magnetic fields systematically affect the presence of SLF variability in massive stars requires the detection of this phenomenon for a large sample of magnetic hot stars, which is currently not available. It has been confirmed that the magnetic field has an effect on stabilizing the atmosphere and suppressing the generation of macroturbulence, which in turn can affect the variability of SLF caused by subsurface convection \citep{2019ApJ...886L..15L, 2013MNRAS.433.2497S}. Exploring the correlation between SLF variability and the magnetic field could provide valuable insights into the underlying mechanisms responsible for the observed SLF variability.

In this work, we classified the variability of 118 magnetic hot stars using short cadence target pixel files (TPFs) observed by TESS from sector 1 to sector 56. We detected the SLF variability for each magnetic hot star within our sample. We investigated the characteristics of SLF variability in magnetic hot stars and explored the correlation of SLF variability and the magnetic field. In Section 2, we presented our sample component of magnetic hot stars. In Section 3, we introduced the methods used to extract the light curves and analyze frequency spectra. We presented our results, and evaluated the asteroseismic potential for multiperiodic pulsators in Section 4. We discussed our results and drew conclusions in Sections 5 and 6, respectively.

\section{Sample Selection}
Magnetic fields in the stars can be detected through the circular polarization and split induced in spectral lines by the Zeeman effect. The first-generation photo-polarimeters carried out the measurements of magnetic fields in the bulk of magnetic Ap/Bp stars \citep[e.g.][]{1980ApJS...42..421B}. Dozens of magnetic hot stars are detected with second generation instruments, such as the low-resolution spectropolarimetric instruments FORS 1 and 2 mounted at the Very Large Telescope (VLT) \citep{1998Msngr..94....1A}, or high-resolution spectrometric spectropolarimeters MUSICOS \citep{1988ESASP.286..665F}, ESPaDOnS \citep{2003ASPC..307...41D}, Narval \citep{2003EAS.....9..105A}, and HARPSpol \citep{2011Msngr.143....7P}.

The low-resolution spectropolarimeters are sensitive to the longitudinal (line-of-sight) field component ($\langle B_{z}\rangle$), whereas high-resolution spectroscopy detecting magnetic fields in stars by observing the Zeeman splitting of spectral lines \citep{2009ARA&A..47..333D}. However, a significant number of magnetic stars claimed by the low-resolution spectropolarimeters have not been confirmed by high-resolution spectroscopy \citep{2009MNRAS.398.1505S, 2012AIPC.1429..110S, 2012AIPC.1429...67G}. \cite{2012A&A...538A.129B} elaborated the polarization measurements data set of the FORS 1 archive, explored the results using various data reduction pipelines, and analyzed the sources of uncertainties. Using the new prescription for low-resolution spectropolarimeteric data analysis, the consistency of the results between the low-resolution spectropolarimeters and high-resolution spectroscopy are improved significantly \citep[e.g.][]{2012ApJ...750....2S}.

\begin{figure}
    \gridline{\fig{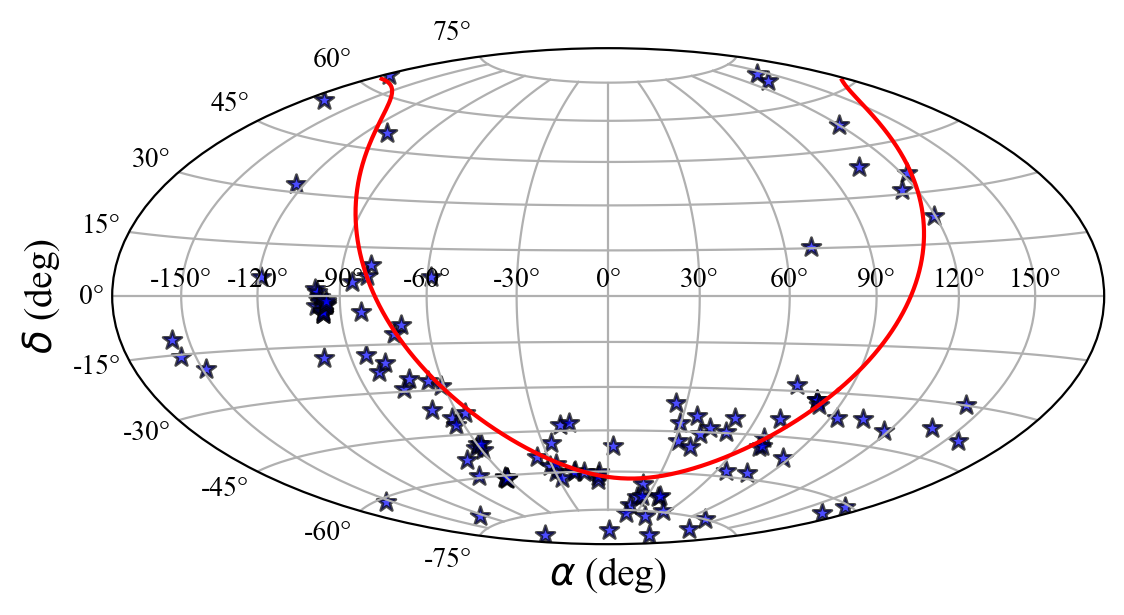}{0.4\textwidth}{}}
    \caption{Location of the stars in our catalogue in celestial coordinates. The galactic plane (in red) is plotted as the reference.\label{location}}
\end{figure}

Therefore, we only considered the results based on these new detection statuses for magnetic hot stars. We use the compilations of the magnetic hot stars including the results of FORS1 \citep{2015A&A...583A.115B}, B fields in OB stars Survey \citep[BOB/FORS2;][]{2014Msngr.157...27M,2015IAUS..307..342M}, and the Magnetism in Massive Stars Survey \citep[MiMeS;][]{2017MNRAS.465.2432G} surveys. In the results of FORS1, the mean longitudinal magnetic field ($\langle B_{z}\rangle$) of each object is measured from the analysis carried out on the H Balmer lines, on the metal lines, and on the combination of the two sets of lines, respectively \citep{2012A&A...538A.129B, 2015A&A...583A.115B}. The BOB/FORS2 and MiMeS catalogs also show the results in a similar format to FORS1. The stars which $|\langle B_{z}\rangle|/\sigma_{\langle B_{z}\rangle} \geq 3$ in at least two sets are selected as the samples to ensure the sigma completeness of our catalog. Other exhaustive studies that identify hot stars with confirmed field detections are complemented in our catalog \citep[e.g.][]{2013MNRAS.429..398P,2018MNRAS.475.5144S,2019MNRAS.485.1508S,2019MNRAS.490..274S,2020MNRAS.499.5379S}. Many magnetic hot stars show chemical abundance peculiarities \citep[e.g.][]{2007A&A...476..911F,2000MNRAS.313..851W,2000A&A...355.1080W}, and their spectral data sets are not always readily available. It is difficult to ensure completeness at the low-temperature boundary. Therefore, we consider all magnetic hot stars with spectral type in the range of A5 and earlier. Out of the 729 stars, 168 met our criteria and were included in the original sample. We cross-matched the catalog consisting of the original sample with the Mikulski Archive for Space Telescopes (MAST\footnote[1]{https://archive.stsci.edu/}) archive, acquired the 360 TPFs of 118 magnetic hot stars, which form the basis of our study. The galactic distribution of 118 magnetic hot stars in our catalog is shown in Figure \ref{location}.

\section{Methodology}
\subsection{TESS light curve extraction\label{Section 3.1}}
The TESS space telescope observed the stars with the field of view of 24 by 96$^{\circ}$. A portion of the sky will be observed for a duration of 27.3 days before TESS  turns to the next position. This is referred to as a sector. To obtain the TESS light curves, we retrieved the short cadence TPFs for Sectors 1-56 from the MAST. We performed the aperture photometry and data reduction using the python package $Lightkurve$ \citep{2018ascl.soft12013L}. The aperture mask for each sector of each star is defined using the thresholding method \footnote[2]{https://docs.lightkurve.org/reference/api/lightkurve.TessTargetPixelFile}. This method selects all pixels which have a flux higher than threshold times standard deviations above a median brightness as the photometric aperture. The standard deviation is estimated in a robust way by multiplying the Median Absolute Deviation (MAD) by $1.4826$. In order to minimize contamination from nearby stars in the light curves, we meticulously optimized the aperture masks. Because the TESS pixels are large (21 arcsec), the photometry will be contaminated by the neighboring stars within our selected aperture masks \citep[e.g.][]{2023AJ....165..239P}. FluxCT \citep{2023RNAAS...7...18S} is a dedicated tool designed to identify contaminating flux in TESS TPFs. It utilizes the Gaia EDR3 database to search for potential sources of contamination within the corresponding regions of the aperture masks, providing a contamination coefficient. Each pixel within the aperture masks was individually assessed using FluxCT to calculate its contamination coefficient. By excluding pixels with higher contamination coefficients, the majority of stars in the sample exhibit a photometric aperture contamination rate below 5\%. For stars in crowded fields, we selected pixels with a flux higher than 5 times the standard deviation above the median brightness as the photometric aperture to minimize the impact of background sources.

The raw time series of the photometry are obtained by summing flux counts for each pixel within photometric apertures. For removing the background noise, we defined background aperture masks using the threshold method. The pixels which are fainter than 1-sigma above the median flux will be used. To remove the background noise, we employed the PLDCorrector \footnote[3]{http://docs.lightkurve.org/reference/api/lightkurve.correctors.PLDCorrector}. The outliers are removed using the sigma-clipping method, which is provided by $Lightkurve$. Based on the quality flags contained in each individual data cadence, the photometric data points marked with labels such as "Argabrightening", and "Cosmic ray in optimal aperture" were ignored in our study to reduce the noise in the light curves. The long-term instrumental trends such as jumps and drifts are removed by means of a linear or low-order polynomial fit to the light curve. We converted the corrected flux to the magnitude and subtracted its average value for each individual TPF. The consecutive sectors for each star are combined as one light curve. We inspected all the combined light curves to ensure that long period trends between each sector were correctly removed. We used Combined Differential Photometric Precision (CDPP) metric to quantify the noise level of the combined light curves, ensuring that no additional noise is increased in the combined process.

\begin{figure}
    \gridline{\fig{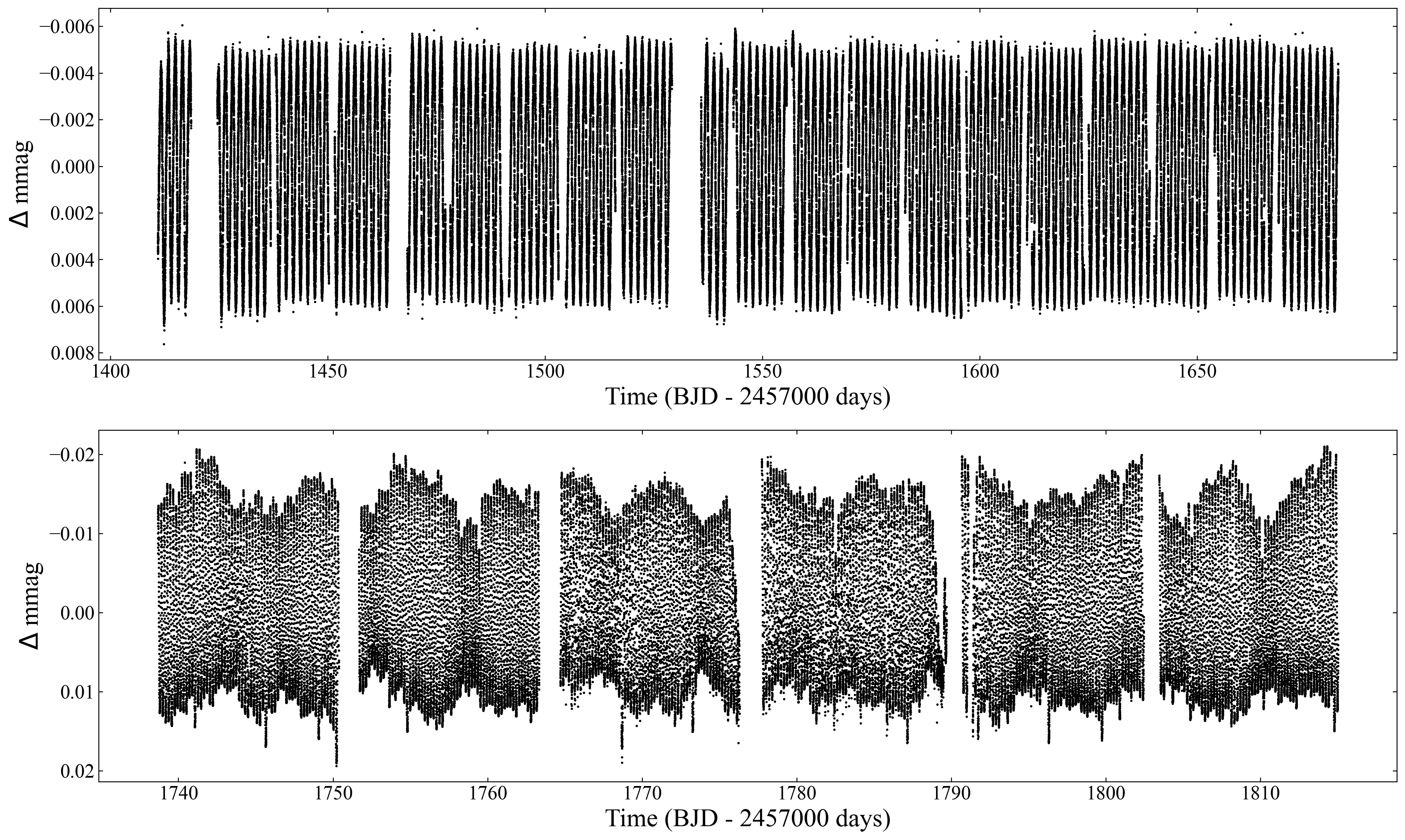}{0.9\textwidth}{}}
    \caption{The light curves of two stars in our sample. The top panel shows the light curve of the binary star HD\;53921, which resulted from combining 10 sectors (sectors 4 to 13). In the bottom panel, we shown the light curve of the pulsating star HD\;205021, which resulted from combining data from 3 sectors (sectors 16 to 18).  \label{ex_lc}}
\end{figure}

In Figure \ref{ex_lc}, we demonstrated our results of the aperture photometry and data reduction using the light curves of HD\;53921 and HD\;205021 as examples. The top panel of the Figure \ref{ex_lc} shows the light curve of the binary star HD\;53921, which resulted from combining 10 sectors (sectors 4 to 13). In the bottom panel of the Figure \ref{ex_lc}, we presented the light curve of the pulsating star HD\;205021, which resulted from combining data from 3 sectors (sectors 16 to 18). From the Figure \ref{ex_lc}, it is evident that the long period trends in the light curves of each sector have been properly removed.

\subsection{Frequency Analysis and Variability Classification}
In order to analyze the coherent pulsations, we used the Discrete Fourier Transform \citep[DFT;][]{1985MNRAS.213..773K} to calculate the amplitude spectra of the combined light curves. The Nyquist frequency refers to the highest frequency that can be correctly resolved by the sampling process. It can be calculated using the $f_{\mathrm{Ny}} = 1/(2\Delta t)$, where $\Delta t$ represents the time interval between consecutive samples. Thanks to the sampling interval of short cadence being that 120 seconds, we determined the $f_{\mathrm{Ny}} = \mathrm{359.9\,d^{-1}}$. The frequencies in the range of $0<f<f_{\mathrm{Ny}}$ are considered in our frequency analysis thanks to they are not undersampled for the sampling rate of TESS short cadence. We extracted the significant frequencies from the amplitude spectra using the software PERIOD 04 \citep{2005CoAst.146...53L}. The parameters of the frequencies derived from the amplitude spectra are calculated and optimized using the following formula:
\begin{equation}\label{eq1}
m = m_0 + \sum_{i}^{N}A_{i}\sin(f_{i}t + \Phi_{i}),
\end{equation} 
where $N$ is the number of the coherent pulsations, $m_0$ is the zero point. $A_i$, $f_i$, and $\Phi_{i}$ represent the amplitude, frequency, and phase of the $i$th coherent pulsation, respectively. According to the study by \cite{2021A&A...656A.158B}, frequencies extracted from space photometry using a signal-to-noise ratio (SNR) criterion of $3.5 < SNR < 4.0$ are indistinguishable from high-amplitude noise. Frequencies between $4.0 \leq SNR \leq 4.6$ have a modest chance of being noise peaks. Therefore, we have adopted a conservative significance threshold of $SNR \geq 4.6$ for determining significant frequencies. \cite{2020A&A...639A..81B} point out that the noise behavior of stars at different evolutionary stages should be accounted for when choosing the window size to calculate the SNR of the significant frequencies. Therefore, we optimized the window size based on the spectral type of each star. We set the window size with a value of 1 d$^{-1}$ for the O-type stars. The window sizes of 5 d$^{-1}$ are set for the early B-type dwarfs, subgiants, and giants (corresponding with the spectral types similar to BV and BVI), while the window size values for the B-type giants, bright giants, and supergiants (corresponding with the spectral types similar to III, II, and I) are set as 1 d$^{-1}$.  For the A-type stars, we set the window size with the value of 2 d$^{-1}$. The window size values of spectroscopic binaries are set at the value of 5 d$^{-1}$. The frequency resolution criterion is defined as $f_{\rm res} = 1.5/T$, where $T$ represents the duration of the observations \citep{1978Ap&SS..56..285L}. If the difference between two frequencies exceeds the $f_{\rm res}$, we conclude that these two frequencies have been resolved. So the combined light curves have the best frequency resolution and afford the greatest precision compared with the light curves derived from the single sector.

Based on visual inspection of the light curves and amplitude spectra, the variability classification of each star is provided. The details of the classification criteria that we applied as follows:
\begin{itemize}
\item[(1)] $\beta$ Cep ($\beta$ Cephei star): coherent $p$ modes with frequencies in the range between 2.5 and 20 d$^{-1}$ \citep{2005ApJS..158..193S, 2020MNRAS.493.5871B};
\item[(2)] SPB (Slowly Pulsating B Star): coherent $g$ modes with frequencies below $\sim$ 2.5 d$^{-1}$ \citep{2020MNRAS.493.5871B, 2020A&A...639A..81B};
\item[(3)] hybrid: a combination of $\beta$ Cep and SPB;
\item[(4)] SLF: stochastic variability with broad excess in amplitude spectrum, has
periods of several hours to days, the amplitudes range from tens to hundreds $\mu$mag \citep{2020A&A...640A..36B};
\item[(5)] rot.: rotational variability caused by spots, stellar wind and/or peculiar abundances of the elements \citep{2017MNRAS.468.2745N, 2019MNRAS.487..304D};
\item[(6)] Multi: eclipsing binary or multiple system;
\item[(7)] cont.: no significant frequency is detected in the amplitude spectrum;
\item[(8)] unknown: thanks to the poor quality data led to we can not identify the classification (i.e. low duty-cycles, and the light curve is contaminated by nearby star or instrumental periodicities).
\end{itemize}

\subsection{Stochastic Low-frequency Variability Analysis \label{Section 3.3}}
\begin{figure}[ht]
    \gridline{\fig{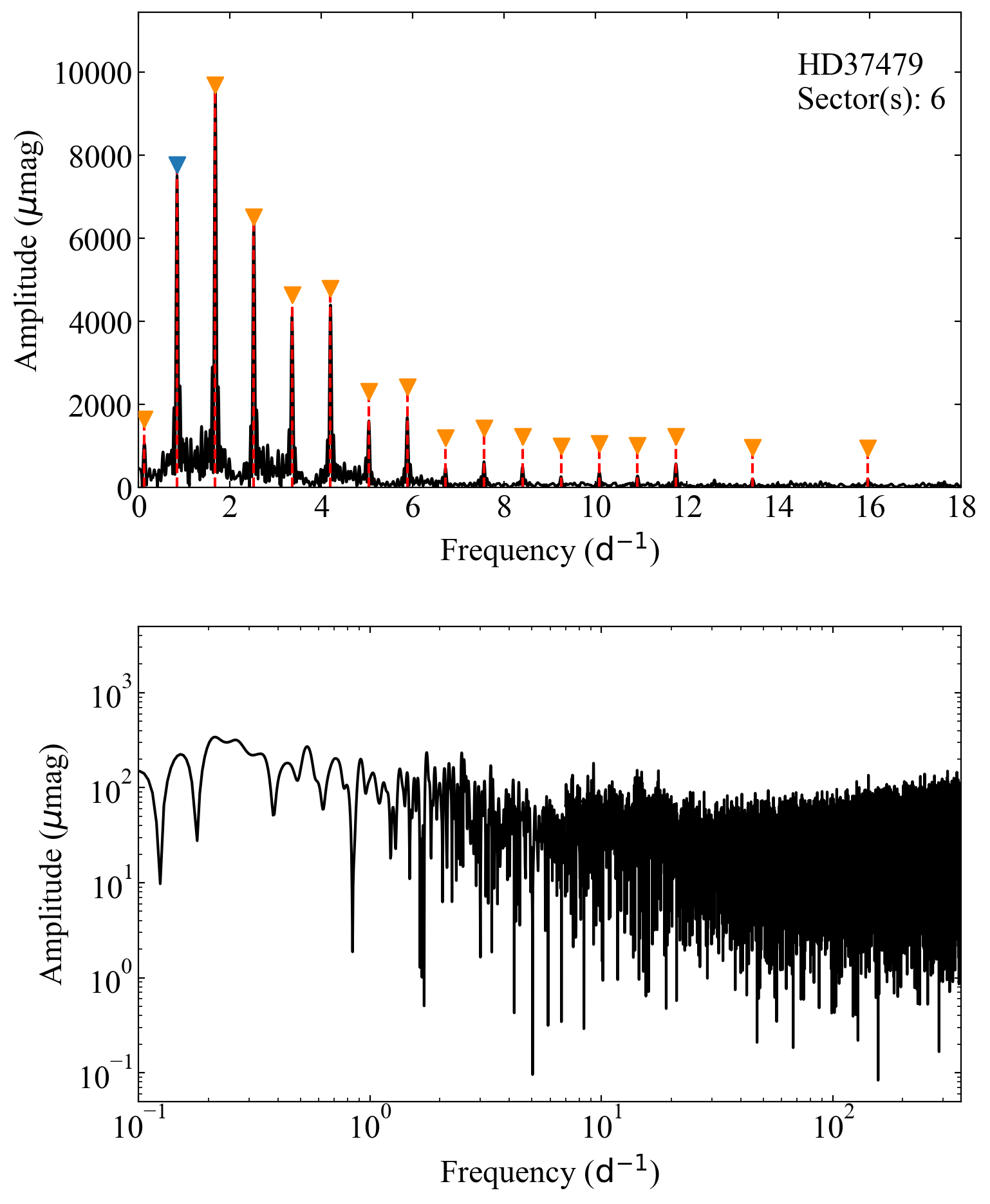}{0.4\textwidth}{}}
    \caption{The frequency analysis result of HD\;37479. In the top panel, the significant peak is marked with a blue inverted triangle, and the harmonics of the dominant frequency are marked with orange inverted triangle. In the bottom panel, the residual amplitude spectrum is shown as the logarithmic diagram. \label{HD37479}}
\end{figure}

The stochastic low-frequency variability is typically manifested in the amplitude spectrum as the red noise component of a stochastic signal \citep[e.g.][]{2019NatAs...3..760B}. To obtain the light curves that only contained the stochastic signal caused by the stochastic low-frequency variability, the coherent pulsations should be removed from the light curves. Inspired by the methodology of \cite{2019A&A...621A.135B}, we employed first a traditional prewhitening for each light curve. The significant peaks are selected and fitted using the eq.(\ref{eq1}). The fitted functions are subtracted from the light curves. Based on the study by \cite{2021A&A...656A.158B}, after removing frequencies with SNR $\geq$ 4.6, the light curves reached the noise level. Therefore, we adopted this criterion (SNR $\geq 4.6$) to identify the significant peaks during the prewhitening process. Using the conservative criterion also ensures that we have reached the noise level. Finally, we derived the residual amplitude spectra from the light curves using the DFT method. In the Figure \ref{HD37479}, the analysis result of coherent pulsation is shown in the top panel. The significant peak is marked with a blue inverted triangle, and the harmonics of the dominant frequency are marked with yellow inverted triangle. In the bottom panel of Figure \ref{HD37479}, the residual amplitude spectrum is shown as the logarithmic diagram. 

It should be noted that one sector TESS light curves are generally sufficient to place constraints on the type of variability, but the low frequency resolution of only one or two sectors makes the identification of rotational splitting, harmonics, and low SNR astrophysical signal difficult, especially in the low frequency regime and in stars with significant SLF variability \citep[e.g.][]{2020A&A...639A..81B, 2020A&A...640A..36B}. For example, the bottom panel of Figure \ref{HD37479} shows the obvious remnant peaks after the prewhitening. The density of the peaks in $0.5 < \mathit{f} < 6\;\text{d}^{-1}$ is extremely high (the top panel of Figure \ref{HD37479}), low frequency resolution of only one sector is not sufficient to resolve the beating and identify close-frequency in this range. As a result, unresolved frequencies appear in the residual amplitude spectrum, compromising the accuracy of the fitting parameters of SLF variability. To address this concern, we utilized combined light curves to improve the frequency resolution in our study.

\begin{figure}[ht]
    \gridline{\fig{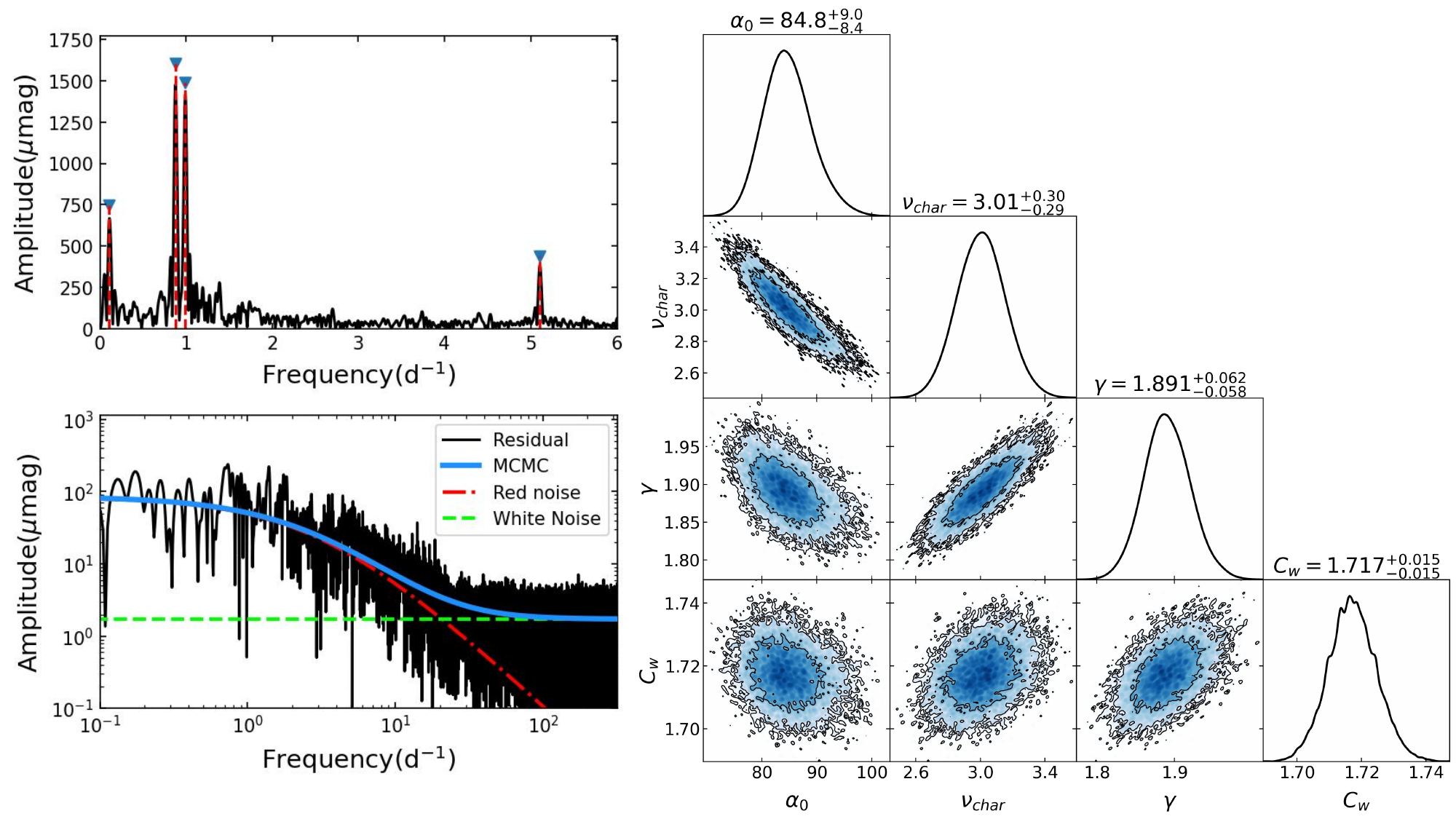}{0.9\textwidth}{}}
    \caption{The analysis results of HD\,121743. The top left panel shows the amplitude spectrum using the DFT method, we marked significant frequencies with blue inverted triangles. The bottom left panel shows the logarithmic amplitude spectrum, and the best-fit result of Eq.$\;$(\ref{eq3}) is shown as a solid blue line. Red dash-dot and green dash lines represent the red noise and white noise components, respectively. The right panel shows the posterior distributions from the MCMC for the parameters of the SLF variability.\label{MCMC}}
\end{figure}

The morphology of the SLF variability in the amplitude spectrum is normally interpreted by the Lorentzian function as the following formula \citep[e.g.][]{2020A&A...640A..36B}:
\begin{equation}\label{eq2}
\alpha(\nu) = \frac{\alpha_{\rm{0}}}{1 + (\frac{\nu}{\nu_{\rm{char}}})^{\gamma}} + C_{\rm{w}},
\end{equation}
where $\alpha_{\rm{0}}$ represents the scaling factor and represents the amplitude at zero frequency, $\gamma$ means the logarithmic amplitude gradient, $\nu_{\rm{char}}$ is the characteristic frequency, and $C_{\rm{w}}$ is a white noise term. Inspired by previous efforts by \cite{2019NatAs...3..760B, 2019A&A...621A.135B, 2020A&A...640A..36B}, in this work, Bayesian Markov Chain Monte Carlo (MCMC) framework with the Python code emcee \citep{2013PASP..125..306F} is applied to derive the morphologies from the residual amplitude spectra as the eq.(\ref{eq2}). Flat priors and 32 parallel chains are used to fit the residual amplitude spectra. In the iteration processing, each parallel chain used to construct a model is evaluated with the likelihood function \citep{1990ApJ...364..699A}:
\begin{equation}\label{eq3}
\ln \mathfrak{L}  = - \sum_{i}^{N} \Bigg\{{\ln [M_{i}(\boldsymbol{\theta})]} + \frac{O_{i}}{[M_{i}(\boldsymbol{\theta})]}\Bigg\},
\end{equation}
where $N$ is the number of total data points contained in the residual amplitude spectrum, $O_{i}$ represents the $i$th point of the residual amplitude spectrum, $M_{i}(\boldsymbol{\theta})$ is the model with parameters $\boldsymbol{\theta}$ of the $i$th point. The criterion from the \cite{1992StaSc...7..457G} is adopted to confirm the convergence of the parameters, which is typically achieved after approximately 1000 iterations. Each star is analyzed by the aforementioned process. As an example, the analysis results of the B2IV star HD\,121743 are shown in Figure \ref{MCMC}. The top left panel of Figure \ref{MCMC} shows the amplitude spectrum using the DFT method, we marked significant frequencies with blue inverted triangles. The bottom left panel of Figure \ref{MCMC} shows the logarithmic amplitude spectrum, and the best-fit result of Eq.$\;$(\ref{eq3}) is shown as a solid blue line. Red dash-dot and green dash lines represent the red noise and white noise components, respectively. We presented the posterior distributions from the MCMC for the parameters of the SLF variability in the right panel of Figure \ref{MCMC}.

\subsection{Astrophysical parameters}
Around half of the stars in our sample have relatively accurate stellar parameters derived from the high resolution spectra \citep[e.g.][]{2012MNRAS.426.2208G, 2015MNRAS.447.2551W, 2019MNRAS.485.1508S}. However, astrophysical parameters are only available for a small number of magnetic chemically peculiar (mCP) stars in our sample that are close to the low-temperature boundary. Due to the abnormal colors of mCP stars, their effective temperatures and luminosities of them cannot be determined using the usual calibrations. We adopted the literature value of the astrophysical parameters, which are corrected by the intensive researches \citep[e.g.][]{2006AA...450..763K, 2008A&A...491..545N}. For the mCP stars that lack of the astrophysical parameters in our sample, we performed the spectral energy distribution (SED) fitting to derive their effective temperatures and luminosities using a wide range of wavelengths. The photometric bands adopted include $Gaia\;GG_{\rm rp}G_{\rm bp}G_{\rm rvs}$ \citep{2022yCat.1355....0G}, $SDSS\;ugriz$ \citep{2015ApJS..219...12A}, $2MASS\;JHK_{S}$ \citep{2006AJ....131.1163S}, and $WISE$ $W_{\rm 1}W_{\rm 2}W_{\rm 3}W_{\rm 4}$ \citep{2010AJ....140.1868W}. We used the ATLAS9 SED Grid \citep{1997A&A...318..841C} for mCP stars with the effective temperatures ranging from 8,000 to 15,000 K and surface gravities log$g$ (cgs) from 3.5 to 5. \cite{2023A&A...674A..27A} have noted that the distance assessments by \cite{2021AJ....161..147B} outperform GSP-Phot distances. Consequently, we have determined the distances to the stars in our sample using the results derived by \cite{2021AJ....161..147B}. We estimated the instellar extinction from the 3D extinction map provided by \cite{2017AstL...43..472G}.
Using the VO Sed Analyzer \citep{2008A&A...492..277B}, we performed the minimum chi-squared method to compare the observational values with SED templates as following:
\begin{equation}\label{eq4}
\chi_r^2=\dfrac{1}{N-n_p}\sum_{i=1}^N\left\{\dfrac{(Y_{i,o}-M_dY_{i,m})^2}{\sigma_{i,o}^2}\right\}
\end{equation}
where $N$ is the number of photometric data points, $n_p$ is the number of the fitted parameters for the model. $Y_{i,o}$ and $\sigma_{i,o}$ represent the the observed flux and observational error in the flux of the $i$th point, respectively. $Y_{i,m}$ is theoretical flux predicted by the model of the $i$th data point. $M_d$ is the dilution factor.

\cite{2008A&A...491..545N} compiled the temperature calibrations for the individual CP groups with the different photometric systems. \cite{2017MNRAS.468.2745N} validated the photometric temperature calibrations concluded by \cite{2008A&A...491..545N}, and pointed out they are consistent with the SED results. In order to investigate the validity of our SED approach, we gathered the magnitudes of Johnson $UBV$ photometric system for the mCP stars, and calculated the astrophysical parameters from the photometric results. We derived the effective temperatures using the Johnson $UBV$ photometric temperature calibrations provided by \cite{2008A&A...491..545N}. The luminosities are derived using the bolometric corrections ($BC_V$) and extinction ($A_\mathrm{V}$) evaluated from the intrinsic colour $(B-V)_{\rm 0}$. The theoretical $BC_V$ values are obtained using the second order fit $BC_V = -5.737+18.685\theta_{\text{eff}}-15.135\theta_{\text{eff}}^2$, where $\theta_{\text{eff}} = 5040/T_{\text{eff}}$ \citep{2008A&A...491..545N}. We used the relationship of $R_V=A_\mathrm{V}/E(B-V) = 3.1$ \citep{1989ApJ...345..245C} to derive the extinction coefficient $A_\mathrm{V}$. Based on the distances ($r$) from the Gaia EDR 3, we obtained the absolute magnitude $M_v = m_V + 5 - 5\text{lg}r$. The luminosity is calculated using $\text{log}(L/L_\odot)=-0.4[M_V-V_\odot-31.572+(\mathrm{BC}_V-\mathrm{BC}_{V,\odot})]$, where $V_\odot = -26.72$ and $\mathrm{BC}_{V,\odot} = -0.19$ are taken from the previous work \citep{2010AJ....140.1158T}.

\begin{figure}[ht]
    \gridline{\fig{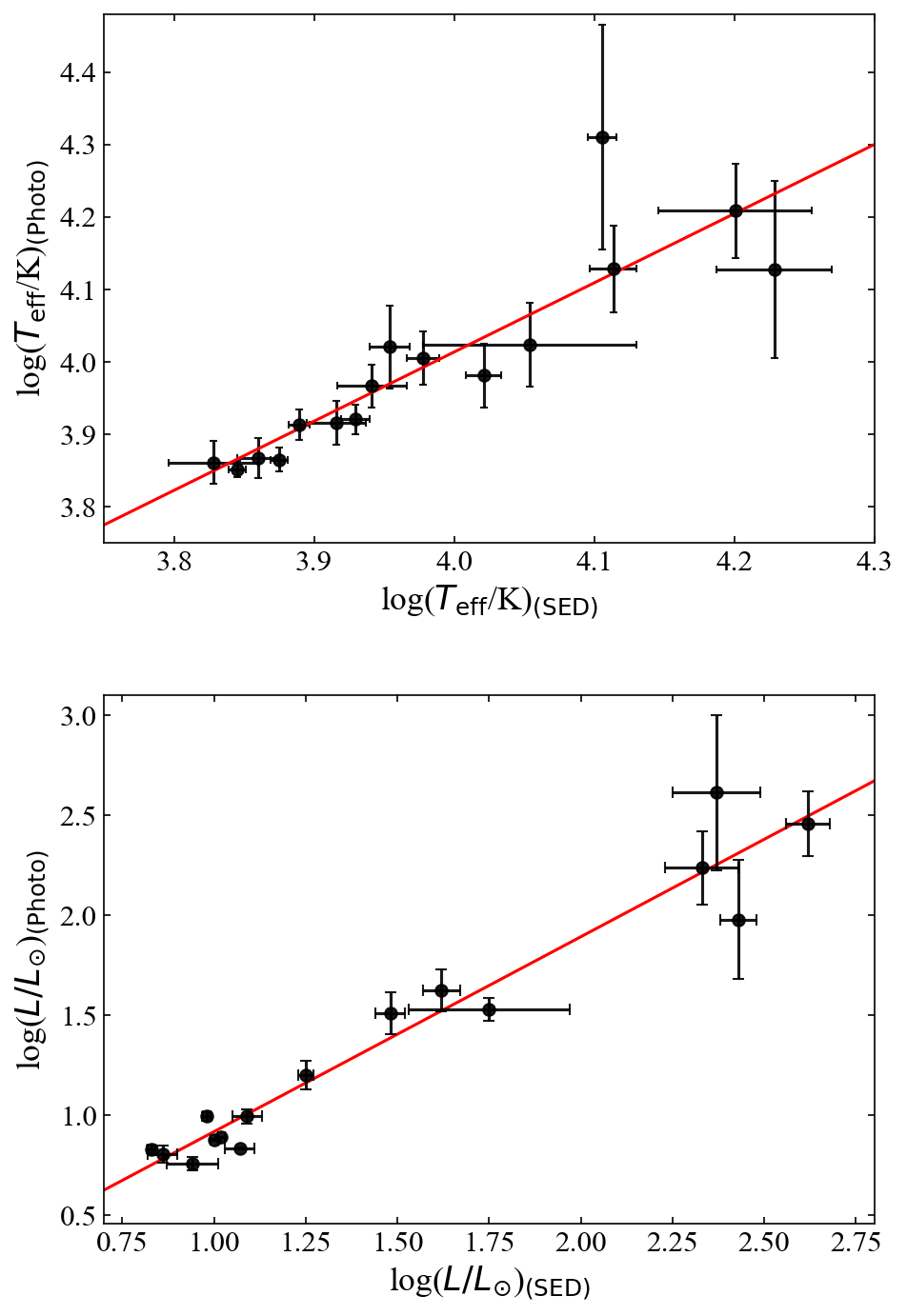}{0.4\textwidth}{}}
    \caption{We compared astrophysical parameters obtained from spectral energy distribution (SED) analysis with those derived from photometry. The top panel presents the effective temperatures derived from both methods, and a linear relation is shown by the red line. In the bottom panel, we compared the luminosities derived from SED analysis with those obtained from photometry, and their correlation is also displayed by the red line. \label{SEDcheek}}
\end{figure}

In the top panel of Figure \ref{SEDcheek}, we presented the effective temperatures derived from the SED analysis and photometry. A linear relation exists between the results based on the SED analysis and those derived from photometry, which can be expressed as $\text{log}(T_\text{eff}/K)_\text{(SED)} = 0.955\text{log}(T_\text{eff}/K)_\text{(Photo)} + 0.194$. Out of the 16 mCP stars, only HD\;146555 shows the discrepancy in the effective temperature that is caused by inappropriate temperature calibrations thanks to the lack of the $U$ band magnitude. In the bottom panel of Figure \ref{SEDcheek}, we compared the luminosities derived from the SED analysis with those obtained from the photometry. Our results indicate that a linear correlation exists between the luminosities obtained using the two methods. This correlation can be expressed as $\text{log}(L/L_\odot)_\text{(SED)} = 0.973\text{log}(L/L_\odot)_\text{(Photo)} - 0.054$. Overall, the results of the SED analysis are consistent with the photometry results, as shown in the comparison of effective temperature and luminosity in Figure \ref{SEDcheek}.

\section{RESULTS}
In our sample, we definitely detected 87 variable stars, 14 constant stars, and 17 unknown variables. All coherent pulsators and multiple systems in our sample are identified by previous studies. 9 new rotating variable stars are detected from our sample. On the other hand, HD\;35912 is classified as a magnetic hot star according to \cite{2005A&A...430.1143B} and \cite{1970ApJ...159..723C}. However, \cite{2018MNRAS.475.5144S} reported an inability to detect a magnetic field in the DAO measurements. Given the inconsistent results from multiple measurements of the magnetic field in HD\;35912, the classification of this star as a magnetic hot star is disputed. As a result, HD\;35912 is removed from our analysis. For the rotating variables, we provided more accurate rotational periods that were only derived via spectroscopy or spectropolarimetry observations. In Table \ref{table: stars}, we summarized the stellar parameters and the classification results for the stars in our sample.

\subsection{Pulsating variables}
Our sample contains 10 coherent pulsators, namely: HD\;121743, HD\;136504, HD\;156424, HD\;205021, HD\;3360, HD\;35912, HD\;43317, HD\;44743, HD\;46328, and HD\;96446. Among them, HD 136504 and HD 156424 are components of the binary systems. HD\;3360 was previously identified as a $\beta$ Cep star in earlier studies \citep[e.g.][]{1989nos..book.....U, 2006A&A...452..945T}. However, our results suggest that HD\;3360 is actually a hybrid star, rather than a $\beta$ Cep star.

\subsubsection{HD\;3360}
HD\;3360 (TIC 240669906) is known as a nitrogen-rich B2IV star \citep[e.g.][]{1992ApJ...387..673G, 2003A&A...406.1019N, 2012A&A...539A.143N}. \cite{2003A&A...406.1019N} discovered the presence of the magnetic fields in the HD\;3360 using the Musicos spectropolarimeter at T$\acute{\rm e}$lescope Bernard Lyot (TBL). The fundamental parameters of HD\;3360 have been determined by the high-resolution spectroscopy in various studies \cite[e.g.][]{2007CoAst.150..183B, 2010A&A...515A..74L, 2012A&A...539A.143N, 2014A&A...566A...7N}. The spectroscopic parameters of HD\;3360 include an effective temperature of $T_{\rm eff} = 20.8 \pm 0.2 k\rm{K}$, a luminosity of $\text{log}(L/L_\odot) = 3.82 \pm 0.06$ and a surface gravity of $\text{log} g = 3.80 \pm 0.05$. \cite{2012A&A...539A.143N} determined the microturbulence velocity of $\xi = 2 \pm 1\;km\;s^{-1}$, the projected rotational velocity of $v\text{sin}i = 20 \pm 2\;km\;s^{-1}$ and macroturbulence velocity of $\zeta = 12 \pm 5\;km\;s^{-1}$ of the HD\;3360.

\begin{figure}[ht]
    \gridline{\fig{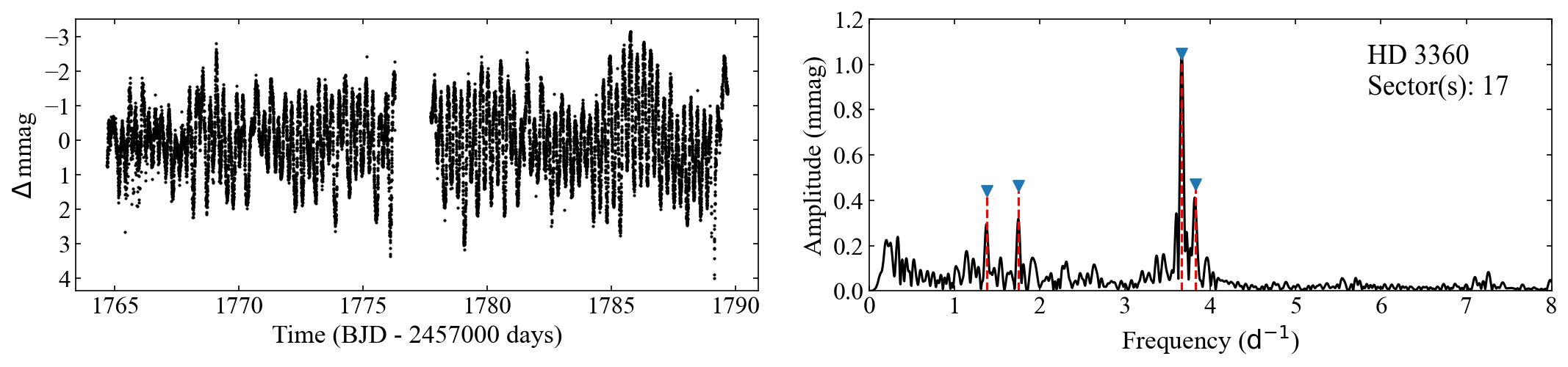}{1\textwidth}{}}
    \caption{The light curve and frequency spectrum of HD\;3360. The light curve of HD\;3360 is shown in the left panel, and the corresponding frequency spectrum is shown in the right panel. No significant variability is observed beyond 8 $\mathrm{d}^{-1}$. In the frequency spectrum, the significant peaks are marked with a blue inverted triangle.\label{pulsating_stars}}
\end{figure}

The non-radial pulsation mode of $l = 2 \pm 1$ at the frequency $f = 0.64\;\text{d}^{-1}$ is detected in the radial velocity amplitude variations \textbf{by} \citep{2003A&A...406.1019N}. The rotational period ($P_{\rm rot} = 5.370447 \pm 0.000078\;\text{days}$) of the HD\;3360 was obtained through a detailed study of time-resolved equivalent width measurements of the C$_{\rm \;IV}$, Si$_{\rm \;IV}$ and N$_{\rm \;V}$ lines, also conducted by \citep{2003A&A...406.1019N}. The light curve and amplitude spectrum of HD\;3360 are presented in the Figure \ref{pulsating_stars}. We detected four significant frequencies in the amplitude spectrum of HD\;3360. The rotational period is not present in the amplitude spectrum. Out of four significant frequencies, two frequencies below 2.5 $\rm d^{-1}$, while the other frequencies are close to 4 $\rm d^{-1}$. These results suggest that there are two types (corresponding to $g$-mode and $p$-mode) of coherent pulsations in HD\;3360. Therefore, the HD\;3360 should be classified as a hybrid star rather than an SPB star.

\subsection{Rotating stars}
Mainly rotating stars in our sample are members of magnetic chemically peculiar (mCP) stars. They generally have slowly rotating velocities, and atmospheric abundances abnormities of some chemical elements due to the effects of radiative diffusion and gravitational settling \citep{1974ARA&A..12..257P,2010A&A...516A..53A,2015MNRAS.454.3143A}. The 54 out of the 62 rotational variables are known as rotating variables prior to our study, we confirmed their variabilities and rotational periods. Additionally, we have identified 8 new rotational variables in our sample: CPD-62\;2124, HD\;89103, HD\;118913, HD\;119308, HD\;127453, HD\;138758, HD\;157751, and HD\;172690. By combining the spectra observed by previous studies with the $TESS$ light curve, we clarified that HD\;67621 is a rotating variable, rather than a young stellar object (YSO). The fundamental frequencies of mCP stars in our sample are distributed in the range of 0.1 $\sim$ 1.6 d$^{-1}$. We determined their rotational periods, rotational modulation, and/or coherent pulsations. The light curves and amplitude spectra of new rotating variables are presented in Figure \ref{rotating_stars} and Figure \ref{rotating_stars2}.

\subsubsection{CPD-62\;2124}
CPD-62\;2124 (TIC 290796936) is an evolved, He-rich star with a spectral type of Of?p \citep{2017A&A...597L...6C}. \cite{2017A&A...597L...6C} reported the detection of an exceptionally strong magnetic field in CPD-62\;2124 with a strength of approximately 5.2 kG, as measured by the FORS2 low-resolution spectropolarimeter and HARPSpol spectropolarimeter. CPD-62\;2124 has an effective temperature of $T_{\rm eff} = 23.6 \pm 0.2 \text{kK}$, a luminosity of $\text{log}(L/L_\odot) = 3.8 \pm 0.2$, a surface gravity of $\text{log} g = 4.05 \pm 0.10$, and a rotational period $P_{\rm rot} = 2.62809 \pm 000005\;\text{days}$ \citep{2017A&A...597L...6C,2019MNRAS.485.1508S}. Using the stellar evolutionary models, these parameters, including the mass $M_{\star} = 10.0 \pm 0.4\;M_{\odot}$, radius $R_{\star} = 5.6\;R_{\odot}$ are determined by \cite{2011A&A...530A.115B} and \cite{2012A&A...537A.146E}. 

The light curve and amplitude spectrum of CPD-62\;2124 are shown in the middle panel of Figure \ref{rotating_stars}. We detected three significant frequencies and a harmonic frequency in the amplitude spectrum. The most significant frequency ($\mathit{f} = 0.3810113 \pm 0.0000027\;\text{d}^{-1}$) is corresponding with the rotational frequency of the CPD-62\;2124. Therefore, we determined the rotational period with the value of $2.62459 \pm 0.00002 \;\text{days}$, which slightly smaller than the result derived from the spectropolarimetry. In previous study, \cite{2020MNRAS.499.5379S} determined the log$R_\text{A}/R_{\rm K} = 0.4 \pm 0.2$, where $R_\text{A}$ is the Alfv$\acute{\rm e}$n radius, and $R_{\rm K}$ is the $Kepler$ radius. The fact that $R_\text{A}>R_{\rm K}$ suggests that CPD-62\;2124 has a centrifugal magnetosphere, which indecates that the rotationally modulated light curve of  CPD-62\;2124 is caused by the material trapped in closed magnetic loops. Additionally, other significant frequencies indicate that CPD-62\;2124 has coherent pulsations, possibly due to the Z-bump effect \citep{1993MNRAS.260..465S}, or other physical processes.

\subsubsection{HD\;67621}
HD\;67621 (TIC 238489837) is a B2IV star, and is a member star of the Vel OB2 association determined by \cite{1999AJ....117..354D}. \cite{2014A&A...567A..28A} reported the presence of magnetic fields, and determined the stellar parameters of HD\;67621 include an effective temperature of $T_{\rm eff} = 21.0 \pm 0.6\;\text{kK}$, a surface gravity of $\text{log} g = 4.18 \pm 0.10$ and a luminosity of $\text{log}(L/L_\odot) = 3.3 \pm 0.1$. \cite{2018MNRAS.475.5144S} confirmed that HD\;67621 has a dipolar magnetic field with the strength of $0.160 \pm 0.004\;\text{kG}$. HD\;67621 has a rotational period $P_{\rm rot} = 3.593 \pm 0.001\;\text{days}$ \citep{2014A&A...567A..28A, 2018MNRAS.475.5144S}. 

The light curve and amplitude spectrum of the HD\;67621 is shown in the second row of Figure \ref{rotating_stars}. We detected a significant frequency and 2 harmonics. The rotational period $P_{\rm rot} = 3.50695 \pm 0.00006\;\text{days}$ is deduced from the significant frequency. The rotational period determined by us is slightly shorter than the results derived by the \cite{2014A&A...567A..28A} and \cite{2018MNRAS.475.5144S}. \cite{2019A&A...621A.115C} determined the HD\;67621 as a YSO using the color-magnitude-diagram (CMD). This determination may have been influenced by the abnormal color index of HD\;67621, which is affected by its chemical peculiarities. However, by combining the $TESS$ light curve with data from previous studies \citep{2014A&A...567A..28A, 2018MNRAS.475.5144S}, we determined that HD\;67621 is a rotating variable. In \cite{2018MNRAS.475.5144S}, the authors attempted to constrain the period using $Hipparcos$ photometry. However, the almost complete absence of variability in the data resulted in a light curve that was dominated by noise. In contrast, the light curve obtained from $TESS$ clearly shows variations, demonstrating the great potential of $TESS$ for detecting variability in magnetic hot stars.

\subsubsection{HD\;89103}
HD\;89103 (TIC 219614759) is a B9V star \citep{2015A&A...583A.115B}. Based on data from the FORS1 archive, \cite{2015A&A...583A.115B} reported the presence of magnetic fields in HD\;89103 with the strength of $\langle B_{z}\rangle = 2.14 \pm 0.07 \;\text{kG}$. The only literature has provided the fundamental parameters including an effective temperature of $\text{log}(T_{\rm eff}/\text{K}) = 4.07 \pm 0.02$ a luminosity of $\text{log}(L/L_\odot) = 1.67 \pm 0.12$, and a mass of $M_{\star} = 2.68 \pm 0.11 M_{\odot}$ \citep{2006AA...450..763K}.

The light curve and amplitude spectrum of HD\;89103 are presented in the third row of Figure \ref{rotating_stars}. 2 significant frequencies and 2 harmonics are detected in the amplitude spectrum. Using the most significant frequency ($\mathit{f} = 0.4600395 \pm 0.0000033\;\text{d}^{-1}$), and using this value, we determined the rotational period of HD\;89103 to be $P_{\rm rot} = 2.17242 \pm 0.00002\;\text{days}$.

\subsubsection{HD\;118913}
HD\;118913 (TIC 342995661) is an A0V star \citep{2015A&A...583A.115B}. \cite{2015A&A...583A.115B} cataloged HD\;118913 as an mCP star, which has the strength of $\langle B_{z}\rangle = 0.52 \pm 0.03 \;\text{kG}$. The only literature has provided the fundamental parameters include an effective temperature of $\text{log}(T_{\rm eff}/\text{K}) = 3.98 \pm 0.01$ a luminosity of $\text{log}(L/L_\odot) = 1.67 \pm 0.14$, and a mass of $M_{\star} = 2.40 \pm 0.14 M_{\odot}$ \citep{2006AA...450..763K}.

The light curve and amplitude spectrum of HD\;118913 are presented in the fourth row of Figure \ref{rotating_stars}. We detected 4 significant frequencies and one harmonic frequency in the amplitude spectrum. We deduced the rotational period $P_{\rm rot} = 2.45892 \pm 0.00002\;\text{days}$ using the frequency of $\mathit{f} = 0.40668 \pm 0.00002\;\text{d}^{-1}$. The presence of three additional significant frequencies suggests that HD\;118913 is not only a rotating variable, but also exhibits coherent pulsations.

\subsubsection{HD\;119308}
HD\;119308 (TIC 305716720) is a A0V star \citep{2015A&A...583A.115B}. The star has a dipole magnetic field with strength of $\langle B_{z}\rangle = -0.29 \pm 0.05 \;\text{kG}$ \citep{2015A&A...583A.115B}. The only literature has provided the fundamental parameters of HD\;119308 include an effective temperature of $\text{log}(T_{\rm eff}/\text{K}) = 4.01 \pm 0.01$, a luminosity of $\text{log}(L/L_\odot) = 1.48 \pm 0.15$, and a mass of $M_{\star} = 2.30 \pm 0.11 M_{\odot}$ for this star \citep{2006AA...450..763K}.

The light curve and amplitude spectrum of HD\;119308 are presented in the fifth row of Figure \ref{rotating_stars}. From the amplitude spectrum, we detected 1 significant frequencies and 2 harmonics. The rotational period of $P_{\rm rot} = 1.90666 \pm 0.00002\;\text{days}$ is derived from the most significant frequency of $\mathit{f} = 0.524476 \pm 0.000005\;\text{d}^{-1}$. 

\subsubsection{HD\;127453}
HD\;127453 (TIC 446846414) is a B8V star \citep{2015A&A...583A.115B}. The star has a magnetic fields with the strength of $\langle B_{z}\rangle = -0.26 \pm 0.07 \;\text{kG}$ \citep{2015A&A...583A.115B}. Using the photometric data, \cite{2006AA...450..763K} provided the fundamental parameters include an effective temperature of $\text{log}(T_{\rm eff}/\text{K}) = 4.08 \pm 0.01$, a luminosity of $\text{log}(L/L_\odot) = 2.25 \pm 0.17$, and a mass of $M_{\star} = 3.33 \pm 0.23 M_{\odot}$ for this star.

\begin{figure}[ht]
    \gridline{\fig{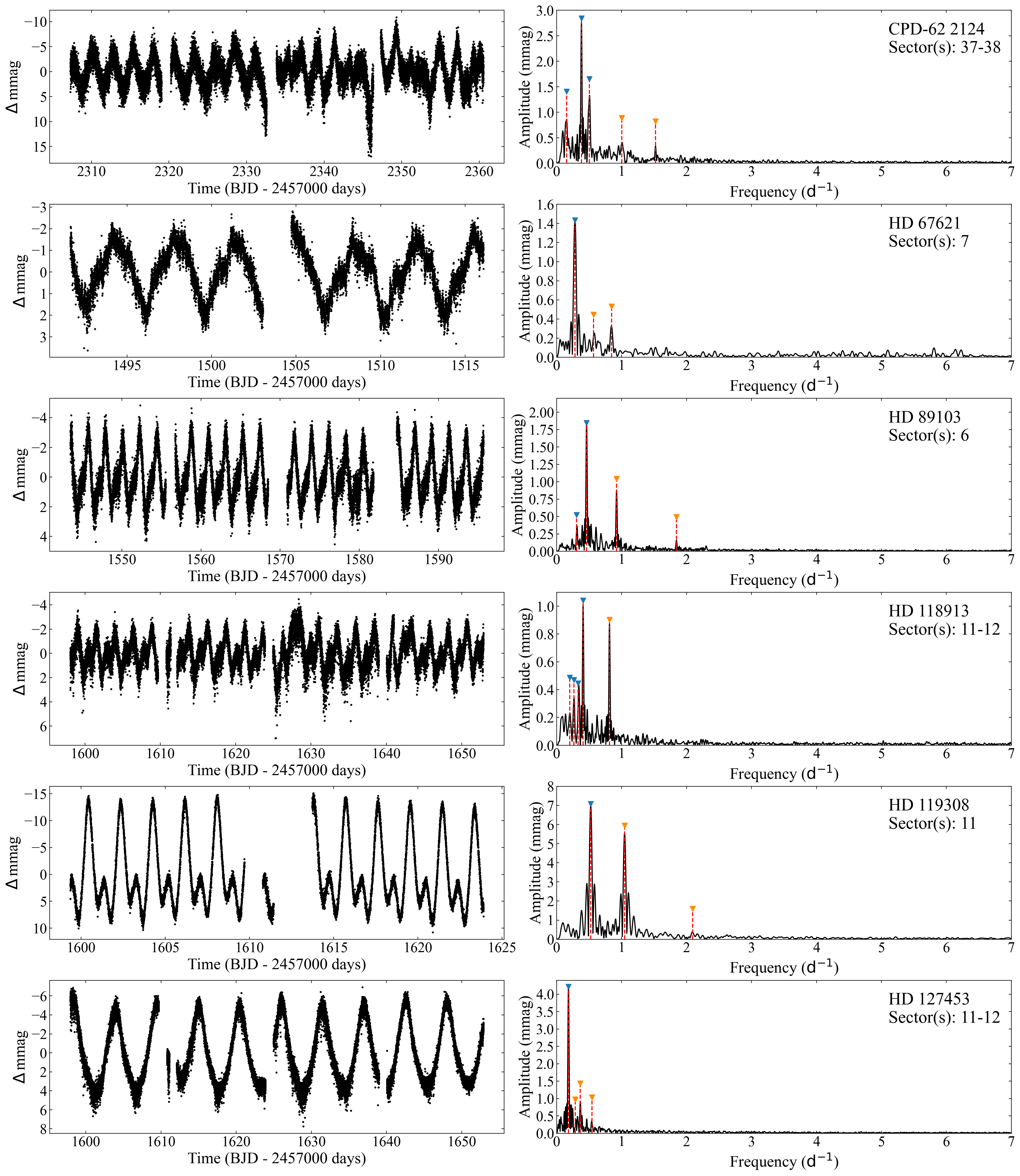}{1\textwidth}{}}
    \caption{The light curves (left panels) and amplitude spectra (right panels) for CPD-62\;2124, HD\;67621, HD\;89103, HD\;118913, HD\;119308, and HD\;127453. In the frequency spectra, the significant peaks are marked with a blue inverted triangle, and the harmonics of the dominant frequency are marked with orange inverted triangles. \label{rotating_stars}}
\end{figure}

The light curve and amplitude spectrum of HD\;127453 are presented in the fifth row of Figure \ref{rotating_stars}. One significant frequency and 3 harmonics are detected in the amplitude spectrum. The rotational period of $P_{\rm rot} = 5.50959 \pm 0.00008\;\text{days}$ is derived from the significant frequency of $\mathit{f} = 0.1815015 \pm 0.0000026\;\text{d}^{-1}$. No coherent pulsation is detected in the amplitude spectrum.

\subsubsection{HD\;138758}
HD\;138758 (TIC 425871448) is a main sequence star with the B9 spectral type \citep{2015A&A...583A.115B}. Using the FORS1 archive data, \cite{2015A&A...583A.115B} reported the presence of magnetic fields in HD\;138758. Using the photometric data, \cite{2006AA...450..763K} provided the fundamental parameters include an effective temperature of $\text{log}(T_{\rm eff}/\text{K}) = 4.02 \pm 0.02$, a luminosity of $\text{log}(L/L_\odot) = 1.68 \pm 0.17$, and a mass of $M_{\star} = 2.51 \pm 0.15 M_{\odot}$ for this star.

The light curve and amplitude spectrum of HD\;138758 is shown in the top row of the Figure \ref{rotating_stars2}. From the amplitude spectrum, we detected a significant frequency and a harmonic. The rotational period $P_{\rm rot} = 1.100073 \pm 0.000005\;\text{days}$ is determined from the significant frequency of $\mathit{f} = 0.909030 \pm 0.000004\;\text{d}^{-1}$.

\subsubsection{HD\;157751}
HD\;157751 (TIC 160367734) has a B9V spectral type, and a large-scale magnetic fields with the strength of $\langle B_{z}\rangle = 4.10 \pm 0.08 \;\text{kG}$ \citep{2015A&A...583A.115B}. Based on the calibration of the Geneva photometric system, \cite{2006AA...450..763K} provided the stellar parameters include an effective temperature of $\text{log}(T_{\rm eff}/\text{K}) = 3.99 \pm 0.01$, a luminosity of $\text{log}(L/L_\odot) = 1.38 \pm 0.15$, and a mass of $M_{\star} = 2.18 \pm 0.10 M_{\odot}$ for this star. 

The light curve and amplitude spectrum of HD\;157751 is shown in the middle row of the Figure \ref{rotating_stars2}. Our frequency analysis detected one significant frequency and three harmonics in the spectrum. The significant frequency $\mathit{f} = 0.370186 \pm 0.000004\;\text{d}^{-1}$, corresponds to the rotational frequency, allowing us to determine a rotational period of $P_{\rm rot} = 2.70134 \pm 0.00003\;\text{days}$ for HD\;157751.  

\subsubsection{HD\;172690}
HD\;172690 (TIC 469692862) has a A0V spectral type, and a dipole magnetic field with the strength of $\langle B_{z}\rangle = -0.25 \pm 0.08 \;\text{kG}$ \citep{2015A&A...583A.115B}. Based on the calibration of the Geneva photometric system, \cite{2006AA...450..763K} provided the stellar parameters include an effective temperature of $\text{log}(T_{\rm eff}/\text{K}) = 4.06 \pm 0.02$, a luminosity of $\text{log}(L/L_\odot) = 2.05 \pm 0.15$, and a mass of $M_{\star} = 3.00 \pm 0.18 M_{\odot}$ for this star. 

\begin{figure}
    \gridline{\fig{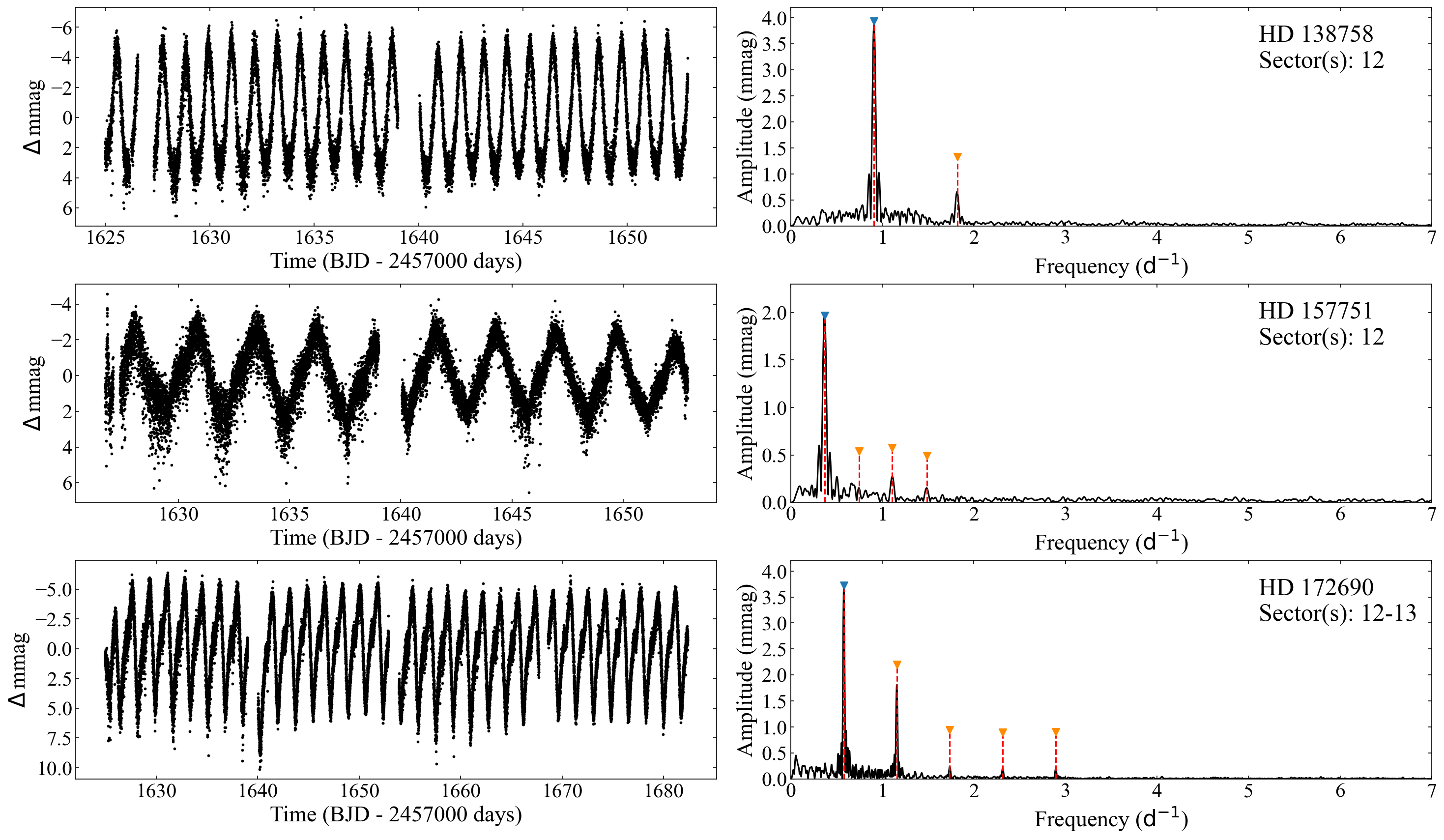}{1\textwidth}{}}
    \caption{The light curves (left panels) and amplitude spectra (right panels) for HD\;138758, HD\;157751, and HD\;172690. The markers in the amplitude spectra have the same meaning as in Figure \ref{rotating_stars}. \label{rotating_stars2}}
\end{figure}

The light curve and amplitude spectrum of HD\;172690 is shown in the bottom row of the Figure \ref{rotating_stars2}. We detected one significant frequency and four harmonics in the amplitude spectrum of the HD\;172690. The most significant frequency of $\mathit{f} = 0.5786305 \pm 0.0000028\;\text{d}^{-1}$ corresponds to the rotational frequency, allowing us to determine a rotational period of $P_{\rm rot} = 1.72822 \pm 0.000008\;\text{days}$ for HD\;172690.

\subsection{Multiple systems and Other types}
\begin{figure}
    \gridline{\fig{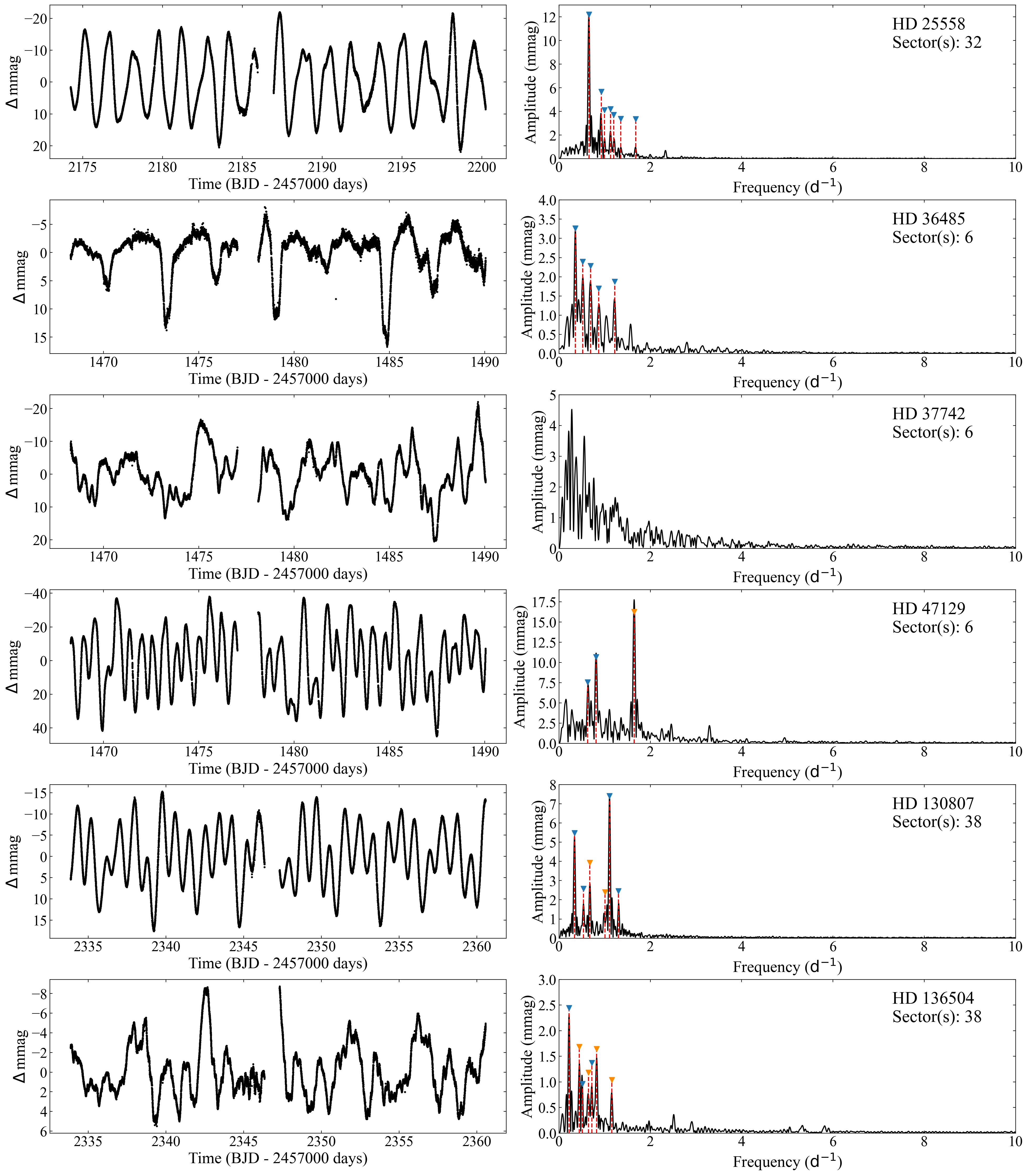}{1\textwidth}{}}
    \caption{The light curves (left panels) and amplitude spectra (right panels) of binary stars. The markers in the amplitude spectra have the same meaning as in Figure \ref{rotating_stars}. \label{binaries}}
\end{figure}

Our sample contains 16 multiple systems, the majority of which have been confirmed as spectroscopic binaries. Among these, 8 multiple systems do not exhibit obvious eclipses in their light curves. HD\;25558, HD\;36485, HD\;37742, HD\;47129, HD\;130807, and HD\;136504 exhibit complex variations in their light curves. We presented the light curves and amplitude spectra for these systems in Figure \ref{binaries}, and our analysis suggests that continued spectroscopic monitoring will advance the discovery of magneto-asteroseismic phenomena in binary stars.

Among the variables in our sample, we were unable to definitively determine the variability of 17 stars. CPD-60 944B and HD\;66318 are severely contaminated by nearby stars, making it impossible to obtain their light curves. HD\;101412 is reported to be a Herbig Ae star \citep{2010AN....331..361H}. The observed variations in its light curve are likely caused by chemical spots and magnetospheric accretion \citep{2016A&A...592A..50S, 2016AN....337..329J}. HD\;200775 is a well-known Herbig Be star \citep{2008MNRAS.385..391A, 2013A&A...555A.113B, 2020MNRAS.494.5851S}, and we detected a significant peak in its amplitude spectrum. However, without additional photometry and spectroscopy data, we are unable to determine the classifications of the remaining stars.

\subsection{Stochastic Low-Frequency (SLF) variability}

\begin{figure*}
    \gridline{\fig{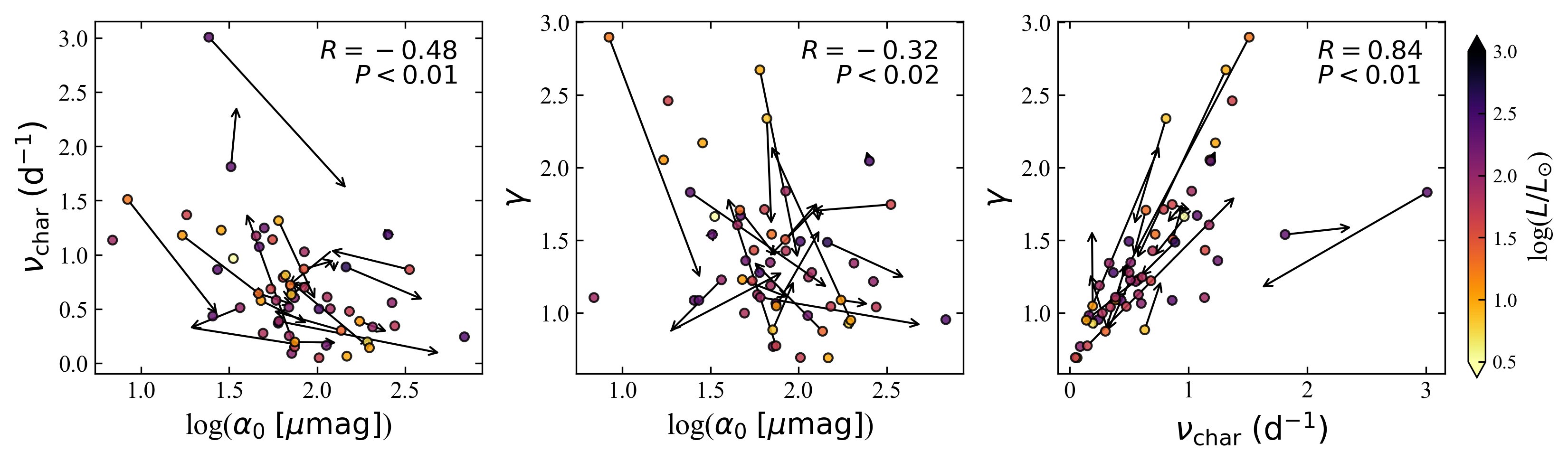}{1\textwidth}{}}
    \gridline{\fig{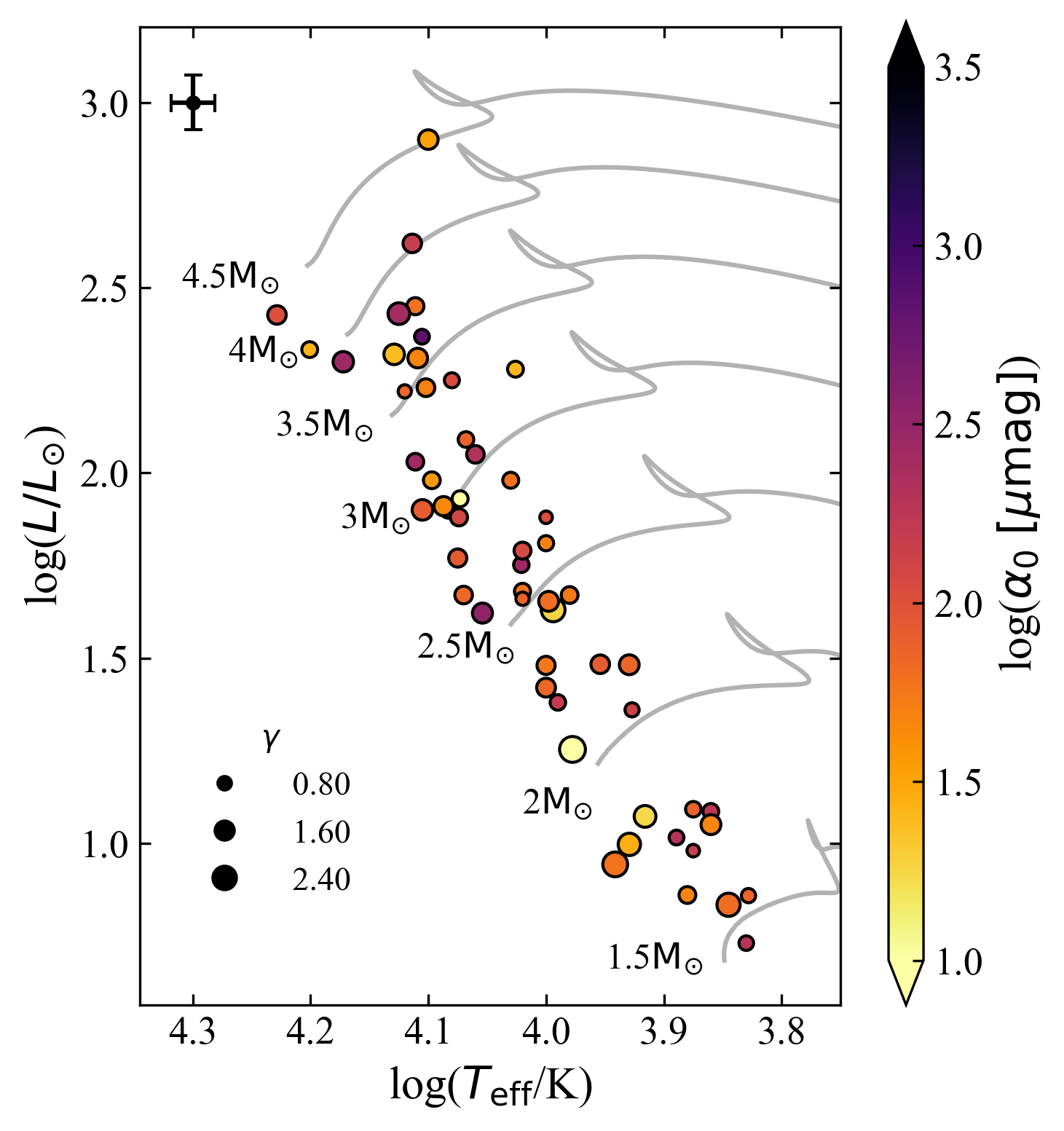}{0.46\textwidth}{}
    \fig{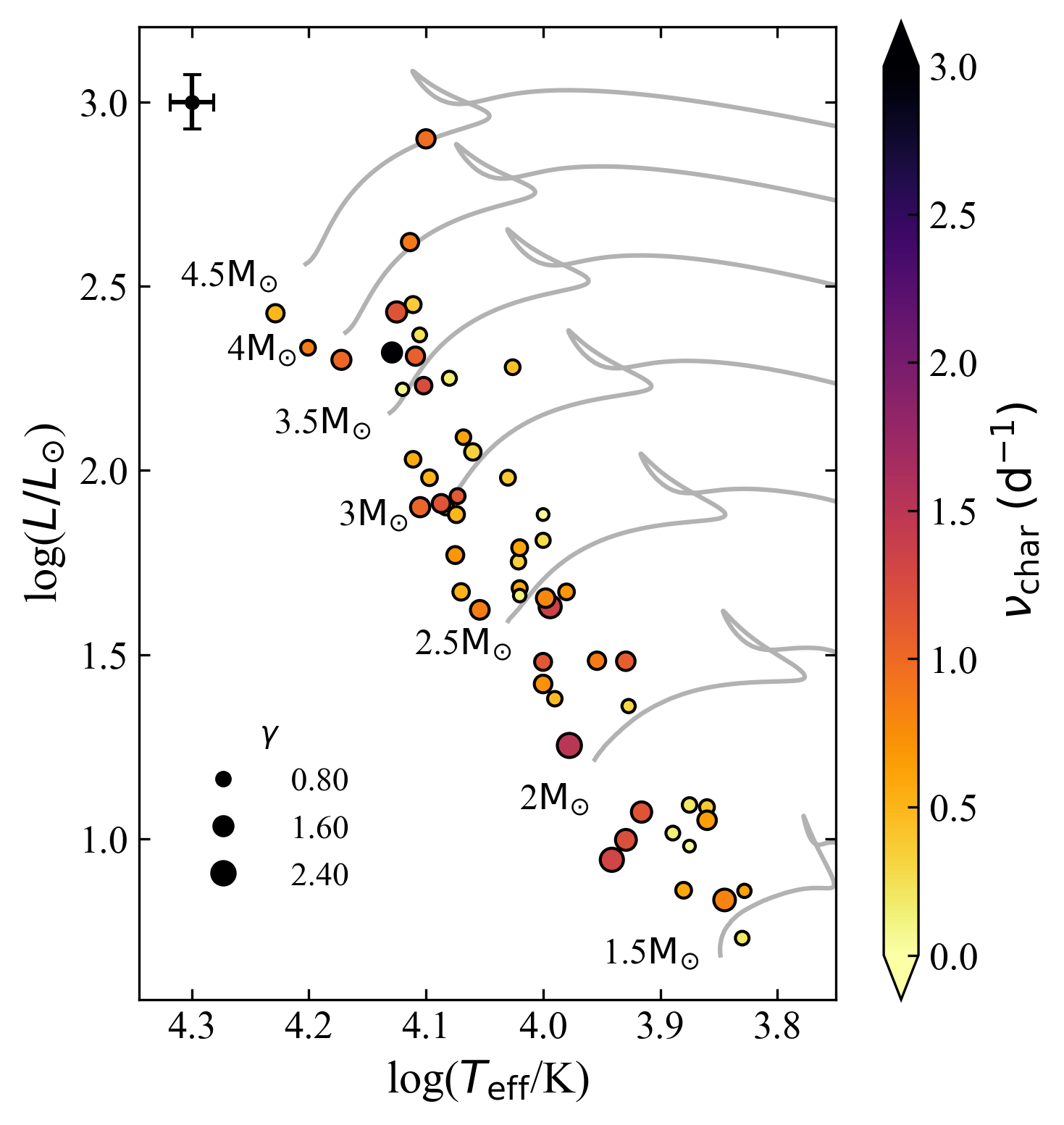}{0.46\textwidth}{}}
    \caption{The SLF variability analysis results of Ap/Bp stars in our sample. The pair-wise relationships between the fitting parameters, $\alpha_{0}$, $\nu_{char}$ and $\gamma$ in Eq. (\ref{eq2}) are presented in top row. Each star is color-coded by the luminosity. The arrows extending from the single colored circle indicate the direction of variation in the fitting parameters of the SLF variabilities, revealing trends in the fitting parameters across the different sectors. Location of the stars within our sample in the HR diagram are presented in bottom row. Each star is color-coded by the fitting parameters $\alpha_{0}$ (left panel) and $\nu_{\rm char}$ (right panel). The symbol sizes are proportional to the $\gamma$ values. Evolutionary tracks are plotted as grey lines. A typical error bar is shown in the top-left corner. \label{fig:apbp_SLF_HRD}}
\end{figure*}

\begin{figure*}
    \gridline{\fig{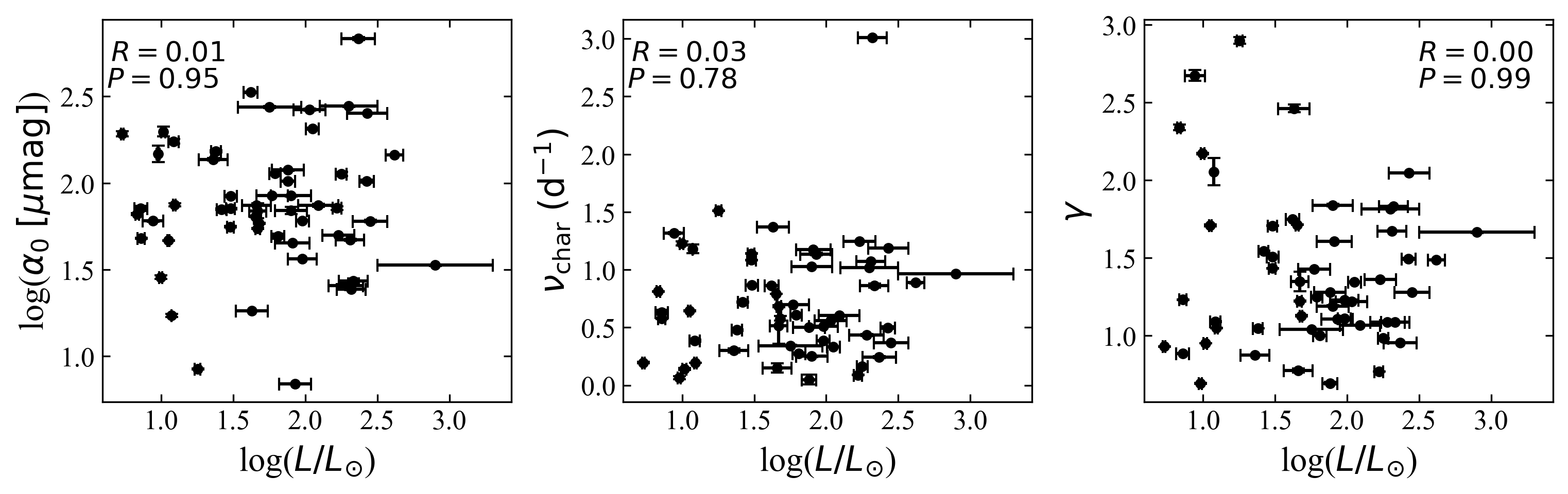}{1\textwidth}{}}
    \caption{The pair-wise relationships between the luminosities and fitting parameters(log$\alpha_{\rm 0}$, $\nu_{\rm char}$, and $\gamma$) for our sample of Ap/Bp stars. The Spearman rank-order correlation coefficients and p-values are provided in the figures accordingly.\label{fig:apbp_logl_SLF_HRD}}
\end{figure*}

\begin{figure*}
    \gridline{\fig{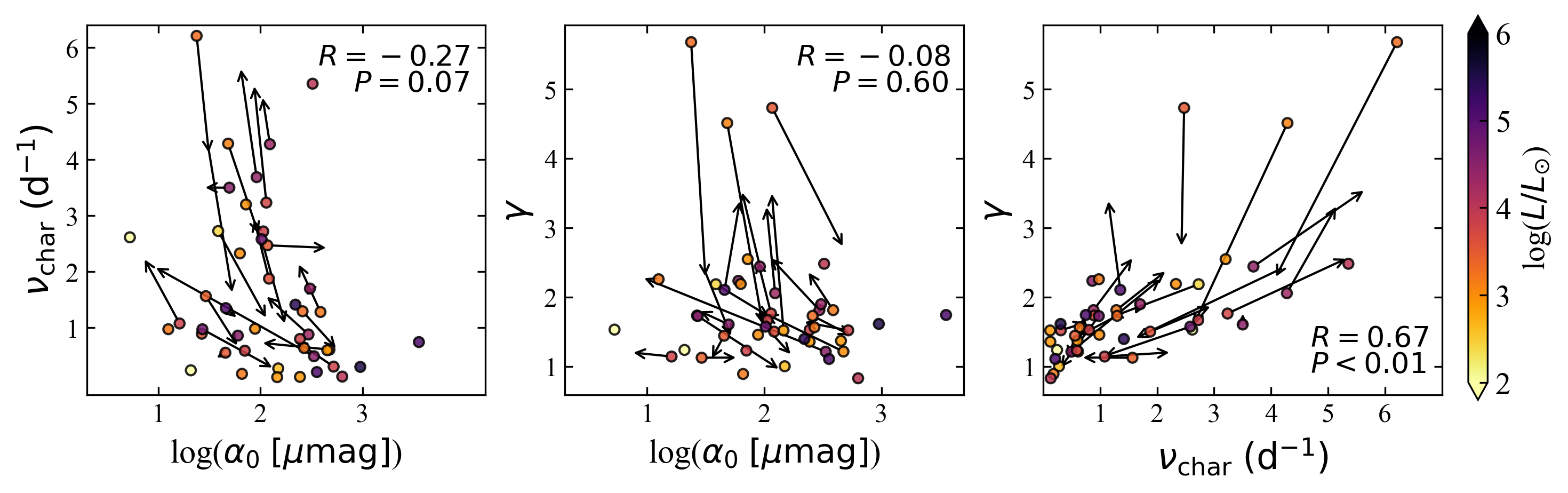}{1\textwidth}{}}
    \gridline{\fig{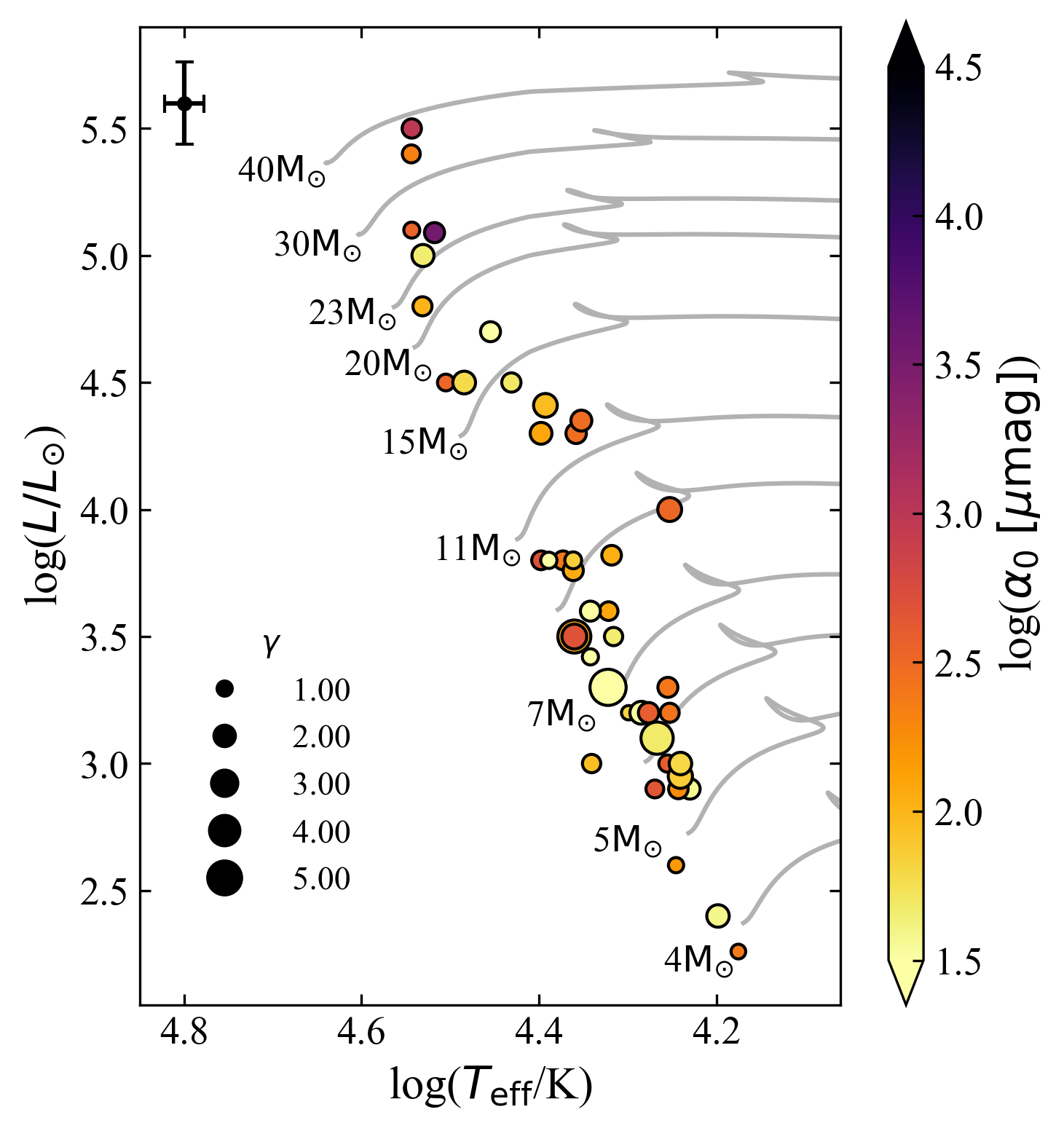}{0.46\textwidth}{}
    \fig{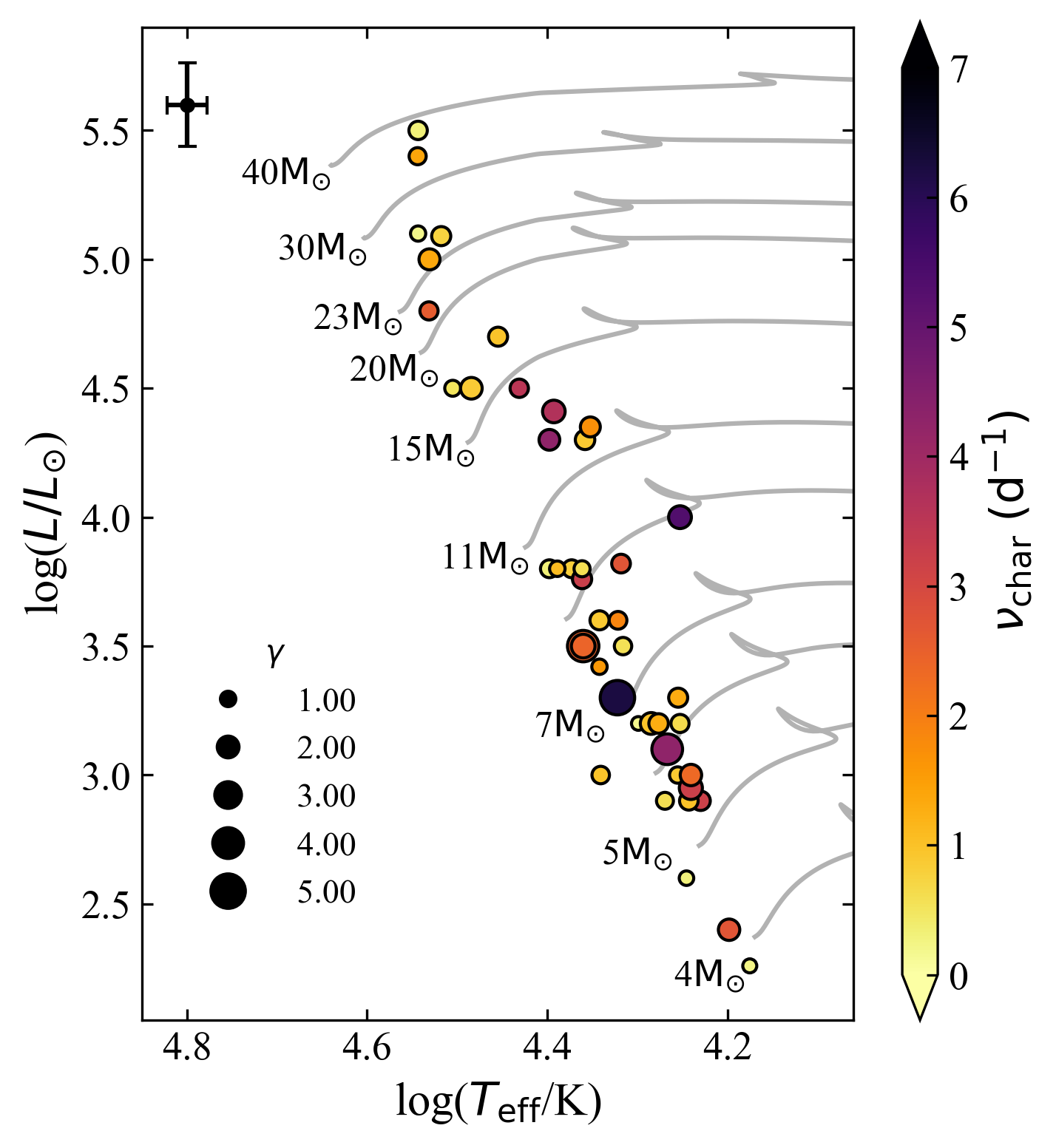}{0.45\textwidth}{}}
    \caption{The SLF variability analysis results of upper main-sequence stars in our sample. The pair-wise relationships between the fitting parameters, $\alpha_{0}$, $\nu_{char}$ and $\gamma$ in Eq. (\ref{eq2}) are presented in top row. Each star is color-coded by the luminosity. The arrows extending from the single colored circle indicate the direction of variation in the fitting parameters of the SLF variabilities, revealing trends in the fitting parameters across the different sectors. Location of the stars within our sample in the HR diagram are presented in bottom row. Each star is color-coded by the fitting parameters $\alpha_{0}$ (left panel) and $\nu_{\rm char}$ (right panel). The symbol sizes are proportional to the $\gamma$ values. Evolutionary tracks are plotted as grey lines.\label{fig:normal_SLF_HRD}}
\end{figure*}

\begin{figure*}
    \gridline{\fig{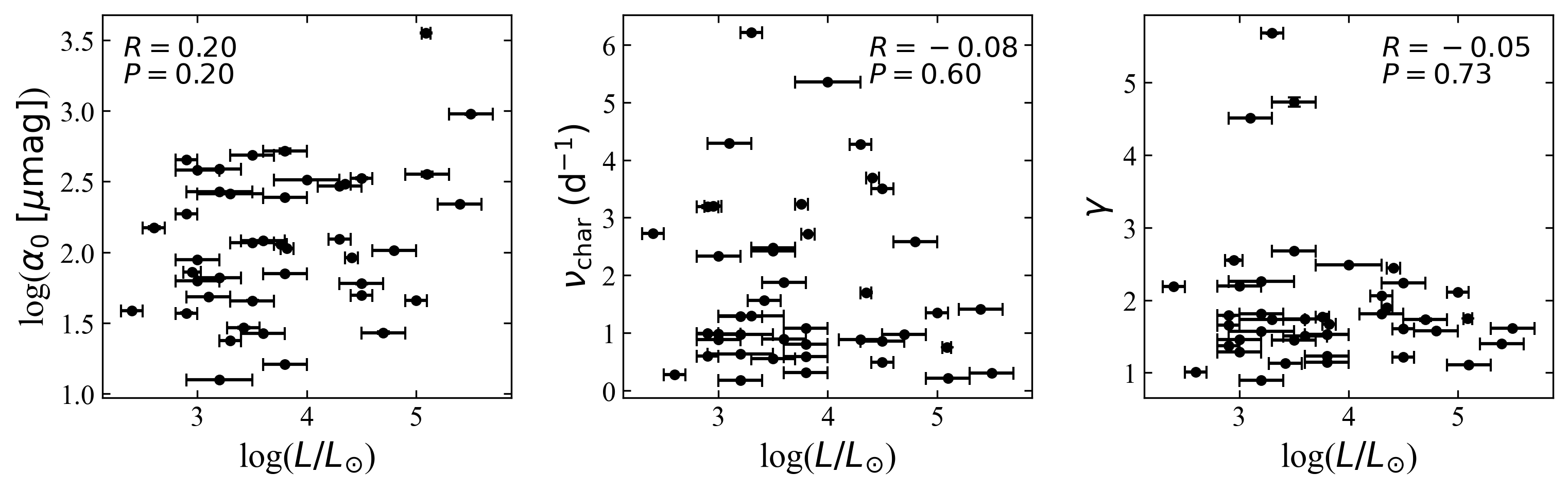}{1\textwidth}{}}
    \caption{The pair-wise relationships between the luminosities and fitting parameters(log$\alpha_{\rm 0}$, $\nu_{\rm char}$, and $\gamma$) for our sample of more massive magnetic hot stars. The Spearman rank-order correlation coefficients and p-values are provided in the figures accordingly.\label{fig:normal_logl_SLF_HRD}}
\end{figure*}

Inspired by previous efforts by \cite{2019NatAs...3..760B, 2019A&A...621A.135B, 2020A&A...640A..36B}, we conducted an analysis of the SLF variability in approximately two hundred light curves from our sample using the MCMC framework. The fitting parameters for SLF variability are listed in Table \ref{table: SLF}, and the logarithmic amplitude spectra and corresponding fitting results are included in Appendix B (ranging from Figure \ref{Appendix C: Fig1} to Figure \ref{Appendix C: Fig11}). As demonstrated by \cite{2019A&A...621A.135B, 2020A&A...640A..36B}, excluding eclipsing binary (EB) systems is warranted in SLF variability analysis, because binary interaction dominates the evolution of massive stars in these systems \citep{2012Sci...337..444S, 2014ApJS..215...15S}. In addition, the significant light contributions from the secondary component in EB systems can cause modulation of the photometric variability detected in the primary star during the binary phase, which limits the scientific inference of SLF variability. Therefore, we excluded EB systems from our sample in SLF variability analysis. We also excluded stars that exhibit strong contamination from our analysis.

On the other hand, our sample includes various types of magnetic hot stars, such as Of?p-type stars, rigidly rotating magnetosphere (RRM) stars, and Ap/Bp stars. To explore SLF variability across a wider mass range in magnetic hot stars, it is necessary to improve the homogeneity of both the stellar and magnetic field structures in our sample. Ap/Bp stars often exhibit chemical peculiarities and slow rotation \citep[e.g.][]{2012A&A...537A.120Z,2017AstL...43..252R,2017MNRAS.468.2745N, 2020A&A...640A..40H}. These stars also display vertical stratification of the abundance of several chemical species, indicating distinct chemical and temperature stratification in their photospheres \citep[e.g.][]{1998AstL...24..516S, 2005asp..book.....Z, 2014IAUS..302..272K}. In contrast to more massive magnetic stars (with spectral types earlier than B5), the magnetic field geometry and evolution of Ap/Bp stars depend on their mass \citep[e.g.][]{1997A&AS..123..353M, 2000ApJ...539..352H, 2007A&A...470..685L, 2017A&A...601A..14M, 2021mfob.book.....H}. Consequently, we grouped Ap/Bp stars with other types of stars based on their spectral type, resulting in two sub-samples for a systematic exploration of SLF variability in magnetic hot stars. The first sub-sample comprises 55 magnetic Ap/Bp stars from our original sample, and the second sub-sample comprises 45 more massive stars from our original sample.

\subsubsection{SLF variability of magnetic Ap/Bp stars\label{sample1}}
In the top row of Figure \ref{fig:apbp_SLF_HRD}, we presented the pair-wise relationships between the fitting parameters of SLF variability for our Ap/Bp sub-sample, each star is color-coded by its luminosity. The arrows extending from the individual colored circles indicate the direction of variation in the fitting parameters of the SLF variability, revealing trends in the fitting parameters across the different sectors. The characteristic frequency, $\nu_{\rm char}$ for magnetic Ap/Bp stars exhibits a primary distribution between 0 and 1.5 $\text{d}^{-1}$. The amplitude of SLF variability, log$\alpha_{\rm 0}$, shows a dominant distribution ranging from 0.8 to 3, while $\gamma$ primarily falls between 0 and 1.5. We calculated Spearman rank-order correlation coefficients $R$ to quantify the relationships between fitting parameters in the sample of Ap/Bp stars and labelled them in the figures accordingly. Due to the smaller sample sizes of Ap/Bp stars in our study, p-values were obtained using a permutation test. A linear regression analysis reveals a strong negative correlation (R = -0.48, p < 0.01) between log$\alpha_{\rm 0}$ and $\nu_{\rm char}$ indicating that stars with larger log$\alpha_{\rm 0}$ tend to have smaller values of $\nu_{\rm char}$. Our statistical analysis also demonstrates a significant correlation between log$\alpha_{\rm 0}$ and $\gamma$ (R = -0.32, p < 0.02), as well as a relationship between $\gamma$ and $\nu_{\rm char}$ (R = 0.84, p < 0.01). These significant correlations among fitting parameters suggest that an Ap/Bp star with larger amplitude of SLF variability is likely to have smaller values of $\nu_{\rm char}$ and $\gamma$.

To more accurately investigate the SLF variability of magnetic Ap/Bp stars, we placed the Ap/Bp stars in the Hertzsprung-Russell diagram (HR diagram). In the bottom-left and bottom-right panels of Figure \ref{fig:apbp_SLF_HRD}, each Ap/Bp star in our sub-sample is represented by a circle, which also be color-coded by the fitting parameters $\alpha_{\rm 0}$ and $\nu_{\rm char}$, respectively. The symbol sizes of the points are proportional to the fitting parameter $\gamma$. To illustrate the distribution of the Ap/Bp stars in our sub-sample, we computed the non-rotating, solar abundance evolutionary tracks for initial masses from 2 to 4.5 $M_{\odot}$ with Modules for Experiments in Stellar Astrophysics \citep[MESA;][]{2019ApJS..243...10P}. During the computation process, the Ledoux criterion is used for convection, \textbf{the mixing-length parameter $\alpha_{\rm MLT}$} and the efficiency parameter $\alpha_{\rm SEM}$ for semi-convection are taken as 1.5 and 1.0, respectively \citep{2011A&A...530A.115B}. These evolutionary tracks are plotted as grey lines in the Figure \ref{fig:apbp_SLF_HRD}.

The relationships between the fitting parameters and the distribution in the HR diagram for stars with masses below 20 $M_{\odot}$ are less clear due to the limited sample size and the variability associated with the opacity mechanism \citep{2020A&A...640A..36B}. It is also challenging to identify any clear dependence of SLF variability on mass from the two HR diagrams presented in the bottom-left and bottom-right panels of Figure \ref{fig:apbp_SLF_HRD}. However, in view of the mass-luminosity relation of the stars, we can indirectly investigate any potential dependence of SLF variability on stellar mass by examining the correlation between the fitting parameters of SLF variability and luminosity. The pair-wise relationships between the luminosities and fitting parameters(log$\alpha_{\rm 0}$, $\nu_{\rm char}$, and $\gamma$) for the sample of Ap/Bp stars are shown in Figure \ref{fig:apbp_logl_SLF_HRD}. The Spearman rank-order correlation coefficients and p-values are provided in the figures accordingly. Our results indicate that no significant correlation exists between the luminosity and the fitting parameters, which suggests that there is no mass-dependence of SLF variability in Ap/Bp stars with masses between approximately 1.5 $M_{\odot}$ < M < 4.5 $M_{\odot}$.

Magnetic Ap/Bp stars constitute approximately 10\% to 15\% of the intermediate-mass and massive main-sequence stars with spectral types ranging from approximately B8 to F0 \citep{2021mfob.book.....H}. However, it is important to note that our sample only covers a limited section of the main sequence of Ap/Bp stars, significantly limiting our ability to explore the dependence of SLF variability on the stellar evolutionary stage. Furthermore, estimating the age of Ap/Bp stars based solely on their positions in the HR diagram typically introduces uncertainties on the order of 25\% of the total main sequence lifetime \citep{2006A&A...450..777B}. Consequently, the ages determined from the HR diagram positions of the Ap/Bp stars in our sample currently have high uncertainties, making them of limited value in constraining SLF variability in relation to stellar age. Therefore, no sufficient constraints on the dependence of SLF variability on stellar age are provided by our sub-sample of Ap/Bp stars.

\subsubsection{SLF variability of more massive magnetic hot stars}

The pair-wise relationships between the fitting parameters of SLF variability for our second sub-sample are presented in the top row of the Figure \ref{fig:normal_SLF_HRD}. We also plotted the HR diagrams for each star in our second sub-sample in the bottom panels of Figure \ref{fig:normal_SLF_HRD}. The markers in Figure \ref{fig:normal_SLF_HRD} have the same meaning as in Figure \ref{fig:apbp_SLF_HRD}. The log$\alpha_{\rm 0}$  and $\gamma$ values in the sample of more massive magnetic hot stars are primarily concentrated between 1 and 3, which is consistent with the distribution range of log$\alpha_{\rm 0}$  and $\gamma$ in the Ap/Bp star sample. However, $\nu_{\rm char}$ of the stars in the sample of more massive magnetic hot stars are primarily distributed in the range from 0 to 7 d$^{-1}$, exhibiting a wider range compared to the Ap/Bp star sample. \cite{2019ApJ...883..106C} provided detailed convective atlas for stars of different masses during the main sequence evolution stage. For stars below 4$M_{\odot}$, the sub-surface convection zone (CZ) predominantly consists of H CZ, HeI CZ, and HeII CZ. On the other hand, for stars above 4$M_{\odot}$, the sub-surface CZ primarily consists of HeII CZ and Fe CZ. If the SLF variability is caused by the stellar envelope convection, the differences in $\nu_{\rm char}$ distribution between the sample of Ap/Bp stars and the sample of more massive magnetic hot stars may be reflect the variations in the number and efficiency of CZ. Furthermore, SLF variability is expected to provide effective constraints on the efficiency of heat transport or the mixing length parameters within the CZ of stars.

We employed the same statistical methodology as described in section \ref{sample1} to analyze the correlation of fitting parameters in the sample of more massive magnetic hot stars. We provided the Spearman rank-order correlation coefficients $R$ and $p$-value for each panel in the top row of Figure \ref{fig:normal_SLF_HRD}. The $\nu_{\rm char}$ and log$\alpha_{\rm 0}$, as well as $\nu_{\rm char}$ and $\gamma$, in the sample of more massive magnetic hot stars exhibit significant correlations. However, there is no correlation between log$\alpha_{\rm 0}$ and $\gamma$. This result contradicts the findings from the sample of Ap/Bp stars. Notably, both in the sample of Ap/Bp stars and the sample of more massive magnetic hot stars, we found a negative correlation between log$\alpha_{\rm 0}$ and $\nu_{\rm char}$. This implies that the relationship between log$\alpha_{\rm 0}$ and $\nu_{\rm char}$ in the SLF variability of stars is universal and independent of stellar mass. The pair-wise relationships between the luminosities and fitting parameters (log$\alpha_{\rm 0}$, $\nu_{\rm char}$, and $\gamma$) for our sample of more massive magnetic hot stars are illustrated in Figure \ref{fig:normal_logl_SLF_HRD}. The Spearman rank-order correlation coefficients $R$ and corresponding $p$-values are provided in the figures. Our results indicate that no significant correlations are observed between the luminosities and fitting parameters, suggesting no clear dependence of SLF variability on stellar mass for our sample of more massive magnetic hot stars.

The surveys conducted on magnetic hot stars have indicated that probably about 7\% of O-type stars with masses exceeding 18 $M_{\odot}$ and about 6\% of early-B and O stars have measurable magnetic fields \citep[e.g.][]{2017MNRAS.465.2432G, 2017A&A...599A..66S}. Currently, observations of magnetic hot stars with masses above 20$M_{\odot}$ are limited to only around ten objects, and the majority of stars with strong magnetic fields are observed in the main sequence stage. This observational bias results in our sample primarily consisting of stars close to the zero-age main sequence early (ZAMS), which indicates that the current sample of magnetic hot stars still suffers from a significant lack of completeness in terms of evolutionary stages. Therefore, with the current sample of magnetic hot stars, our ability to investigate the relationship between SLF variability and stellar age is limited. The implication of previous study is that up to about 30\%-50\% of all massive stars might be magnetic, and only the tip of the iceberg has been seen \citep[][]{2004Natur.431..819B, 2016PASA...33...11S}.Therefore, an extended sample of magnetic hot stars can be expected to be utilized in order to investigate the dependence between SLF variability and stellar physical parameters in future.

\section{DISCUSSION}
As previously mentioned in section \ref{Section 3.1}, we gathered all short cadence data for stars in our sample from TESS sectors 1 to 56. Thanks to $TESS$ repeated observations of the entire sky, the majority of stars in our sample were observed in different sectors that were 26 sectors apart (approximately 710 days), except for stars that overlapped consecutive sectors. This allows us to explore variations in SLF variability over time, ranging from months to years. In the top panels of Figure \ref{fig:apbp_SLF_HRD} and Figure \ref{fig:normal_SLF_HRD}, the arrows extending from the single point indicate the direction of variation in the fitting parameters of the SLF variability, revealing trends in the fitting parameters across the different sectors. The fitting parameters of SLF variability for the same star exhibit the significant variations between the different sectors observed by $TESS$, suggesting that the morphology of SLF variability occurred significant changes in the time. This is consistent with the findings previously reported by \cite{2011A&A...533A...4B} and \cite{2019A&A...621A.135B}, who noted that the dominant peaks of SLF variability in stars have short lifetimes on the order of hours and days.

The steepness $\gamma$ and characteristic frequency $\nu_{\rm char}$ are crucial features for understanding the physical mechanism behind SLF variability \citep{2019A&A...621A.135B,2019NatAs...3..760B}. Our analysis reveals that the magnetic hot stars in our sample have $\gamma < 5.5$ with the vast majority having $1 \leq \gamma \leq 3$. This is consistent with the predictions of numerical simulations of IGWs  \citep[][]{2013ApJ...772...21R, 2015ApJ...815L..30R}. The $\nu_{\rm char}$ for the stars in our sample is primarily in the ranges of $0\;\text{d}^{-1} < \nu_{\rm char} < 6.3\;\text{d}^{-1}$. These results are consistent with the findings of \cite{2019A&A...621A.135B, 2020A&A...640A..36B}. Previous studies have suggested that such a broad frequency range can be explained by the presence of an entire spectrum of IGWs \citep{2019ApJ...876....4E,2020A&A...641A..18H}.  

\cite{2019NatAs...3..760B} studied the SLF variability in 167 OB-type stars observed by the K2 and TESS. They pointed out that the $\nu_{\rm char}$ and $\gamma$ are insensitive to the metallicity of the star, while a significant correlation is discovered between the intrinsic brightness of a star and $\nu_{\rm char}$. Investigating the systematic effects of magnetic fields on the occurrence of SLF variability in massive stars also requires the detection of this phenomenon in a larger sample of magnetic hot stars. To accurately investigate the correlation between the SLF variability and the magnetic field strength, we selected 37 single stars within our sample of more massive magnetic hot stars. This sample consists of the magnetic hot stars identified by \cite{2020MNRAS.499.5379S} and \cite{2013MNRAS.429..398P}, for which sufficient spectroscopic and spectropolarimetric data are available to evaluate their magnetic properties.

\begin{figure*}
    \gridline{\fig{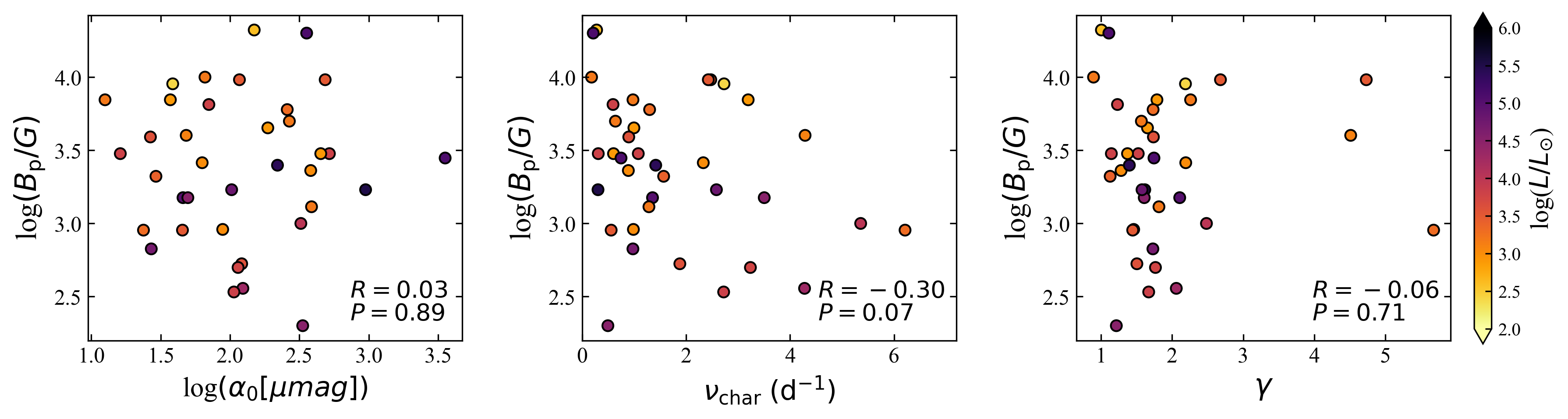}{1\textwidth}{}}
    \gridline{\fig{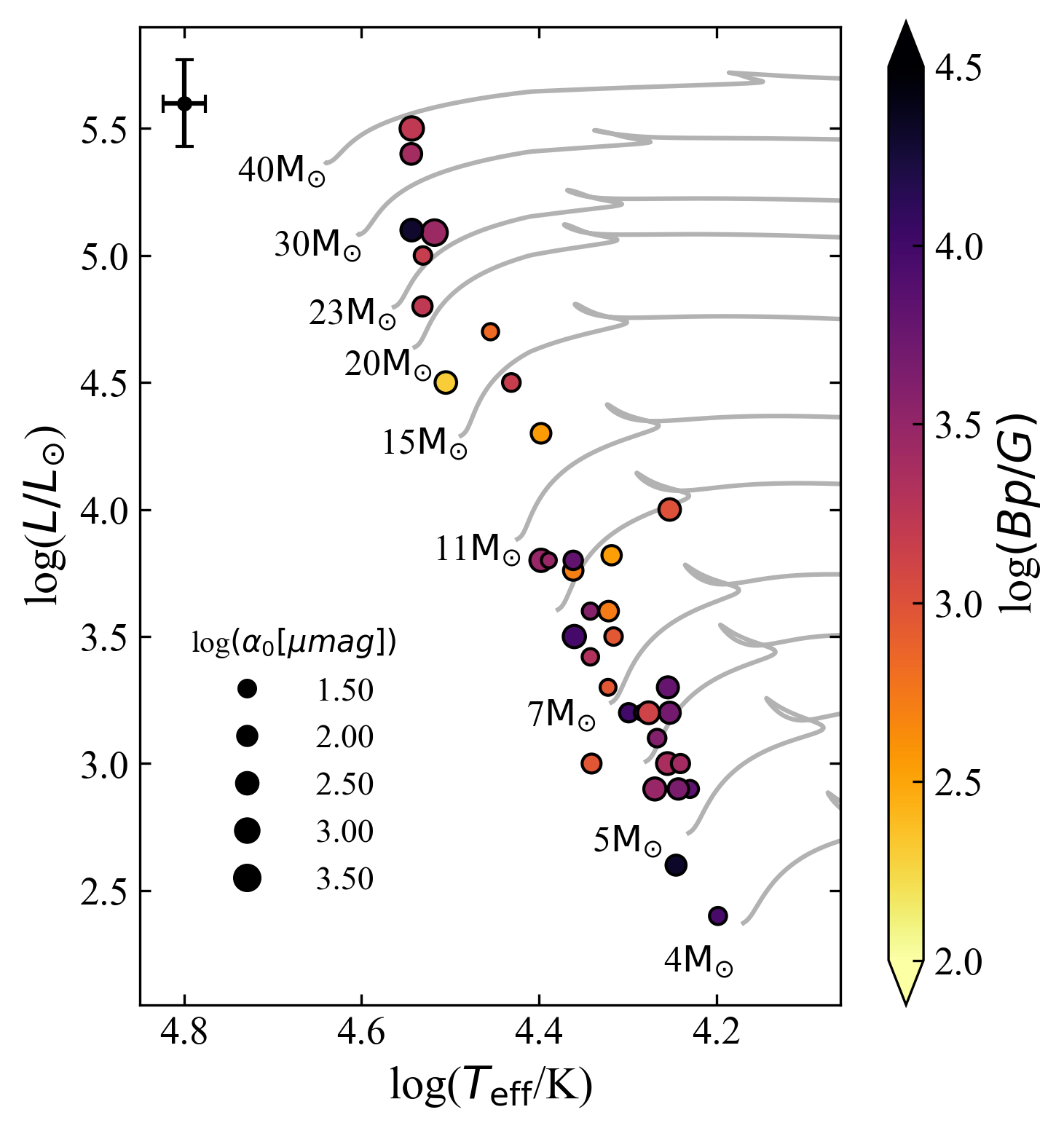}{0.46\textwidth}{}
    \fig{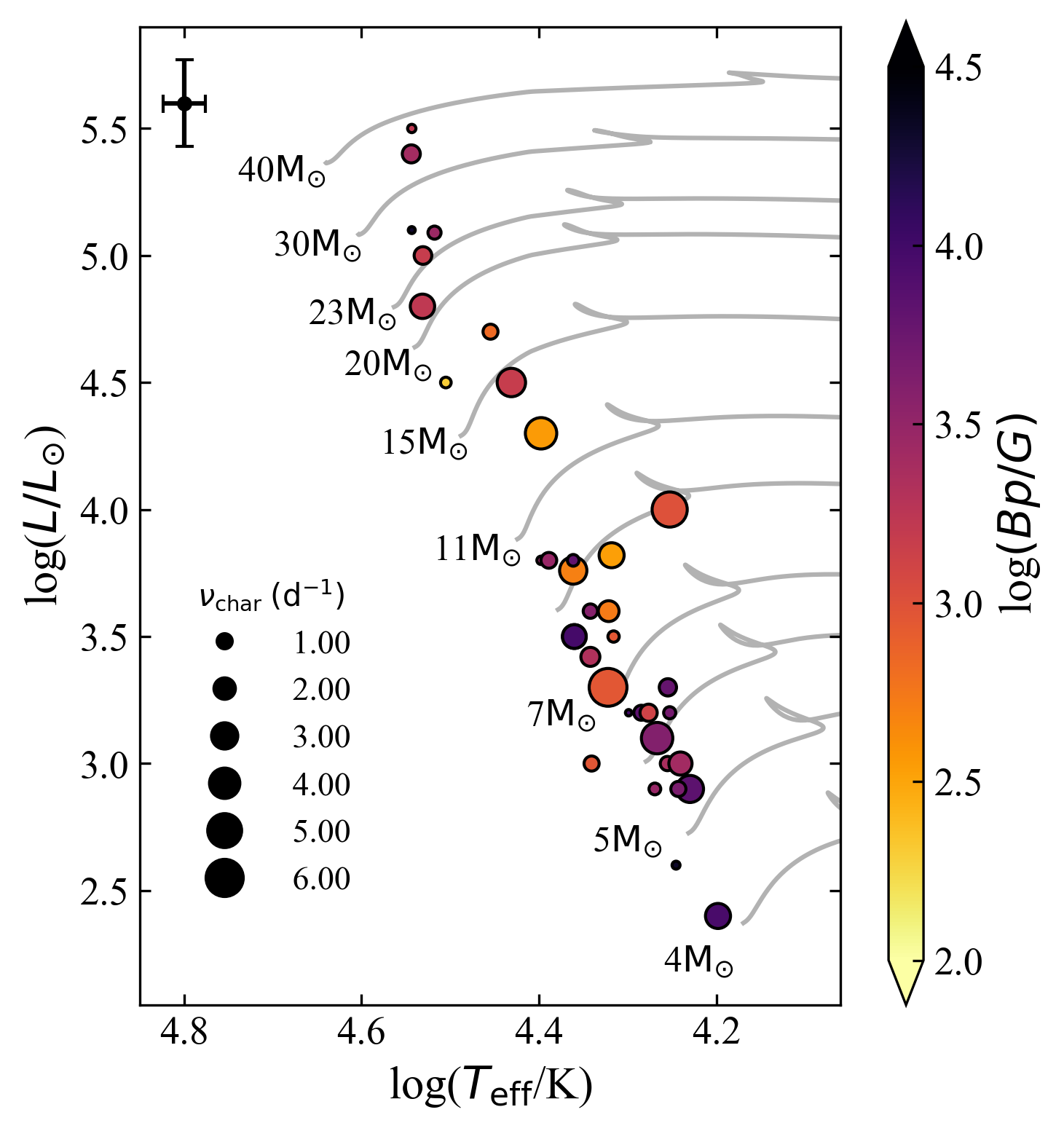}{0.47\textwidth}{}}
    \caption{The relationship between the fitting parameters ($\alpha_{0},\;\nu_{\rm char},\; \text{and}\;\gamma$) and the polar strength ($B_{\rm p}$) of the surface dipole magnetic fields are presented in the top panels. Each star is colour-coded by luminosity. The Spearman's rank correlation coefficient, R, and the corresponding p-value are also provided. The HR diagrams of 37 magnetic hot stars are shown in bottom panels. Each star is represented by a circle, which also be color-coded by the polar strength of the surface dipole magnetic fields ($B_{\rm p}$). The symbol sizes of the points are proportional to the fitting parameter log$\alpha_{\rm 0}$ (bottom-left panel) and $\nu_{\rm char}$ (bottom-right panel), respectively. \label{fig:Bp_SLF}}
\end{figure*}

In the top row of Figure \ref{fig:Bp_SLF}, we presented the relationship between the fitting parameters ($\alpha_{0},\;\nu_{\rm char},\; \text{and}\;\gamma$) and the polar strength of the surface dipole magnetic fields ($B_{\rm p}$). Each star is represented by a circle and color-coded according to its luminosity. The Spearman's rank correlation coefficient, $R$, and the corresponding $p$-value are provided in the panels. In our study, we found a significant negative correlation between the $B_{\rm p}$ and $\nu_{char}$ ($R = -0.30, p = 0.07$). Previous studies reported a positive correlation between macroturbulent ($\nu_{\rm macro} $) and the fitting parameters of SLF variability \citep[e.g.][]{2015ApJ...806L..33A, 2015ApJ...808L..31G, 2017A&A...597A..22S, 2020A&A...640A..36B}. Additionally, the magnetic field is confirmed to have an effect to stabilize the atmosphere and suppress the generation of macroturbulence \citep{2019ApJ...886L..15L, 2013MNRAS.433.2497S}. Therefore, the suppression effect of magnetic fields on $\nu_{\rm char}$ may be a result of their inhibition of macroturbulence. We plotted the HR diagrams in the bottom panels of Figure \ref{fig:Bp_SLF}. Each star is represented by a circle, which also be color-coded by the polar strength of the surface dipole magnetic fields ($B_{\rm p}$). The symbol sizes of the points are proportional to the fitting parameter log$\alpha_{\rm 0}$ (bottom-left panel) and $\nu_{\rm char}$ (bottom-right panel), respectively. Inferring the relationships among the fit parameters and the strength of the magnetic fields in the HR diagram for stars are less clear.

\section{CONCLUSIONS}
Using sectors 1-56 $TESS$ short cadence photometric data, we provided a variability study for 118 magnetic hot stars, which was confirmed using spectropolarimeter or high-resolution spectrograph \citep[e.g.][]{2015A&A...583A.115B,2014Msngr.157...27M,2015IAUS..307..342M,2017MNRAS.465.2432G}. Based on the light curves and amplitude spectra, we determined the photometric variability classification of the stars in our sample. As the results, we detected 87 variable stars, 14 constant stars, and 17 unknown variables. We identified 9 new rotating variable stars in our sample. We discussed the new variable stars aided by the available spectroscopic parameters, spectropolarimeteric parameters, and variability. The details of our results are listed in Table \ref{table: stars}. 

Using a Bayesian MCMC method inspired by \cite{2019NatAs...3..760B,2019A&A...621A.135B}, we derived the fitting parameters of SLF variability from the prewhitening light curves for the stars within our sample. We concluded that the SLF variability is ubiquitous in magnetic hot stars, which are consistent with the findings of \cite{2019A&A...621A.135B, 2020A&A...640A..36B}. The fitting parameters for SLF variabilities are listed in Table \ref{table: SLF}. We also included all logarithmic amplitude spectra, along with the corresponding fitting results, in Appendix B (From the Figure \ref{Appendix C: Fig1} to Figure \ref{Appendix C: Fig11}). To systematically explore SLF variability in magnetic hot stars, we divided our sample into two groups. Our analysis reveals that the magnetic hot stars in our sample have $\gamma < 5.5$ with the vast majority having $1 \leq \gamma \leq 3$. The $\nu_{\rm char}$ for the stars in our sample is primarily in the ranges of $0\;\text{d}^{-1} < \nu_{\rm char} < 6.3\;\text{d}^{-1}$. These results are consistent with the findings of \cite{2019A&A...621A.135B, 2020A&A...640A..36B}. Furthermore, no significant correlations are observed between the luminosities and fitting parameters, suggesting no clear dependence of SLF variability on stellar mass for our sample of magnetic hot stars. To accurately investigate the correlation between SLF variability and magnetic field strength, we selected 37 single stars from our sample of more massive magnetic hot stars. In our study, we found a significant negative correlation between the $B_{\rm p}$ and $\nu_{char}$ ($R = -0.30, p = 0.07$). This suppression effect of magnetic fields on $\nu_{\rm char}$ may be a result of their inhibition of macroturbulence. It should be noted that our results are affected by observational selection effects. Therefore, future studies should employ an expanded sample of magnetic hot stars to investigate the relationship between SLF variability and stellar physical parameters in more detail.

Finally, our results are valuable for guiding future magneto-asteroseismic studies of massive stars, such as the multiple systems, hybrid, and $\beta$ Cep. In the future, we will expand the photometric variables catalog for the magnetic hot stars, and combine it with the spectroscopic databases, such as the IACOB, Large Sky Area Multi-Object Fiber Spectroscopy Telescope (LAMOST), and Sloan Digital Sky Survey (SDSS) to explore the variability more accurately.


\appendices
\renewcommand\thefigure{\Alph{section}\arabic{figure}}
\section[]{Appendix A}
\startlongtable
\begin{deluxetable*}{lccccccc}
\tablecaption{\label{table: stars}Identification numbers, physical parameters and
  variability classifications of the 118 magnetic hot stars.}
%    \tablecaption{Observable Characteristics of Galactic/Magellanic Cloud novae with X-ray observations\label{tab:object}}
%    \tablewidth{500pt}
%    \tabletypesize{\scriptsize}
    \tablehead{
    \colhead{Name} & \colhead{TIC} &
    \colhead{Sp.type} & \colhead{log($\rm L/L_{\odot}$)} &
    \colhead{log($T_{\rm eff}/\text{K}$)} & \colhead{log$\;g$} &
    \colhead{Reference} & \colhead{Variability}
    }
\startdata
    ALS~15218 & 458859193 & O8.5Vp & 5.00 $\pm$ 0.10 & 4.53 $\pm$ 0.03 & 4.00 $\pm$ 0.20 & R1& cont. + SLF \\
    ALS~3694 & 37053664 & B1 & 3.80 $\pm$0.20 & 4.40 $\pm$ 0.02 & 4.00 $\pm$ 0.10 & R2& rot. + SLF \\
    CD-59~3239 & 390668263 & B0.5V & 4.30 $\pm$ 0.20 & 4.36 $\pm$ 0.06 & 3.60 $\pm$ 0.20 & R1& cont. + SLF \\
    CPD-28~2561 & 129361218 & O6Ifp & 5.50 $\pm$ 0.20 & 4.54 $\pm$ 0.02 & 4.00 $\pm$ 0.20 & R1& unknown + SLF \\
    CPD-60~944B & 358467049 & B8III & 2.26 $\pm$ 0.02 & 4.18 $\pm$ 0.03 & 4.75 $\pm$ 0.25 & this work & unknown \\
    CPD-62~2124 & 290796936 & B2IV & 3.80 $\pm$ 0.20 & 4.37 $\pm$ 0.00 & 4.05 $\pm$ 0.10 & R2& rot. + SLF \\
    HD~101412 & 320547523 & B9 & 1.65 $\pm$ 0.01 & 4.00 $\pm$ 0.04 & 4.10 $\pm$ 0.40 & this work & unknown + SLF \\
    HD~105382 & 333670665 & B6III & 3.00 $\pm$ 0.20 & 4.26 $\pm$ 0.01 & 4.13 $\pm$ 0.07 & R1& rot. + SLF \\
    HD~108 & 406893156 & O6f?p-O8fp & 5.70 $\pm$ 0.10 & 4.54 $\pm$ 0.02 & 3.50 $\pm$ 0.20 & R1& Multi + SLF \\
    HD~121743 & 359762838 & B2IV & 3.60 $\pm$ 0.20 & 4.32 $\pm$ 0.03 & 4.02 $\pm$ 0.12 & R1& hybird + SLF \\
    HD~125823 & 167480388 & B7IIIp & 3.20 $\pm$ 0.20 & 4.28 $\pm$ 0.05 & 4.14 $\pm$ 0.12 & R1& rot. + SLF \\
    HD~127381 & 457686485 & B1V & 3.76 $\pm$ 0.06 & 4.36 $\pm$ 0.02 & 4.02 $\pm$ 0.10 & R1& rot. + SLF \\
    HD~130807 & 455309397 & B5 & 2.70 $\pm$ 0.20 & 4.23 $\pm$ 0.03 & 4.25 $\pm$ 0.10 & R1& Multi + SLF \\
    HD~136504 & 147226597 & B2IV-V & 3.70 $\pm$ 0.20 & 4.31 $\pm$ 0.01 & 3.97 $\pm$ 0.15 & R1& Multi + $\beta$ Cep  + SLF \\
    HD~149277 & 229331994 & B2IV/V & 3.50 $\pm$ 0.20 & 4.30 $\pm$ 0.04 & 3.75 $\pm$ 0.15 & R3& Multi + SLF \\
    HD~149438 & 205175750 & B0.2V & 4.50 $\pm$ 0.10 & 4.50 $\pm$ 0.01 & 4.00 $\pm$ 0.10 & R1& unknown + SLF \\
    HD~156324 & 153805844 & B2V & 3.80 $\pm$ 0.60 & 4.34 $\pm$ 0.06 & 4.00 $\pm$ 0.30 & R2& Multi + SLF \\
    HD~156424 & 154190068 & B2V & 3.50 $\pm$ 0.40 & 4.30 $\pm$ 0.07 & 3.99 $\pm$ 0.10 & R4& Multi + $\beta$ Cep + SLF \\
    HD~175362 & 113272022 & B5V & 2.60 $\pm$ 0.10 & 4.25 $\pm$ 0.01 & 4.24 $\pm$ 0.10 & R1& rot. + SLF \\
    HD~176582 & 120355394 & B5IV & 2.90 $\pm$ 0.10 & 4.23 $\pm$ 0.03 & 4.00 $\pm$ 0.10 & R2& rot. + SLF \\
    HD~189775 & 303247444 & B5V & 2.90 $\pm$ 0.10 & 4.24 $\pm$ 0.01 & 4.12 $\pm$ 0.08 & R2& rot. + SLF \\
    HD~191612 & 378273410 & O6f?p & 5.40 $\pm$ 0.20 & 4.54 $\pm$ 0.01 & 3.50 $\pm$ 1.00 & R1& unknown + SLF \\
    HD~200775 & 387813185 & B1 & 4.00 $\pm$ 0.30 & 4.25 $\pm$ 0.05 & 3.40 $\pm$ 0.20 & R1& unknown + SLF\\
    HD~205021 & 321818578 & B1IV & 4.30 $\pm$ 0.10 & 4.40 $\pm$ 0.02 & 3.80 $\pm$ 0.15 & R1& $\beta$ Cep + SLF \\
    HD~23478 & 26432264 & B3IV & 3.20 $\pm$ 0.20 & 4.30 $\pm$ 0.04 & 4.20 $\pm$ 2.00 & R2& rot. + SLF \\
    HD~25558 & 407615454 & B3V & 2.80 $\pm$ 0.30 & 4.23 $\pm$ 0.02 & 4.20 $\pm$ 0.20 & R1& Multi + SLF \\
    HD~306795 & 290682018 & B2V & 3.20 $\pm$ 0.30 & 4.25 $\pm$ 0.05 & 3.90 $\pm$ 0.20 & R1& unknown + SLF \\
    HD~3360 & 240669906 & B2IV & 3.82 $\pm$ 0.06 & 4.32 $\pm$ 0.00 & 3.80 $\pm$ 0.05 & R1& $\beta$ Cep + SLF \\
    HD~345439 & 424048289 & B2IV & 4.00 $\pm$ 0.30 & 4.36 $\pm$ 0.04 & 4.29 $\pm$ 0.19 & R2& rot. + SLF \\
    HD~35298 & 269118344 & B3Vw & 2.40 $\pm$ 0.10 & 4.20 $\pm$ 0.02 & 4.25 $\pm$ 0.12 & R1& rot. + SLF \\
    HD~35502 & 4289780 & B5V & 3.00 $\pm$ 0.10 & 4.26 $\pm$ 0.01 & 4.30 $\pm$ 0.20 & R2& Multi + SLF \\
    HD~35912 & 464839773 & B1.5Ve & 3.30 $\pm$ 0.30 & 4.25 $\pm$ 0.02 & 4.00 $\pm$ 0.10 & R1& SPB + SLF \\
    HD~36485 & 50743458 & B2Ve & 3.50 $\pm$ 0.10 & 4.30 $\pm$ 0.04 & 4.00 $\pm$ 0.10 & R2& Multi + SLF \\
    HD~36526 & 50787573 & B8Vp & 2.30 $\pm$ 0.20 & 4.17 $\pm$ 0.06 & 4.10 $\pm$ 0.14 & R1& rot. + SLF \\
    HD~36982 & 427395300 & B1.5Vp & 3.00 $\pm$ 0.20 & 4.34 $\pm$ 0.04 & 4.40 $\pm$ 0.20 & R1& unknown + SLF\\
    HD~37017 & 427393058 & B1.5-2.5IV-Vp & 3.40 $\pm$ 0.20 & 4.32 $\pm$ 0.04 & 4.10 $\pm$ 0.20 & R2& Multi + SLF \\
    HD~37022 & 427394772 & O6V & 5.30 $\pm$ 0.10 & 4.59 $\pm$ 0.01 & 4.10 $\pm$ 0.10 & R1& Multi + SLF \\
    HD~37058 & 427393350 & B3VpC & 2.90 $\pm$ 0.10 & 4.27 $\pm$ 0.01 & 4.17 $\pm$ 0.07 & R1& unknown + SLF \\
    HD~37479 & 11286209 & B2Vp & 3.50 $\pm$ 0.20 & 4.36 $\pm$ 0.04 & 4.20 $\pm$ 0.20 & R2& rot. + SLF \\
    HD~37742 & 11360636 & O8f?p & 5.60 $\pm$ 0.10 & 4.46 $\pm$ 0.01 & 3.20 $\pm$ 0.10 & R1& Multi + SLF \\
    HD~37776 & 11400909 & B2Vp & 3.30 $\pm$ 0.20 & 4.34 $\pm$ 0.02 & 4.25 $\pm$ 0.20 & R2& rot. + SLF \\
    HD~43317 & 265080539 & B3IV & 2.95 $\pm$ 0.08 & 4.24 $\pm$ 0.00 & 4.07 $\pm$ 0.10 & R5 & hybird + SLF \\
    HD~44743 & 34590771 & B1II/III & 4.41 $\pm$ 0.06 & 4.39 $\pm$ 0.01 & 3.78 $\pm$ 0.08 & R6 & $\beta$ Cep + SLF \\
    HD~46328 & 47763235 & B1III & 4.50 $\pm$ 0.10 & 4.43 $\pm$ 0.02 & 3.78 $\pm$ 0.07 & R1& $\beta$ Cep + SLF \\
    HD~47129 & 220197273 & O7Vp & 5.09 $\pm$ 0.04 & 4.52 $\pm$ 0.03 & 4.10 $\pm$ 0.10 & R1& Multi + SLF \\
    HD~47777 & 220322778 & B3V & 3.42 $\pm$ 0.15 & 4.34 $\pm$ 0.02 & 4.20 $\pm$ 0.10 & R1& rot. + SLF \\
    HD~52089 & 63198307 & B1.5II & 4.35 $\pm$ 0.05 & 4.35 $\pm$ 0.01 & 3.40 $\pm$ 0.08 & R6 & unknown + SLF \\
    HD~54879 & 177860391 & O9.7V & ... & ... & ... & ... & unknown + SLF \\
    HD~55522 & 65403216 & B2IV/V & 3.00 $\pm$ 0.20 & 4.24 $\pm$ 0.01 & 3.95 $\pm$ 0.06 & R1& rot. + SLF \\
    HD~57682 & 187458882 & O7.5III & 4.80 $\pm$ 0.20 & 4.53 $\pm$ 0.01 & 4.00 $\pm$ 0.20 & R1& unknown + SLF \\
    HD~58260 & 284735369 & B3Vp & 3.20 $\pm$ 0.30 & 4.28 $\pm$ 0.03 & 4.20 $\pm$ 0.20 & R1& cont. + SLF \\
    HD~61556 & 110798652 & B5V & 3.10 $\pm$ 0.20 & 4.27 $\pm$ 0.02 & 4.10 $\pm$ 0.15 & R1& rot. + SLF \\
    HD~63425 & 175599737 & B0.5V & 4.49 $\pm$ 0.07 & 4.47 $\pm$ 0.01 & 4.00 $\pm$ 0.10 & R1& cont. + SLF? \\
    HD~64740 & 268971806 & B1.5Vp & 3.80 $\pm$ 0.20 & 4.39 $\pm$ 0.02 & 4.01 $\pm$ 0.09 & R2& rot. + SLF \\
    HD~66522 & 238314043 & B2III & 3.50 $\pm$ 0.20 & 4.32 $\pm$ 0.04 & 3.88 $\pm$ 0.23 & R1& cont.? + SLF \\
    HD~66665 & 443848410 & B0.5V & 4.70 $\pm$ 0.20 & 4.45 $\pm$ 0.02 & 3.90 $\pm$ 0.10 & R1& cont. + SLF \\
    HD~66765 & 238381092 & B1/B2V & 3.40 $\pm$ 0.20 & 4.30 $\pm$ 0.04 & 4.13 $\pm$ 0.20 & R2& Multi + SLF \\
    HD~67621 & 238489837 & B2IV & 3.30 $\pm$ 0.10 & 4.32 $\pm$ 0.01 & 4.18 $\pm$ 0.10 & R1& rot. + SLF \\
    HD~9289 & 136842396 & A3V & 0.94 $\pm$ 0.07 & 3.94 $\pm$ 0.02 & 4.50 $\pm$ 0.20 & this work & rot. + SLF \\
    HD~96446 & 466675723 & B1IVp/B2Vp & 3.80 $\pm$ 0.20 & 4.36 $\pm$ 0.02 & 3.74 $\pm$ 0.10 & R1& $\beta$ Cep + SLF \\
    NGC 1624 2 & 354879267 & O7V & 5.10 $\pm$ 0.20 & 4.54 $\pm$ 0.02 & 4.00 $\pm$ 0.20 & R1& unknown + SLF \\
    HD~184927 & 41099689 & B2Vp & 3.60 $\pm$ 0.20 & 4.34 $\pm$ 0.02 & 3.90 $\pm$ 0.23 & R1& rot. + SLF \\
    HD~105770 & 357310008 & B9V & 2.22 $\pm$ 0.11 & 4.12 $\pm$ 0.01 & ... & R7 & rot. + SLF \\
    HD~115226 & 394272819 & A3V & 0.86 $\pm$ 0.12 & 3.88 $\pm$ 0.01 & ... & R7 & rot. + SLF \\
    HD~117025 & 314265097 & A2V & 1.42 $\pm$ 0.05 & 3.95 $\pm$ 0.02 & ... & R7 & rot. + SLF \\
    HD~118913 & 342995661 & A0V & 1.67 $\pm$ 0.14 & 3.98 $\pm$ 0.01 & ... & R7 & rot. + SLF \\
    HD~119308 & 305716720 & A0V & 1.48 $\pm$ 0.15 & 4.0 $\pm$ 0.01 & ... & R7 & rot. + SLF \\
    HD~122983 & 329012708 & B9 & 1.75 $\pm$ 0.22 & 4.02 $\pm$ 0.01 & 4.00 $\pm$ 0.34 & this work & rot. + SLF \\
    HD~127453 & 446846414 & B8V & 2.25 $\pm$ 0.17 & 4.08 $\pm$ 0.01 & ... & R7 & rot. + SLF \\
    HD~12932 & 268751602 & A5V & 1.07 $\pm$ 0.04 & 3.92  $\pm$  0.02 & 3.50  $\pm$  0.30 & this work & cont. \\
    HD~136933 & 148822705 & A0V & ... & ... & ... & ... & rot. + SLF \\
    HD~138758 & 425871448 & B9V & 1.68 $\pm$ 0.17 & 4.02 $\pm$ 0.02 & ... & R7 & rot. + SLF \\
    HD~146555 & 208684639 & A0V & 2.37 $\pm$ 0.12 & 4.11 $\pm$ 0.01 & 4.00 $\pm$ 0.34 & this work & rot. + SLF \\
    HD~150562 & 44827786 & A5V & 0.86 $\pm$ 0.04 & 3.83 $\pm$ 0.03 & 3.50 $\pm$ 0.50 & this work & cont. + SLF \\
    HD~154708 & 173372645 & A2V & 0.73 $\pm$ 0.14 & 3.83 $\pm$ 0.01 & ... & R7 & rot. + SLF \\
    HD~157751 & 160367734 & B9V & 1.38 $\pm$ 0.15 & 3.99 $\pm$ 0.01 & ... & R7 & rot. + SLF \\
    HD~161459 & 363716787 & A2V & 1.09 $\pm$ 0.04 & 3.86 $\pm$ 0.01 & 3.50 $\pm$ 0.50 & this work & rot. + SLF \\
    HD~166473 & 368866492 & A5V & 0.98 $\pm$ 0.01 & 3.88 $\pm$ 0.01 & 4.50 $\pm$ 0.24 & this work & cont. + SLF \\
    HD~172690 & 469692862 & A0V & 2.05 $\pm$ 0.15 & 4.06 $\pm$ 0.02 & ... & R7 & rot. + SLF \\
    HD~176196 & 387132889 & B9V & 1.81 $\pm$ 0.15 & 4.00 $\pm$ 0.01 & ... & R7 & cont. + SLF \\
    HD~187474 & 300279840 & A0V & 1.88 $\pm$ 0.08 & 4.00 $\pm$ 0.01 & ... & R7 & unknown \\
    HD~190290 & 318007796 & A0V & 0.83 $\pm$ 0.01 & 3.85 $\pm$ 0.01 & 4.00 $\pm$ 0.30 & this work & rot. + SLF \\
    HD~19918 & 348717688 & A5V & 1.09 $\pm$ 0.01 & 3.87 $\pm$ 0.01 & 4.00 $\pm$ 0.10 & this work & cont. + SLF \\
    HD~201018 & 115150623 & A2V & ... & ... & ... & ... & Multi + SLF \\
    HD~203932 & 211404370 & A5V & 1.02 $\pm$ 0.01 & 3.89 $\pm$ 0.01 & 4.50 $\pm$ 0.41 & this work & unknown + SLF \\
    HD~218495 & 237336864 & A2V & 1.00 $\pm$ 0.01 & 3.93 $\pm$ 0.01 & 0.00 $\pm$ 0.00 & this work & rot. + SLF \\
    HD~45583 & 42884620 & B9V & 2.33 $\pm$ 0.10 & 4.20 $\pm$ 0.05 & 3.75 $\pm$ 0.25 & this work & rot. + SLF \\
    HD~49299 & 49314893 & A0V & 2.43 $\pm$ 0.05 & 4.23 $\pm$ 0.04 & 4.00 $\pm$ 0.10 & this work & unknown + SLF \\
    HD~53921 & 766092864 & B9III+B8V & ... & ... & ... & ... & Multi + SLF \\
    HD~56350 & 344175577 & BV & 1.79 $\pm$ 0.08 & 4.02 $\pm$ 0.02 & ... & R7 & rot. + SLF \\
    HD~65712 & 358467700 & A0V & 1.48 $\pm$ 0.04 & 3.95 $\pm$ 0.01 & 4.00 $\pm$ 0.30 & this work & rot. + SLF \\
    HD~65987 & 372913684 & B9V & 2.62 $\pm$ 0.06 & 4.11 $\pm$ 0.02 & 4.00 $\pm$ 0.20 & this work & rot. + SLF \\
    HD~66295 & 410451777 & B8V & 1.62 $\pm$ 0.05 & 4.05 $\pm$ 0.08 & 4.00 $\pm$ 0.20 & this work & rot. + SLF \\
    HD~66318 & 410451752 & A0V & 1.48 $\pm$ 0.03 & 3.93 $\pm$ 0.01 & 4.00 $\pm$ 0.10 & this work & unknown + SLF? \\
    HD~87240 & 462162948 & B9V & 2.9 $\pm$ 0.4 & 4.1 & 3.6 & R8 & rot. + SLF \\
    HD~89103 & 219614759 & B9V & 1.67 $\pm$ 0.12 & 4.07 $\pm$ 0.02 & ... & R7 & rot. + SLF \\
    HD~92499 & 146715928 & A2V & 1.05 $\pm$ 0.17 & 3.86 $\pm$ 0.01 & ... & R7 & cont. + SLF \\
    HD~94660 & 147622676 & A0V & 1.98 $\pm$ 0.09 & 4.03 $\pm$ 0.02 & ... & R7 & rot. + SLF \\
    HD~9672 & 54003409 & A1V & 1.25 $\pm$ 0.02 & 3.98 $\pm$ 0.01 & 4.00 $\pm$ 0.20 & this work & cont. + SLF \\
    HD~10840 & 231844926 & B3V & 1.93 $\pm$ 0.11 & 4.07 $\pm$ 0.02 & 4.22 $\pm$ 0.20 & R9 & cont. + SLF \\
    HD~115440 & 394308408 & B9V & 1.90 $\pm$ 0.11 & 4.08 $\pm$ 0.02 & 4.29 $\pm$ 0.20 & R9 & rot. + SLF \\
    HD~116890 & 340963903 & B9V & 2.45 $\pm$ 0.12 & 4.11 $\pm$ 0.02 & 3.97 $\pm$ 0.21 & R9 & rot. + SLF \\
    HD~127575 & 421348273 & B9V & 1.77 $\pm$ 0.11 & 4.08 $\pm$ 0.02 & 4.36 $\pm$ 0.20 & R9 & rot. + SLF \\
    HD~128775 & 128886984 & B9V & 2.09 $\pm$ 0.14 & 4.07 $\pm$ 0.02 & 4.07 $\pm$ 0.24 & R9 & rot. + SLF \\
    HD~129899 & 402517183 & A0V & 2.28 $\pm$ 0.12 & 4.03 $\pm$ 0.02 & 3.74 $\pm$ 0.22 & R9 & rot. + SLF \\
    HD~149764 & 83824229 & A0V & 2.03 $\pm$ 0.11 & 4.11 $\pm$ 0.02 & 4.31 $\pm$ 0.20 & R9 & rot. + SLF \\
    HD~149822 & 148207172 & B9V & 1.66 $\pm$ 0.10 & 4.02 $\pm$ 0.02 & 4.21 $\pm$ 0.20 & R9 & rot. + SLF \\
    HD~212385 & 278804454 & A3V & 1.36 $\pm$ 0.10 & 3.93 $\pm$ 0.03 & 4.05 $\pm$ 0.22 & R9 & rot. + SLF \\
    HD~24188 & 32035258 & A0V & 1.98 $\pm$ 0.10 & 4.10 $\pm$ 0.02 & 4.29 $\pm$ 0.18 & R9 & rot. + SLF \\
    HD~34797 & 408217716 & B8V & 2.43 $\pm$ 0.14 & 4.12 $\pm$ 0.02 & 4.04 $\pm$ 0.26 & R9 & rot. + SLF \\
    HD~37633 & 11295912 & B9V & 1.90 $\pm$ 0.14 & 4.11 $\pm$ 0.02 & 4.39 $\pm$ 0.23 & R9 & rot. + SLF \\
    HD~63401 & 175604551 & B8V & 2.32 $\pm$ 0.10 & 4.13 $\pm$ 0.02 & 4.15 $\pm$ 0.18 & R9 & rot. + SLF \\
    HD~69067 & 182362832 & B8V & 1.91 $\pm$ 0.12 & 4.09 $\pm$ 0.02 & 4.30 $\pm$ 0.21 & R9 & rot. + SLF \\
    HD~83625 & 440537642 & A0V & 1.88 $\pm$ 0.11 & 4.07 $\pm$ 0.02 & 4.27 $\pm$ 0.20 & R9 & rot. + SLF \\
    HD~86199 & 469253078 & B9V & 2.23 $\pm$ 0.11 & 4.10 $\pm$ 0.02 & 4.11 $\pm$ 0.20 & R9 & rot. + SLF \\
    HD~88158 & 375583232 & B8V & 2.31 $\pm$ 0.10 & 4.11 $\pm$ 0.02 & 4.07 $\pm$ 0.19 & R9 & rot. + SLF \\
    HD~88385 & 462877348 & A0V & 1.63 $\pm$ 0.11 & 3.99 $\pm$ 0.02 & 4.12 $\pm$ 0.22 & R9 & rot. + SLF \\
    HD~218994 & 2055137574 & ApSr & ... & ... & ... & R10 & $\delta$ Sct \\
\enddata
\tablecomments{The variability classification and stellar parameters of the 118 magnetic hot stars. The first column is the identification names of the stars. The TESS Input Catalog (TIC) numbers are listed in the second column. The luminosities, effective temperatures, and surface gravity are listed in columns 4, 5, and 6, respectively. References for spectroscopic analyses are provided in the seventh column. In the column 7, R1 = \cite{2013MNRAS.429..398P}, R2 =\cite{2020MNRAS.499.5379S}, R3 = \cite{2018MNRAS.481L..30G}, R4 = \cite{2021MNRAS.504.4850S}, R5 = \cite{2019MNRAS.485.1508S}, R6 = \cite{2018MNRAS.475.5144S}, R7 = \cite{2006AA...450..763K}, R8 = \cite{2005RMxAA..41..415S}, R9 = \cite{2017MNRAS.468.2745N}, R10 = \cite{2008MNRAS.386.1750K}, and "this work" represents the stellar parameters derived from the SED using the VOSA. In the last column, we reported the variability classification result for each star. SPB = Slowly Pulsating B star; $\beta$ Cep = $\beta$ Cepheid variable; rot. = rotating variable; cont. = constant; SLF = stochastic low-frequency variable; $\delta$ Sct = $\delta$ Scuti; Multi = binary or multiple star. Unknown represents we can not identify the classification using the amplitude spectrum and light curve.}
\end{deluxetable*}

\startlongtable
\begin{longrotatetable}
\begin{deluxetable*}{lccccclccccc}
\tablecaption{\label{table: SLF}The SLF fitting parameters of the magnetic hot stars.}
%    \tablecaption{Observable Characteristics of Galactic/Magellanic Cloud novae with X-ray observations\label{tab:object}}
%    \tablewidth{500pt}
%    \tabletypesize{\scriptsize}
    \tablehead{
    \colhead{Name} & \colhead{Sector(s)} & \colhead{$\alpha_{\rm 0}$} & \colhead{$\nu_{\rm chat}$} & \colhead{$\gamma$}& \colhead{$C_{\rm w}$} & 
    \colhead{Name} & \colhead{Sector(s)} & \colhead{$\alpha_{\rm 0}$} & \colhead{$\nu_{\rm chat}$} & \colhead{$\gamma$}& \colhead{$C_{\rm w}$}\\
    \colhead{}     & \colhead{}          & \colhead{($\mu\;$mag)}     & \colhead{}                 & \colhead{}        & \colhead{($\mu\;$mag)} &\colhead{}     & \colhead{}          & \colhead{($\mu\;$mag)}     & \colhead{}                 & \colhead{}        &\colhead{($\mu\;$mag)}
    }
\startdata
ALS~15218 & 10~-~11 & ${45.80}_{-3.16}^{+3.35}$ & ${1.35}_{-0.11}^{+0.12}$ & ${2.11}_{-0.11}^{+0.12}$ & ${9.84}_{-0.02}^{+0.02}$ & HD~127381 & 11 & ${114.22}_{-5.55}^{+5.87}$ & ${3.24}_{-0.16}^{+0.17}$ & ${1.77}_{-0.03}^{+0.03}$ & ${2.18}_{-0.01}^{+0.01}$ \\
ALS~15218 & 36 & ${62.76}_{-6.71}^{+8.26}$ & ${1.13}_{-0.12}^{+0.14}$ & ${3.45}_{-0.34}^{+0.43}$ & ${20.17}_{-0.06}^{+0.07}$ & HD~127381 & 38 & ${86.25}_{-3.23}^{+3.40}$ & ${5.39}_{-0.20}^{+0.19}$ & ${2.58}_{-0.05}^{+0.05}$ & ${2.45}_{-0.01}^{+0.01}$ \\
ALS~3694 & 12 & ${520.15}_{-82.72}^{+130.37}$ & ${0.31}_{-0.07}^{+0.07}$ & ${1.52}_{-0.07}^{+0.08}$ & ${26.08}_{-0.09}^{+0.10}$ & HD~127453 & 11~-~12 & ${112.81}_{-15.08}^{+20.59}$ & ${0.16}_{-0.03}^{+0.04}$ & ${0.98}_{-0.03}^{+0.03}$ & ${4.37}_{-0.02}^{+0.02}$ \\ 
ALS~3694 & 39 & ${202.62}_{-22.18}^{+27.70}$ & ${0.81}_{-0.12}^{+0.12}$ & ${1.65}_{-0.09}^{+0.10}$ & ${18.06}_{-0.06}^{+0.06}$ & HD~127575 & 11~-~12 & ${84.77}_{-6.57}^{+6.33}$ & ${0.70}_{-0.06}^{+0.07}$ & ${1.43}_{-0.04}^{+0.04}$ & ${5.05}_{-0.02}^{+0.01}$ \\ 
CD-59~3239 & 10~-~11 & ${295.17}_{-20.65}^{+22.82}$ & ${0.88}_{-0.07}^{+0.07}$ & ${1.82}_{-0.06}^{+0.06}$ & ${24.56}_{-0.06}^{+0.06}$ & HD~$128775$ & 11 & ${74.36}_{-9.29}^{+12.05}$ & ${0.60}_{-0.13}^{+0.14}$ & ${1.07}_{-0.04}^{+0.04}$ & ${5.77}_{-0.04}^{+0.03}$ \\
CD-59~3239 & 36~-~37 & ${107.13}_{-6.55}^{+7.33}$ & ${1.61}_{-0.11}^{+0.12}$ & ${2.61}_{-0.16}^{+0.18}$ & ${21.54}_{-0.05}^{+0.05}$ & HD~12932 & 3 & ${17.17}_{-3.22}^{+3.95}$ & ${1.18}_{-0.24}^{+0.32}$ & ${2.05}_{-0.34}^{+0.52}$ & ${16.21}_{-0.07}^{+0.06}$ \\ 
CPD-28~2561 & 34 & ${948.91}_{-135.20}^{+155.43}$ & ${0.30}_{-0.04}^{+0.05}$ & ${1.62}_{-0.05}^{+0.06}$ & ${21.34}_{-0.08}^{+0.08}$ & HD~12932 & 30 & ${52.11}_{-7.77}^{+10.55}$ & ${0.58}_{-0.10}^{+0.09}$ & ${3.68}_{-1.11}^{+1.64}$ & ${15.40}_{-0.05}^{+0.05}$ \\ 
CPD-62~2124 & 10~-~11 & ${243.83}_{-17.35}^{+19.21}$ & ${0.80}_{-0.07}^{+0.07}$ & ${1.53}_{-0.04}^{+0.04}$ & ${15.55}_{-0.04}^{+0.04}$ & HD~129899 & 12 & ${25.58}_{-3.89}^{+5.40}$ & ${0.43}_{-0.12}^{+0.14}$ & ${1.09}_{-0.07}^{+0.09}$ & ${4.41}_{-0.02}^{+0.02}$ \\ 
CPD-62~2124 & 37~-~38 & ${245.56}_{-21.23}^{+24.68}$ & ${0.54}_{-0.06}^{+0.06}$ & ${1.45}_{-0.04}^{+0.05}$ & ${13.35}_{-0.04}^{+0.04}$ & HD~130807 & 11 & ${310.59}_{-27.16}^{+29.37}$ & ${0.84}_{-0.07}^{+0.07}$ & ${1.76}_{-0.04}^{+0.04}$ & ${4.92}_{-0.02}^{+0.02}$ \\ 
HD~101412 & 37~-~38 & ${64.08}_{-4.78}^{+6.09}$ & ${0.79}_{-0.08}^{+0.08}$ & ${1.71}_{-0.07}^{+0.08}$ & ${7.53}_{-0.02}^{+0.02}$ & HD~130807 & 38 & ${270.44}_{-21.20}^{+22.70}$ & ${1.00}_{-0.06}^{+0.07}$ & ${2.06}_{-0.04}^{+0.04}$ & ${2.30}_{-0.01}^{+0.01}$ \\
HD~$105382$ & 10 & ${473.411}_{-32.79}^{+38.40}$ & ${0.61}_{-0.09}^{+0.10}$ & ${1.22}_{-0.02}^{+0.02}$ & ${3.51}_{-0.02}^{+0.03}$ & HD~136504 & 38 & ${109.04}_{-5.54}^{+5.87}$ & ${2.73}_{-0.12}^{+0.12}$ & ${2.28}_{-0.04}^{+0.04}$ & ${1.70}_{-0.01}^{+0.01}$ \\
HD~$105382$ & 37 & ${95.67}_{-6.61}^{+7.22}$ & ${0.73}_{-0.04}^{+0.05}$ & ${1.79}_{-0.06}^{+0.06}$ & ${1.57}_{-0.01}^{+0.01}$ & HD~136933 & 38 & ${5.28}_{-0.51}^{+0.61}$ & ${2.62}_{-0.41}^{+0.43}$ & ${1.53}_{-0.11}^{+0.13}$ & ${2.19}_{-0.01}^{+0.01}$ \\
HD~105770 & 11~-~13 & ${71.74}_{-11.38}^{+14.86}$ & ${0.09}_{-0.03}^{+0.03}$ & ${0.77}_{-0.02}^{+0.02}$ & ${3.66}_{-0.02}^{+0.02}$ & HD~138758 & 12 & ${58.47}_{-8.14}^{+9.25}$ & ${0.58}_{-0.12}^{+0.16}$ & ${1.13}_{-0.05}^{+0.07}$ & ${7.39}_{-0.04}^{+0.04}$ \\
HD~108 & 17~-~18 & ${390.77}_{-30.39}^{+33.72}$ & ${0.57}_{-0.05}^{+0.05}$ & ${1.28}_{-0.02}^{+0.02}$ & ${5.07}_{-0.02}^{+0.02}$ & HD~146555 & 39 & ${683.85}_{-112.35}^{+140.65}$ & ${0.24}_{-0.06}^{+0.08}$ & ${0.95}_{-0.04}^{+0.04}$ & ${32.49}_{-0.24}^{+0.22}$ \\
HD~10840 & 1~-~2 & ${6.91}_{-0.63}^{+0.75}$ & ${1.14}_{-0.19}^{+0.21}$ & ${1.11}_{-0.05}^{+0.06}$ & ${2.25}_{-0.01}^{+0.01}$ & HD~149277 & 12 & ${921.01}_{-155.22}^{+212.71}$ & ${0.20}_{-0.04}^{+0.05}$ & ${1.23}_{-0.03}^{+0.04}$ & ${14.11}_{-0.07}^{+0.07}$ \\ 
HD~115226 & 11~-~12 & ${47.96}_{-4.49}^{+5.30}$ & ${0.58}_{-0.08}^{+0.09}$ & ${1.23}_{-0.04}^{+0.05}$ & ${5.73}_{-0.02}^{+0.02}$ & HD~149277 & 39 & ${593.11}_{-90.88}^{+120.29}$ & ${0.24}_{-0.05}^{+0.05}$ & ${1.22}_{-0.03}^{+0.04}$ & ${10.63}_{-0.05}^{+0.05}$ \\
HD~115226 & 38~-~39 & ${91.59}_{-9.34}^{+12.72}$ & ${0.35}_{-0.06}^{+0.06}$ & ${1.10}_{-0.03}^{+0.04}$ & ${5.67}_{-0.02}^{+0.02}$ & HD~149438 & 12 & ${334.07}_{-38.59}^{+49.14}$ & ${0.49}_{-0.08}^{+0.08}$ & ${1.22}_{-0.04}^{+0.03}$ & ${14.80}_{-0.07}^{+0.07}$ \\
HD~115440 & 11~-~12 & ${69.41}_{-12.59}^{+19.23}$ & ${0.25}_{-0.07}^{+0.08}$ & ${1.19}_{-0.05}^{+0.06}$ & ${5.04}_{-0.02}^{+0.02}$ & HD~149438 & 39 & ${8.71}_{-0.57}^{+0.68}$ & ${2.11}_{-0.16}^{+0.16}$ & ${2.30}_{-0.13}^{+0.14}$ & ${1.42}_{-0.01}^{+0.00}$ \\
HD~116890 & 11~-~12 & ${60.17}_{-5.98}^{+7.91}$ & ${0.37}_{-0.06}^{+0.06}$ & ${1.28}_{-0.04}^{+0.05}$ & ${3.27}_{-0.01}^{+0.01}$ & HD~149764 & 39 & ${265.65}_{-30.82}^{+38.91}$ & ${0.56}_{-0.10}^{+0.09}$ & ${1.22}_{-0.04}^{+0.03}$ & ${7.92}_{-0.04}^{+0.04}$ \\
HD~117025 & 11 & ${70.38}_{-7.69}^{+9.05}$ & ${0.72}_{-0.10}^{+0.11}$ & ${1.54}_{-0.06}^{+0.07}$ & ${4.36}_{-0.01}^{+0.01}$ & HD~149822 & 25 & ${74.46}_{-20.72}^{+20.23}$ & ${0.15}_{-0.06}^{+0.14}$ & ${0.77}_{-0.04}^{+0.06}$ & ${4.87}_{-0.05}^{+0.07}$ \\
HD~118913 & 11~-~12 & ${54.60}_{-4.78}^{+5.32}$ & ${0.68}_{-0.08}^{+0.10}$ & ${1.22}_{-0.04}^{+0.04}$ & ${4.92}_{-0.02}^{+0.02}$ & HD~149822 & 52 & ${39.05}_{-3.61}^{+4.19}$ & ${1.42}_{-0.17}^{+0.17}$ & ${1.82}_{-0.10}^{+0.10}$ & ${3.95}_{-0.01}^{+0.01}$ \\
HD~119308 & 11 & ${55.83}_{-5.73}^{+7.00}$ & ${1.14}_{-0.17}^{+0.18}$ & ${1.43}_{-0.08}^{+0.08}$ & ${7.52}_{-0.03}^{+0.03}$ & HD~150562 & 12 & ${71.36}_{-11.20}^{+16.45}$ & ${0.63}_{-0.20}^{+0.24}$ & ${0.88}_{-0.06}^{+0.06}$ & ${16.22}_{-0.14}^{+0.14}$ \\
HD~121743 & 11 & ${121.32}_{-7.31}^{+7.83}$ & ${1.88}_{-0.13}^{+0.14}$ & ${1.51}_{-0.02}^{+0.02}$ & ${2.01}_{-0.01}^{+0.01}$ & HD~150562 & 39 & ${95.96}_{-11.57}^{+15.96}$ & ${0.78}_{-0.17}^{+0.16}$ & ${1.24}_{-0.07}^{+0.08}$ & ${13.45}_{-0.07}^{+0.06}$ \\
HD~121743 & 38 & ${84.79}_{-4.35}^{+4.37}$ & ${3.00}_{-0.14}^{+0.16}$ & ${1.89}_{-0.03}^{+0.03}$ & ${1.72}_{-0.01}^{+0.01}$ & HD~154708 & 12 & ${192.66}_{-30.73}^{+43.30}$ & ${0.20}_{-0.06}^{+0.07}$ & ${0.93}_{-0.04}^{+0.04}$ & ${10.69}_{-0.07}^{+0.06}$ \\
HD~122983 & 38 & ${275.42}_{-39.12}^{+46.60}$ & ${0.34}_{-0.07}^{+0.08}$ & ${1.04}_{-0.03}^{+0.04}$ & ${13.36}_{-0.07}^{+0.08}$ & HD~154708 & 39 & ${200.04}_{-34.65}^{+42.06}$ & ${0.23}_{-0.06}^{+0.08}$ & ${0.97}_{-0.03}^{+0.04}$ & ${10.26}_{-0.06}^{+0.06}$ \\
HD~125823 & 11 & ${386.76}_{-27.39}^{+29.48}$ & ${1.28}_{-0.08}^{+0.08}$ & ${1.82}_{-0.03}^{+0.03}$ & ${3.39}_{-0.01}^{+0.01}$ & HD~156324 & 12 & ${157.58}_{-19.68}^{+24.18}$ & ${0.62}_{-0.12}^{+0.14}$ & ${1.15}_{-0.05}^{+0.05}$ & ${14.95}_{-0.08}^{+0.07}$ \\
HD~125823 & 38 & ${227.20}_{-11.39}^{+12.49}$ & ${2.20}_{-0.09}^{+0.09}$ & ${2.41}_{-0.04}^{+0.04}$ & ${2.48}_{-0.01}^{+0.01}$ & HD~156324 & 39 & ${151.88}_{-15.33}^{+18.55}$ & ${0.78}_{-0.10}^{+0.10}$ & ${1.76}_{-0.09}^{+0.09}$ & ${11.53}_{-0.04}^{+0.04}$ \\
HD~156424 & 12 & ${295.24}_{-53.82}^{+76.45}$ & ${0.21}_{-0.07}^{+0.07}$ & ${0.93}_{-0.04}^{+0.04}$ & ${12.91}_{-0.09}^{+0.10}$ & HD~218495 & 27~-~28 & ${28.57}_{-2.88}^{+4.24}$ & ${1.23}_{-0.21}^{+0.17}$ & ${2.17}_{-0.23}^{+0.23}$ & ${8.21}_{-0.01}^{+0.02}$ \\
HD~156424 & 39 & ${218.65}_{-42.91}^{+56.77}$ & ${0.14}_{-0.05}^{+0.07}$ & ${0.81}_{-0.03}^{+0.03}$ & ${10.47}_{-0.09}^{+0.09}$ & HD~$23478$ & 42~-~44 & ${65.96}_{-7.69}^{+9.41}$ & ${0.18}_{-0.03}^{+0.04}$ & ${0.90}_{-0.02}^{+0.02}$ & ${2.44}_{-0.01}^{+0.01}$ \\
HD~157751 & 12 & ${152.40}_{-19.09}^{+23.34}$ & ${0.48}_{-0.10}^{+0.11}$ & ${1.04}_{-0.04}^{+0.04}$ & ${8.38}_{-0.06}^{+0.05}$ & HD~24188 & 12~-~13 & ${18.06}_{-2.48}^{+3.14}$ & ${0.31}_{-0.08}^{+0.10}$ & ${0.85}_{-0.04}^{+0.04}$ & ${3.07}_{-0.01}^{+0.01}$ \\
HD~161459 & 13 & ${174.10}_{-26.92}^{+37.96}$ & ${0.39}_{-0.09}^{+0.11}$ & ${1.09}_{-0.05}^{+0.06}$ & ${18.52}_{-0.09}^{+0.09}$ & HD~24188 & 36~-~37 & ${84.80}_{-11.10}^{+14.24}$ & ${0.16}_{-0.03}^{+0.03}$ & ${1.29}_{-0.04}^{+0.04}$ & ${2.51}_{-0.01}^{+0.01}$ \\
HD~161459 & 39 & ${261.43}_{-45.61}^{+63.56}$ & ${0.28}_{-0.07}^{+0.09}$ & ${1.06}_{-0.04}^{+0.05}$ & ${17.09}_{-0.09}^{+0.09}$ & HD~$24188$ & 6~-~7 & ${36.53}_{-3.62}^{+4.04}$ & ${0.51}_{-0.07}^{+0.08}$ & ${1.23}_{-0.04}^{+0.04}$ & ${2.72}_{-0.01}^{+0.01}$ \\
HD~166473 & 13 & ${147.25}_{-42.07}^{+76.56}$ & ${0.06}_{-0.03}^{+0.06}$ & ${0.69}_{-0.03}^{+0.04}$ & ${7.52}_{-0.10}^{+0.08}$ & HD~25558 & 32 & ${509.66}_{-40.68}^{+45.85}$ & ${0.84}_{-0.06}^{+0.06}$ & ${1.90}_{-0.03}^{+0.03}$ & ${3.61}_{-0.01}^{+0.01}$ \\ 
HD~$172690$ & 12~-~13 & ${206.16}_{-19.18}^{+23.94}$ & ${0.33}_{-0.04}^{+0.04}$ & ${1.34}_{-0.03}^{+0.03}$ & ${4.85}_{-0.01}^{+0.01}$ & HD~25558 & 5 & ${839.96}_{-92.28}^{+100.57}$ & ${0.45}_{-0.04}^{+0.05}$ & ${1.58}_{-0.02}^{+0.02}$ & ${3.74}_{-0.02}^{+0.02}$ \\
HD~$175362$ & 13 & ${149.36}_{-22.39}^{+29.97}$ & ${0.28}_{-0.06}^{+0.07}$ & ${1.01}_{-0.03}^{+0.02}$ & ${3.72}_{-0.03}^{+0.03}$ & HD~306795 & 37~-~38 & ${267.72}_{-21.98}^{+26.17}$ & ${0.64}_{-0.07}^{+0.07}$ & ${1.57}_{-0.05}^{+0.06}$ & ${20.80}_{-0.05}^{+0.05}$ \\ 
HD~176196 & 13 & ${49.35}_{-8.24}^{+11.59}$ & ${0.28}_{-0.08}^{+0.10}$ & ${1.00}_{-0.06}^{+0.07}$ & ${5.31}_{-0.03}^{+0.03}$ & HD~3360 & 17 & ${106.62}_{-5.74}^{+6.38}$ & ${2.72}_{-0.16}^{+0.17}$ & ${1.67}_{-0.03}^{+0.03}$ & ${2.40}_{-0.01}^{+0.01}$ \\
HD~$176582$ & 40~-~41 & ${243.64}_{-34.52}^{+46.18}$ & ${0.12}_{-0.02}^{+0.01}$ & ${1.36}_{-0.05}^{+0.07}$ & ${3.12}_{-0.01}^{+0.01}$ & HD~345439 & 54 & ${631.71}_{-131.74}^{+185.84}$ & ${0.13}_{-0.05}^{+0.06}$ & ${0.83}_{-0.03}^{+0.03}$ & ${28.82}_{-0.22}^{+0.21}$ \\
HD~$176582$ & 53~-~54 & ${212.71}_{-35.60}^{+40.10}$ & ${0.13}_{-0.01}^{+0.02}$ & ${1.37}_{-0.08}^{+0.09}$ & ${3.48}_{-0.01}^{+0.01}$ & HD~34797 & 32 & ${252.18}_{-18.48}^{+21.02}$ & ${1.19}_{-0.07}^{+0.08}$ & ${2.05}_{-0.04}^{+0.05}$ & ${4.06}_{-0.01}^{+0.01}$ \\
HD~184927 & 40~-~41 & ${26.67}_{-2.06}^{+2.19}$ & ${0.90}_{-0.09}^{+0.09}$ & ${1.74}_{-0.07}^{+0.09}$ & ${3.87}_{-0.01}^{+0.01}$ & HD~34797 & 37 & ${238.71}_{-16.17}^{+18.03}$ & ${1.24}_{-0.07}^{+0.07}$ & ${2.14}_{-0.05}^{+0.05}$ & ${4.05}_{-0.01}^{+0.01}$ \\ 
HD~187474 & 13 & ${102.49}_{-44.39}^{+46.02}$ & ${0.05}_{-0.03}^{+0.11}$ & ${0.69}_{-0.03}^{+0.04}$ & ${3.45}_{-0.03}^{+0.05}$ & HD~35298 & 32 & ${119.13}_{-10.82}^{+12.93}$ & ${1.11}_{-0.12}^{+0.13}$ & ${1.69}_{-0.06}^{+0.07}$ & ${6.72}_{-0.03}^{+0.03}$ \\
HD~187474 & 27 & ${3.94}_{-0.49}^{+0.64}$ & ${0.83}_{-0.10}^{+0.09}$ & ${4.83}_{-1.34}^{+2.21}$ & ${2.14}_{-0.01}^{+0.01}$ & HD~35298 & 6 & ${38.62}_{-2.81}^{+3.25}$ & ${2.73}_{-0.26}^{+0.26}$ & ${2.19}_{-0.14}^{+0.15}$ & ${6.31}_{-0.02}^{+0.02}$ \\
HD~$189775$ & 16 & ${7.46}_{-0.37}^{+0.46}$ & ${8.86}_{-0.9}^{+0.78}$ & ${1.62}_{-0.04}^{+0.04}$ & ${4.50}_{-0.02}^{+0.02}$ & HD~35502 & 32 & ${138.09}_{-18.61}^{+25.19}$ & ${0.33}_{-0.07}^{+0.07}$ & ${1.13}_{-0.03}^{+0.03}$ & ${4.93}_{-0.02}^{+0.02}$ \\
HD~$189775$ & 41 & ${145.915}_{-16.50}^{+19.95}$ & ${0.12}_{-0.01}^{+0.01}$ & ${1.52}_{-0.23}^{+0.24}$ & ${4.55}_{-0.01}^{+0.01}$ & HD~35502 & 6 & ${242.10}_{-53.85}^{+91.39}$ & ${0.12}_{-0.04}^{+0.05}$ & ${1.02}_{-0.03}^{+0.03}$ & ${5.14}_{-0.03}^{+0.03}$ \\
HD~$189775$ & 56 & ${115.202}_{-22.89}^{+32.14}$ & ${0.12}_{-0.03}^{+0.03}$ & ${3.56}_{-0.24}^{+0.27}$ & ${2.88}_{-0.01}^{+0.01}$ & HD~35912 & 32 & ${587.49}_{-60.48}^{+68.52}$ & ${0.58}_{-0.06}^{+0.07}$ & ${1.43}_{-0.02}^{+0.02}$ & ${5.07}_{-0.02}^{+0.03}$ \\
HD~190290 & 27 & ${66.11}_{-7.75}^{+9.29}$ & ${0.81}_{-0.11}^{+0.12}$ & ${2.34}_{-0.20}^{+0.23}$ & ${12.17}_{-0.04}^{+0.04}$ & HD~35912 & 6 & ${258.42}_{-19.31}^{+22.00}$ & ${1.30}_{-0.10}^{+0.10}$ & ${1.73}_{-0.03}^{+0.03}$ & ${4.08}_{-0.02}^{+0.02}$ \\
HD~190290 & 39 & ${70.21}_{-10.16}^{+14.70}$ & ${0.54}_{-0.12}^{+0.12}$ & ${1.58}_{-0.12}^{+0.15}$ & ${12.47}_{-0.05}^{+0.04}$ & HD~36485 & 32 & ${196.60}_{-10.88}^{+11.72}$ & ${2.13}_{-0.09}^{+0.10}$ & ${3.12}_{-0.10}^{+0.10}$ & ${7.16}_{-0.02}^{+0.02}$ \\
HD~191612 & 14~-~15 & ${219.82}_{-12.01}^{+13.16}$ & ${1.41}_{-0.10}^{+0.10}$ & ${1.40}_{-0.02}^{+0.02}$ & ${6.52}_{-0.03}^{+0.03}$ & HD~36485 & 6 & ${469.22}_{-34.26}^{+39.16}$ & ${1.35}_{-0.09}^{+0.09}$ & ${2.04}_{-0.04}^{+0.04}$ & ${6.31}_{-0.03}^{+0.03}$ \\
HD~19918 & 12~-~13 & ${74.73}_{-9.73}^{+12.49}$ & ${0.19}_{-0.04}^{+0.05}$ & ${1.05}_{-0.05}^{+0.05}$ & ${7.82}_{-0.03}^{+0.03}$ & HD~$36526$ & 32 & ${35.093}_{-2.11}^{+2.17}$ & ${2.40}_{-0.13}^{+0.12}$ & ${1.59}_{-0.05}^{+0.05}$ & ${7.22}_{-0.03}^{+0.03}$ \\
HD~19918 & 27~-~28 & ${135.09}_{-20.17}^{+26.39}$ & ${0.19}_{-0.04}^{+0.04}$ & ${1.59}_{-0.08}^{+0.09}$ & ${7.27}_{-0.02}^{+0.02}$ & HD~$36526$ & 6 & ${32.46}_{-2.32}^{+2.56}$ & ${1.81}_{-0.12}^{+0.12}$ & ${1.54}_{-0.06}^{+0.06}$ & ${7.59}_{-0.03}^{+0.03}$ \\
HD~200775 & 24~-~25 & ${323.94}_{-7.97}^{+8.53}$ & ${5.36}_{-0.11}^{+0.11}$ & ${2.48}_{-0.03}^{+0.03}$ & ${6.22}_{-0.02}^{+0.02}$ & HD~36982 & 32 & ${88.53}_{-8.67}^{+10.25}$ & ${0.98}_{-0.13}^{+0.13}$ & ${1.46}_{-0.05}^{+0.05}$ & ${6.62}_{-0.02}^{+0.03}$ \\
HD~201018 & 1 & ${224.77}_{-32.90}^{+48.64}$ & ${0.34}_{-0.09}^{+0.09}$ & ${0.97}_{-0.03}^{+0.03}$ & ${8.36}_{-0.05}^{+0.06}$ & HD~$37017$ & 32 & ${73.08}_{-7.23}^{+7.87}$ & ${1.20}_{-0.16}^{+0.18}$ & ${1.33}_{-0.04}^{+0.04}$ & ${4.27}_{-0.02}^{+0.02}$ \\
HD~203932 & 1 & ${197.86}_{-43.58}^{+70.31}$ & ${0.14}_{-0.05}^{+0.07}$ & ${0.95}_{-0.04}^{+0.04}$ & ${7.79}_{-0.05}^{+0.04}$ & HD~$37017$ & 6 & ${14.71}_{-1.00}^{+1.15}$ & ${2.06}_{-0.20}^{+0.20}$ & ${1.90}_{-0.15}^{+0.17}$ & ${4.66}_{-0.02}^{+0.02}$ \\
HD~203932 & 28 & ${68.52}_{-8.69}^{+10.27}$ & ${0.77}_{-0.11}^{+0.12}$ & ${2.17}_{-0.20}^{+0.23}$ & ${9.41}_{-0.03}^{+0.03}$ & HD~37022 & 32 & ${332.46}_{-25.45}^{+31.89}$ & ${1.04}_{-0.09}^{+0.09}$ & ${1.47}_{-0.02}^{+0.03}$ & ${4.74}_{-0.02}^{+0.02}$ \\
HD~$205021$ & 16~-~18 & ${123.75}_{-3.07}^{+3.11}$ & ${4.28}_{-0.10}^{+0.11}$ & ${2.06}_{-0.02}^{+0.02}$ & ${3.35}_{-0.01}^{+0.01}$ & HD~37022 & 6 & ${333.26}_{-28.51}^{+34.96}$ & ${0.88}_{-0.06}^{+0.06}$ & ${2.26}_{-0.06}^{+0.06}$ & ${3.34}_{-0.01}^{+0.01}$ \\
HD~$205021$ & 24~-~25 & ${103.70}_{-2.69}^{+2.99}$ & ${5.18}_{-0.10}^{+0.10}$ & ${3.36}_{-0.06}^{+0.06}$ & ${6.61}_{-0.01}^{+0.01}$ & HD~37058 & 32 & ${450.78}_{-47.60}^{+58.22}$ & ${0.60}_{-0.08}^{+0.08}$ & ${1.37}_{-0.03}^{+0.03}$ & ${10.15}_{-0.05}^{+0.05}$ \\ 
HD~212385 & 1 & ${136.83}_{-23.70}^{+28.17}$ & ${0.30}_{-0.08}^{+0.12}$ & ${0.87}_{-0.02}^{+0.03}$ & ${4.44}_{-0.04}^{+0.04}$ & HD~$37479$ & 6 & ${116.88}_{-8.26}^{+9.13}$ & ${2.47}_{-0.15}^{+0.15}$ & ${4.73}_{-0.66}^{+0.79}$ & ${36.19}_{-0.13}^{+0.13}$ \\ 
HD~212385 & 28 & ${53.92}_{-7.04}^{+9.88}$ & ${0.49}_{-0.10}^{+0.10}$ & ${1.37}_{-0.07}^{+0.07}$ & ${4.51}_{-0.02}^{+0.02}$ & HD~$37479$ & 32 & ${485.00}_{-35.63}^{+39.43}$ & ${2.43}_{-0.17}^{+0.18}$ & ${2.68}_{-0.11}^{+0.11}$ & ${85.59}_{-0.29}^{+0.28}$ \\ 
HD~37633 & 6 & ${84.55}_{-9.49}^{+11.23}$ & ${1.03}_{-0.15}^{+0.16}$ & ${1.84}_{-0.12}^{+0.13}$ & ${9.71}_{-0.04}^{+0.04}$ & HD~65712 & 1 & ${84.14}_{-9.32}^{+11.40}$ & ${0.87}_{-0.14}^{+0.14}$ & ${1.51}_{-0.07}^{+0.07}$ & ${10.58}_{-0.03}^{+0.04}$ \\
HD~37742 & 6 & ${3405.39}_{-383.38}^{+469.66}$ & ${0.46}_{-0.05}^{+0.05}$ & ${1.53}_{-0.02}^{+0.02}$ & ${9.96}_{-0.05}^{+0.05}$ & HD~65712 & 4 & ${202.06}_{-22.19}^{+28.10}$ & ${0.80}_{-0.12}^{+0.12}$ & ${1.55}_{-0.06}^{+0.06}$ & ${11.76}_{-0.04}^{+0.05}$ \\
HD~43317 & 33 & ${77.00}_{-4.06}^{+4.52}$ & ${2.98}_{-0.15}^{+0.15}$ & ${2.54}_{-0.09}^{+0.10}$ & ${5.33}_{-0.02}^{+0.02}$ & HD~65712 & 7~-~8 & ${272.54}_{-16.59}^{+17.43}$ & ${1.11}_{-0.07}^{+0.08}$ & ${1.60}_{-0.03}^{+0.03}$ & ${8.15}_{-0.03}^{+0.02}$ \\
HD~43317 & 6 & ${72.25}_{-4.84}^{+4.93}$ & ${3.20}_{-0.24}^{+0.20}$ & ${2.55}_{-0.15}^{+0.13}$ & ${6.45}_{-0.03}^{+0.03}$ & HD~65712 & 10~-~11 & ${30.70}_{-3.09}^{+3.67}$ & ${0.74}_{-0.11}^{+0.12}$ & ${1.37}_{-0.07}^{+0.07}$ & ${8.14}_{-0.02}^{+0.02}$ \\
HD~44743 & 33 & ${63.54}_{-2.19}^{+2.23}$ & ${5.69}_{-0.15}^{+0.14}$ & ${3.57}_{-0.10}^{+0.10}$ & ${2.19}_{-0.01}^{+0.01}$ & HD~65712 & 27~-~28 & ${131.41}_{-9.32}^{+10.75}$ & ${0.96}_{-0.08}^{+0.09}$ & ${1.78}_{-0.05}^{+0.06}$ & ${8.07}_{-0.02}^{+0.02}$ \\
HD~44743 & 6 & ${91.75}_{-5.53}^{+6.01}$ & ${3.69}_{-0.25}^{+0.26}$ & ${2.44}_{-0.12}^{+0.12}$ & ${7.12}_{-0.03}^{+0.03}$ & HD~65712 & 34~-~38 & ${124.55}_{-5.66}^{+6.18}$ & ${0.80}_{-0.04}^{+0.04}$ & ${1.60}_{-0.02}^{+0.03}$ & ${5.10}_{-0.01}^{+0.01}$ \\
HD~45583 & 6 & ${27.20}_{-3.96}^{+5.68}$ & ${0.86}_{-0.23}^{+0.27}$ & ${1.09}_{-0.07}^{+0.07}$ & ${6.59}_{-0.04}^{+0.04}$ & HD~65987 & 1 & ${145.66}_{-15.70}^{+15.25}$ & ${0.89}_{-0.11}^{+0.12}$ & ${1.49}_{-0.04}^{+0.05}$ & ${8.32}_{-0.03}^{+0.03}$ \\
HD~$46328$ & 33 & ${26.14}_{-1.32}^{+1.52}$ & ${3.50}_{-0.22}^{+0.21}$ & ${1.83}_{-0.04}^{+0.04}$ & ${1.39}_{-0.01}^{+0.01}$ & HD~65987 & 4 & ${417.27}_{-49.46}^{+60.34}$ & ${0.57}_{-0.09}^{+0.08}$ & ${1.23}_{-0.02}^{+0.02}$ & ${7.03}_{-0.04}^{+0.04}$ \\
HD~$46328$ & 6 & ${49.65}_{-3.01}^{+3.33}$ & ${3.50}_{-0.27}^{+0.28}$ & ${1.61}_{-0.04}^{+0.04}$ & ${2.88}_{-0.02}^{+0.01}$ & HD~65987 & 7 & ${54.64}_{-6.44}^{+7.54}$ & ${0.94}_{-0.15}^{+0.17}$ & ${1.43}_{-0.07}^{+0.08}$ & ${7.45}_{-0.03}^{+0.03}$ \\
HD~47129 & 6 & ${3553.01}_{-328.89}^{+389.57}$ & ${0.75}_{-0.06}^{+0.06}$ & ${1.75}_{-0.02}^{+0.02}$ & ${12.10}_{-0.05}^{+0.06}$ & HD~65987 & 9~-~11 & ${48.90}_{-4.05}^{+4.37}$ & ${0.54}_{-0.06}^{+0.07}$ & ${1.25}_{-0.03}^{+0.04}$ & ${4.41}_{-0.01}^{+0.01}$ \\
HD~47777 & 33 & ${62.53}_{-8.74}^{+12.66}$ & ${0.62}_{-0.15}^{+0.17}$ & ${1.13}_{-0.06}^{+0.06}$ & ${6.69}_{-0.04}^{+0.04}$ & HD~65987 & 34~-~37 & ${79.06}_{-4.70}^{+5.12}$ & ${0.63}_{-0.04}^{+0.05}$ & ${1.48}_{-0.03}^{+0.03}$ & ${3.62}_{-0.01}^{+0.01}$ \\
HD~47777 & 6 & ${29.23}_{-3.40}^{+4.36}$ & ${1.57}_{-0.34}^{+0.36}$ & ${1.13}_{-0.07}^{+0.08}$ & ${7.24}_{-0.05}^{+0.04}$ & HD~66295 & 4 & ${333.98}_{-31.11}^{+37.51}$ & ${0.86}_{-0.09}^{+0.09}$ & ${1.75}_{-0.06}^{+0.06}$ & ${14.92}_{-0.06}^{+0.05}$ \\
HD~49299 & 33 & ${102.40}_{-16.07}^{+20.27}$ & ${0.50}_{-0.11}^{+0.13}$ & ${1.49}_{-0.12}^{+0.13}$ & ${19.14}_{-0.07}^{+0.07}$ & HD~66295 & 7 & ${113.58}_{-11.45}^{+14.18}$ & ${1.05}_{-0.14}^{+0.14}$ & ${1.70}_{-0.09}^{+0.10}$ & ${11.97}_{-0.05}^{+0.04}$ \\
HD~52089 & 33~-~34 & ${284.71}_{-13.85}^{+14.73}$ & ${1.57}_{-0.08}^{+0.08}$ & ${1.70}_{-0.02}^{+0.02}$ & ${3.46}_{-0.01}^{+0.01}$ & HD~66295 & 9~-~10 & ${65.89}_{-5.92}^{+7.04}$ & ${0.69}_{-0.09}^{+0.10}$ & ${1.37}_{-0.05}^{+0.06}$ & ${8.24}_{-0.02}^{+0.02}$ \\
HD~52089 & 6~-~7 & ${304.99}_{-16.69}^{+29.75}$ & ${1.70}_{-0.11}^{+0.09}$ & ${1.90}_{-0.02}^{+0.02}$ & ${2.38}_{-0.01}^{+0.01}$ & HD~66318 & 34~-~37 & ${71.15}_{-6.87}^{+8.24}$ & ${1.08}_{-0.13}^{+0.13}$ & ${1.71}_{-0.08}^{+0.09}$ & ${7.91}_{-0.03}^{+0.03}$ \\
HD~53921 & 4~-~13 & ${20.85}_{-1.27}^{+1.27}$ & ${0.24}_{-0.02}^{+0.02}$ & ${1.24}_{-0.02}^{+0.02}$ & ${0.82}_{-0.00}^{+0.00}$ & HD~66318 & 35~-~36 & ${116.46}_{-14.35}^{+19.14}$ & ${0.63}_{-0.10}^{+0.10}$ & ${1.55}_{-0.07}^{+0.07}$ & ${7.91}_{-0.03}^{+0.03}$ \\
HD~53921 & 30~-~37 & ${12.14}_{-0.65}^{+0.76}$ & ${0.41}_{-0.03}^{+0.03}$ & ${1.27}_{-0.02}^{+0.02}$ & ${0.94}_{-0.00}^{+0.00}$ & HD~66318 & 7~-~9 & ${149.87}_{-8.98}^{+9.72}$ & ${0.80}_{-0.05}^{+0.05}$ & ${1.55}_{-0.03}^{+0.03}$ & ${5.06}_{-0.01}^{+0.01}$ \\
HD~54879 & 33 & ${33.42}_{-3.56}^{+4.34}$ & ${0.94}_{-0.11}^{+0.12}$ & ${2.23}_{-0.18}^{+0.20}$ & ${4.98}_{-0.02}^{+0.02}$ & HD~66318 & 8~-~11 & ${68.11}_{-3.35}^{+3.41}$ & ${1.10}_{-0.06}^{+0.06}$ & ${1.78}_{-0.03}^{+0.04}$ & ${5.25}_{-0.01}^{+0.01}$ \\ 
HD~54879 & 7 & ${60.12}_{-6.47}^{+8.03}$ & ${0.86}_{-0.11}^{+0.10}$ & ${2.24}_{-0.16}^{+0.17}$ & ${5.80}_{-0.02}^{+0.02}$ & HD~66522 & 34~-~36 & ${34.23}_{-3.06}^{+3.74}$ & ${0.43}_{-0.07}^{+0.07}$ & ${1.07}_{-0.03}^{+0.03}$ & ${3.72}_{-0.01}^{+0.01}$ \\
HD~55522 & 33~-~34 & ${62.97}_{-2.65}^{+2.76}$ & ${2.33}_{-0.09}^{+0.10}$ & ${2.19}_{-0.04}^{+0.05}$ & ${2.54}_{-0.01}^{+0.01}$ & HD~66522 & 7~-~9 & ${45.40}_{-3.26}^{+4.43}$ & ${0.56}_{-0.06}^{+0.05}$ & ${1.45}_{-0.04}^{+0.04}$ & ${3.48}_{-0.01}^{+0.01}$ \\
HD~56350 & 7~-~9 & ${113.86}_{-7.78}^{+8.19}$ & ${0.61}_{-0.05}^{+0.05}$ & ${1.25}_{-0.02}^{+0.01}$ & ${2.69}_{-0.01}^{+0.01}$ & HD~66665 & 34 & ${142.57}_{-22.02}^{+25.39}$ & ${0.25}_{-0.06}^{+0.08}$ & ${0.94}_{-0.04}^{+0.04}$ & ${8.88}_{-0.06}^{+0.06}$ \\
HD~57682 & 34 & ${177.79}_{-18.20}^{+24.94}$ & ${1.00}_{-0.17}^{+0.18}$ & ${1.13}_{-0.03}^{+0.03}$ & ${7.63}_{-0.05}^{+0.05}$ & HD~66665 & 7 & ${27.05}_{-3.80}^{+5.01}$ & ${0.97}_{-0.20}^{+0.21}$ & ${1.73}_{-0.17}^{+0.20}$ & ${7.43}_{-0.03}^{+0.03}$ \\
HD~57682 & 7 & ${102.70}_{-6.42}^{+7.05}$ & ${2.58}_{-0.20}^{+0.21}$ & ${1.58}_{-0.04}^{+0.04}$ & ${3.94}_{-0.02}^{+0.02}$ & HD~66765 & 34 & ${80.66}_{-5.93}^{+6.82}$ & ${1.83}_{-0.18}^{+0.18}$ & ${1.55}_{-0.05}^{+0.05}$ & ${4.82}_{-0.02}^{+0.02}$ \\
HD~58260 & 33~-~34 & ${12.55}_{-0.93}^{+1.16}$ & ${0.98}_{-0.09}^{+0.09}$ & ${2.26}_{-0.17}^{+0.18}$ & ${3.27}_{-0.01}^{+0.01}$ & HD~66765 & 7 & ${41.15}_{-2.38}^{+2.51}$ & ${3.73}_{-0.25}^{+0.28}$ & ${1.87}_{-0.06}^{+0.07}$ & ${4.59}_{-0.02}^{+0.02}$ \\
HD~61556 & 34 & ${98.41}_{-5.34}^{+6.08}$ & ${2.61}_{-0.16}^{+0.16}$ & ${1.55}_{-0.02}^{+0.02}$ & ${2.19}_{-0.01}^{+0.01}$ & HD~66765 & 9 & ${86.43}_{-5.97}^{+6.75}$ & ${2.12}_{-0.17}^{+0.19}$ & ${1.66}_{-0.04}^{+0.05}$ & ${4.22}_{-0.02}^{+0.02}$ \\
HD~61556 & 7 & ${15.81}_{-0.56}^{+0.54}$ & ${3.70}_{-0.10}^{+0.10}$ & ${3.32}_{-0.11}^{+0.13}$ & ${1.67}_{-0.01}^{+0.01}$ & HD~67621 & 34 & ${53.74}_{-4.65}^{+5.48}$ & ${1.56}_{-0.20}^{+0.22}$ & ${1.39}_{-0.06}^{+0.06}$ & ${4.46}_{-0.02}^{+0.02}$ \\
HD~$63401$ & 34~-~35 & ${151.30}_{-7.80}^{+8.84}$ & ${1.59}_{-0.12}^{+0.12}$ & ${1.16}_{-0.01}^{+0.01}$ & ${3.22}_{-0.02}^{+0.02}$ & HD~67621 & 7 & ${23.77}_{-0.83}^{+0.89}$ & ${6.21}_{-0.15}^{+0.14}$ & ${5.68}_{-0.40}^{+0.42}$ & ${3.97}_{-0.01}^{+0.01}$ \\
HD~$63401$ & 7~-~8 & ${24.30}_{-1.13}^{+1.20}$ & ${3.01}_{-0.18}^{+0.19}$ & ${1.83}_{-0.05}^{+0.06}$ & ${2.84}_{-0.01}^{+0.01}$ & HD~67621 & 9 & ${31.54}_{-1.78}^{+1.95}$ & ${4.05}_{-0.28}^{+0.29}$ & ${2.24}_{-0.12}^{+0.13}$ & ${4.05}_{-0.02}^{+0.02}$ \\
HD~$64740$ & 34~-~35 & ${7.02}_{-0.84}^{+0.67}$ & ${2.29}_{-0.34}^{+0.54}$ & ${1.21}_{-0.05}^{+0.07}$ & ${1.03}_{-0.01}^{+0.01}$ & HD~69067 & 7~-~8 & ${45.04}_{-3.21}^{+3.73}$ & ${1.17}_{-0.12}^{+0.12}$ & ${1.61}_{-0.06}^{+0.06}$ & ${5.63}_{-0.01}^{+0.02}$ \\ 
HD~$64740$ & 7~-~9 & ${16.18}_{-1.12}^{+1.28}$ & ${1.08}_{-0.12}^{+0.13}$ & ${1.15}_{-0.03}^{+0.03}$ & ${1.64}_{-0.01}^{+0.01}$ & HD~83625 & 9~-~10 & ${118.77}_{-10.79}^{+12.79}$ & ${0.50}_{-0.06}^{+0.07}$ & ${1.28}_{-0.03}^{+0.03}$ & ${4.18}_{-0.01}^{+0.01}$ \\
HD~86199 & 9~-~10 & ${50.11}_{-3.18}^{+3.62}$ & ${1.25}_{-0.11}^{+0.11}$ & ${1.36}_{-0.03}^{+0.03}$ & ${3.65}_{-0.01}^{+0.01}$ & HD~9289 & 30 & ${99.73}_{-14.86}^{+20.03}$ & ${0.56}_{-0.14}^{+0.15}$ & ${1.35}_{-0.11}^{+0.11}$ & ${14.85}_{-0.06}^{+0.06}$ \\
HD~87240 & 36~-~37 & ${33.47}_{-3.16}^{+3.96}$ & ${0.96}_{-0.13}^{+0.14}$ & ${1.66}_{-0.11}^{+0.12}$ & ${9.83}_{-0.02}^{+0.02}$ & HD~94660 & 36 & ${518.92}_{-115.25}^{+126.49}$ & ${0.09}_{-0.02}^{+0.04}$ & ${0.92}_{-0.02}^{+0.02}$ & ${2.91}_{-0.03}^{+0.03}$ \\
HD~88158 & 9~-~10 & ${46.95}_{-4.33}^{+5.68}$ & ${1.07}_{-0.15}^{+0.14}$ & ${1.67}_{-0.09}^{+0.10}$ & ${5.59}_{-0.02}^{+0.02}$ & HD~94660 & 9~-~10 & ${60.63}_{-6.24}^{+7.45}$ & ${0.39}_{-0.06}^{+0.06}$ & ${1.11}_{-0.03}^{+0.03}$ & ${2.81}_{-0.01}^{+0.01}$ \\
HD~88385 & 36~-~37 & ${18.20}_{-1.38}^{+1.44}$ & ${1.37}_{-0.13}^{+0.13}$ & ${2.46}_{-0.28}^{+0.33}$ & ${4.99}_{-0.01}^{+0.01}$ & HD~96446 & 10~-~11 & ${70.38}_{-5.76}^{+6.89}$ & ${0.59}_{-0.07}^{+0.07}$ & ${1.23}_{-0.03}^{+0.03}$ & ${4.65}_{-0.01}^{+0.01}$ \\
HD~89103 & 9~-~10 & ${69.11}_{-23.99}^{+11.02}$ & ${0.52}_{-0.09}^{+0.41}$ & ${1.35}_{-0.05}^{+0.18}$ & ${4.63}_{-0.02}^{+0.02}$ & HD~9672 & 3 & ${8.41}_{-0.76}^{+0.88}$ & ${1.51}_{-0.14}^{+0.13}$ & ${2.90}_{-0.28}^{+0.33}$ & ${2.15}_{-0.01}^{+0.01}$ \\
HD~92499 & 36 & ${76.89}_{-11.00}^{+13.85}$ & ${0.53}_{-0.11}^{+0.12}$ & ${1.37}_{-0.09}^{+0.13}$ & ${10.05}_{-0.04}^{+0.04}$ & HD~9672 & 30 & ${28.12}_{-4.62}^{+6.16}$ & ${0.42}_{-0.11}^{+0.13}$ & ${1.22}_{-0.08}^{+0.09}$ & ${3.58}_{-0.02}^{+0.01}$ \\
HD~92499 & 9~-~10 & ${46.48}_{-4.30}^{+4.64}$ & ${0.64}_{-0.07}^{+0.08}$ & ${1.71}_{-0.10}^{+0.11}$ & ${6.66}_{-0.02}^{+0.02}$ & NGC~1624-2 & 19 & ${356.63}_{-66.66}^{+96.56}$ & ${0.21}_{-0.06}^{+0.07}$ & ${1.11}_{-0.05}^{+0.06}$ & ${15.86}_{-0.08}^{+0.07}$ \\ 
HD~9289 & 3 & ${60.37}_{-6.13}^{+7.32}$ & ${1.31}_{-0.13}^{+0.15}$ & ${2.67}_{-0.25}^{+0.32}$ & ${11.65}_{-0.04}^{+0.04}$ &&&&&&\\ 
\enddata
\tablecomments{The fitting parameters of SLF for the 118 magnetic hot stars. We reported the result of the fitting parameters of SLF. The Sectors using to derive the parameters are listed behind the identification names.}
\end{deluxetable*}
\end{longrotatetable}
%%%
\section{Appendix B}
\setcounter{figure}{0}
\begin{figure}[ht]
\gridline{\fig{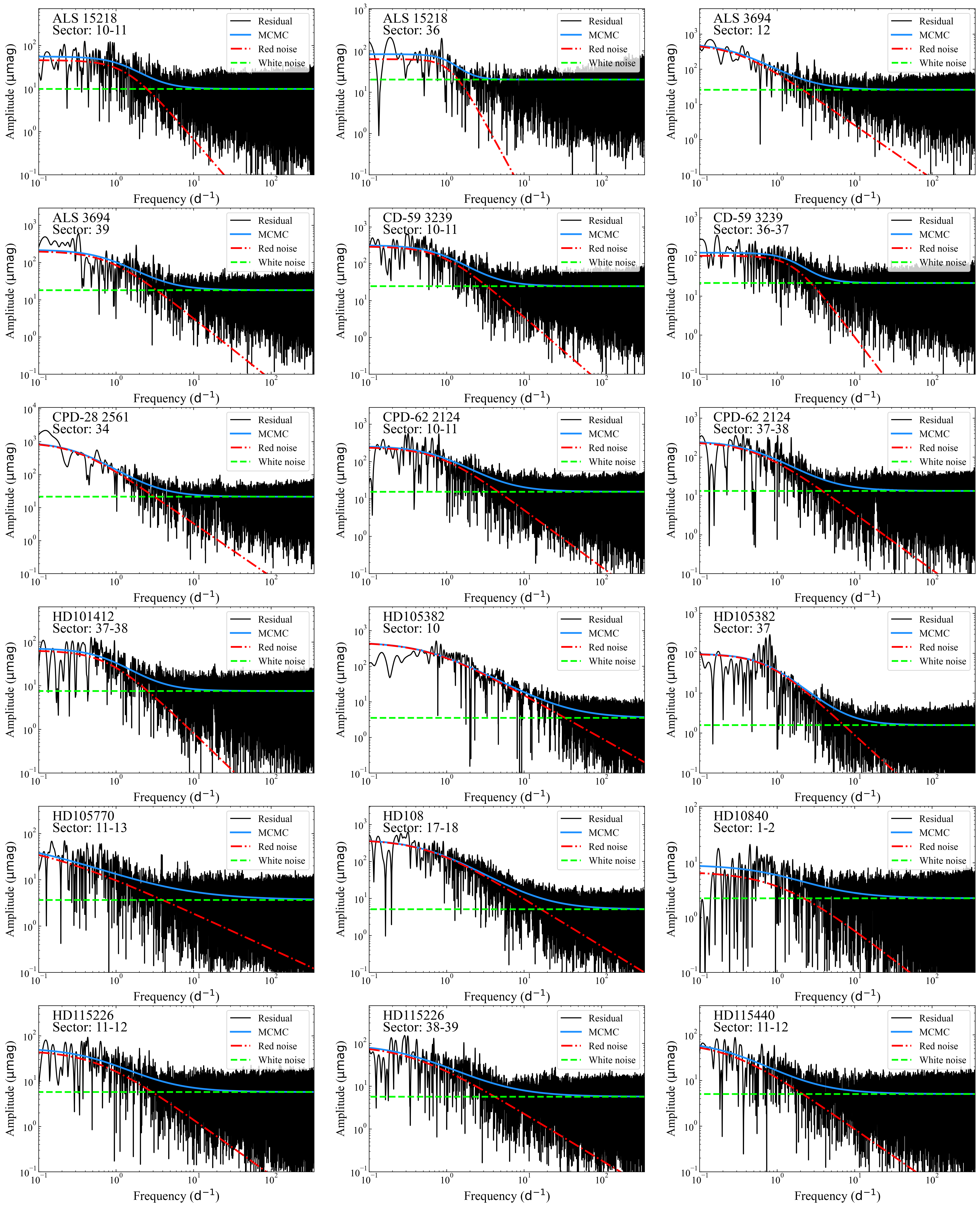}{1\textwidth}{}}
\caption{The logarithmic amplitude spectra for each star in our sample.  Line styles and colours are the same as in Figure~\ref{MCMC}. \label{Appendix C: Fig1}}
\end{figure}

\begin{figure}[ht]
    \gridline{\fig{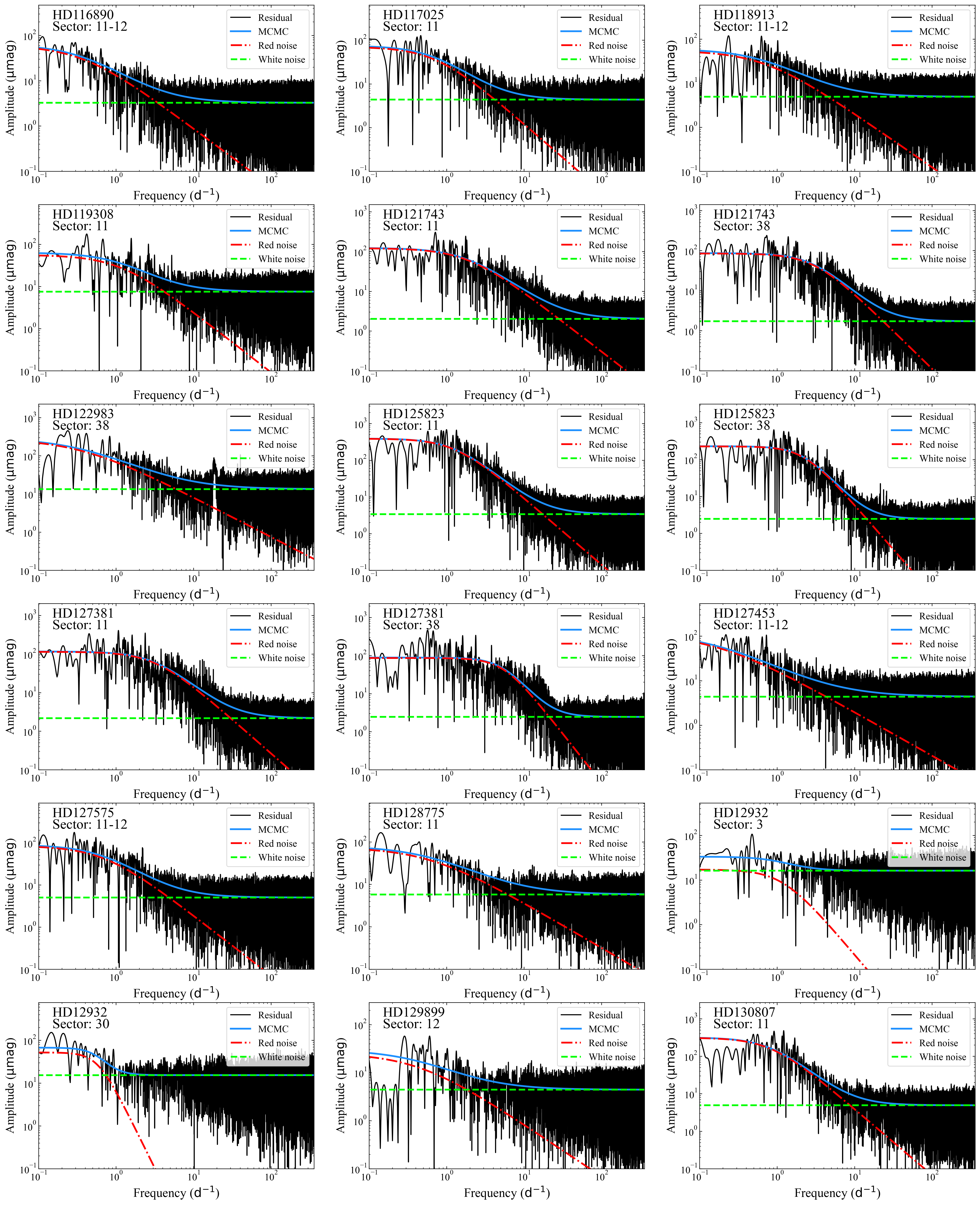}{1\textwidth}{}}
    \caption{continued.\label{Appendix C: Fig2}}
\end{figure}

\begin{figure}[ht]
    \gridline{\fig{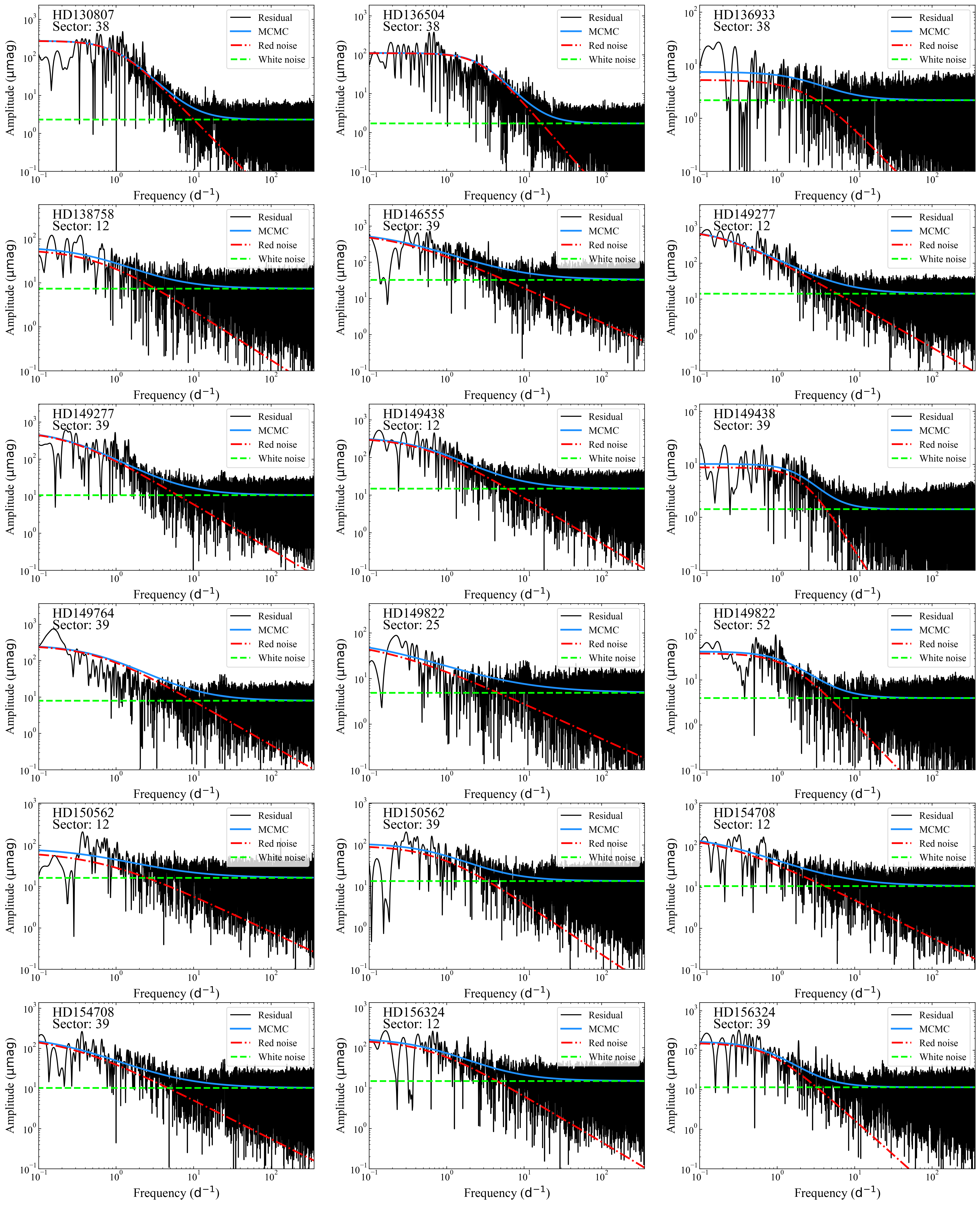}{1\textwidth}{}}
    \caption{continued.\label{Appendix C: Fig3}}
\end{figure}

\begin{figure}[ht]
    \gridline{\fig{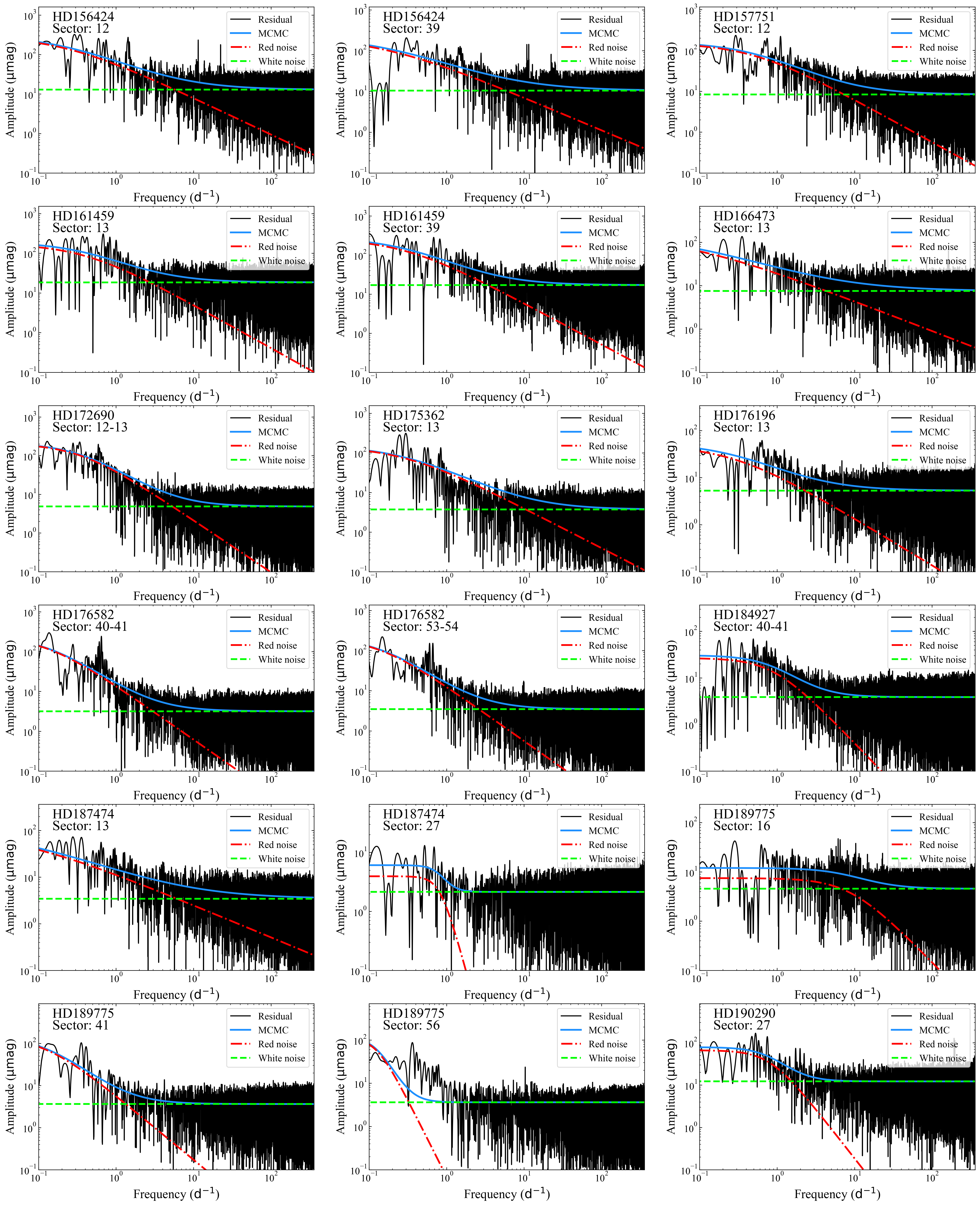}{1\textwidth}{}}
    \caption{continued.\label{Appendix C: Fig4}}
\end{figure}

\begin{figure}[ht]
    \gridline{\fig{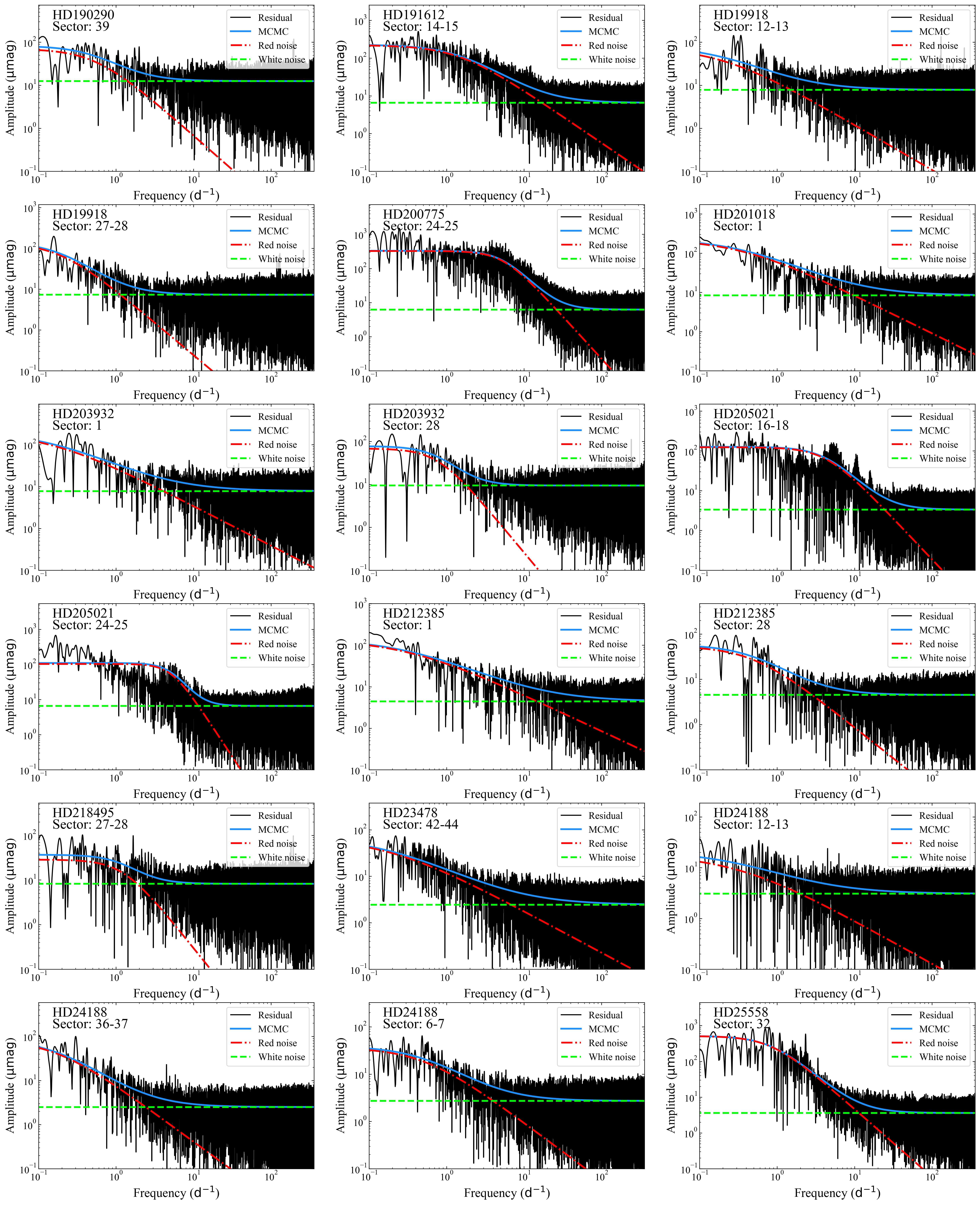}{1\textwidth}{}}
    \caption{continued.\label{Appendix C: Fig5}}
\end{figure}

\begin{figure}[ht]
    \gridline{\fig{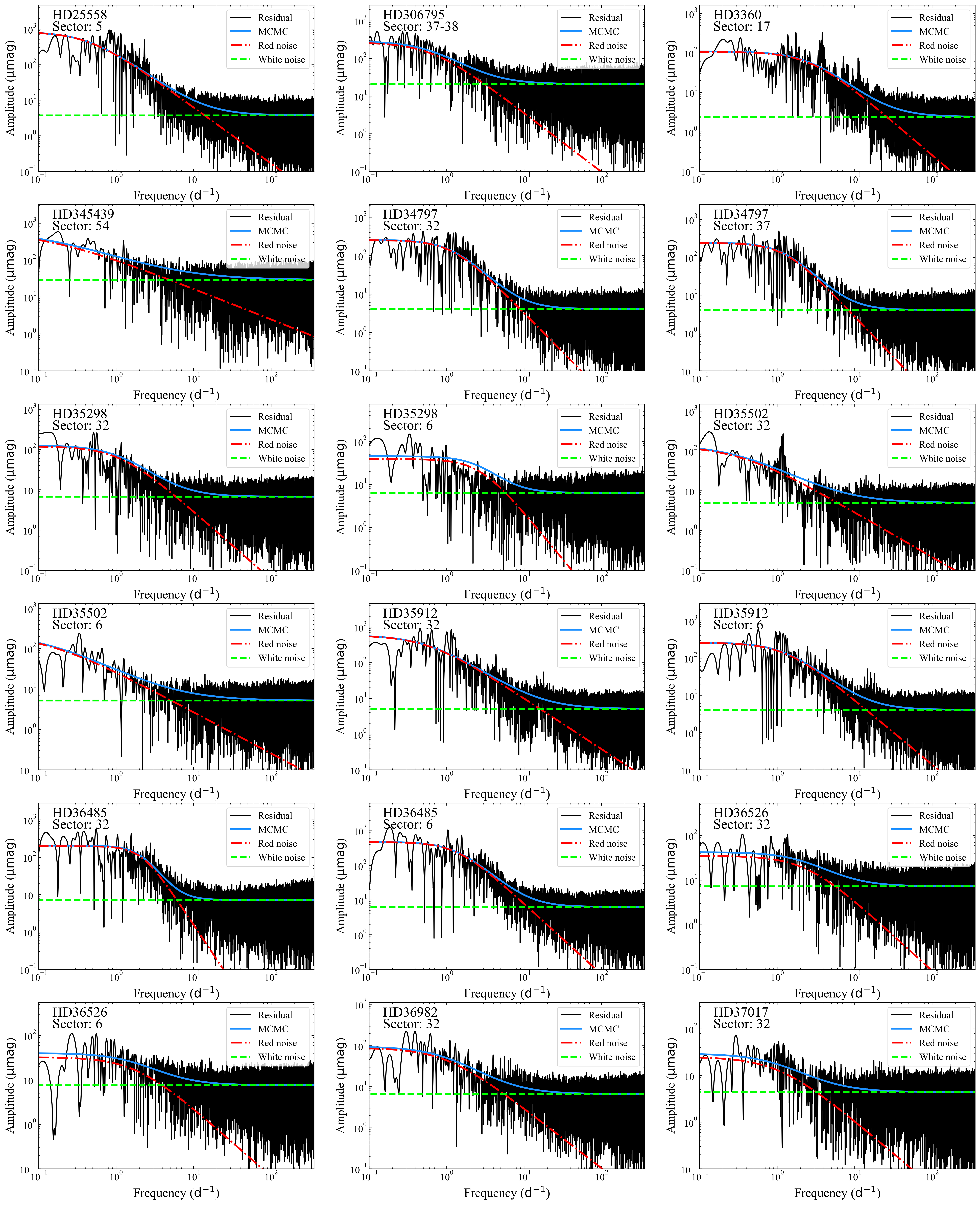}{1\textwidth}{}}
    \caption{continued.\label{Appendix C: Fig6}}
\end{figure}

\begin{figure}[ht]
    \gridline{\fig{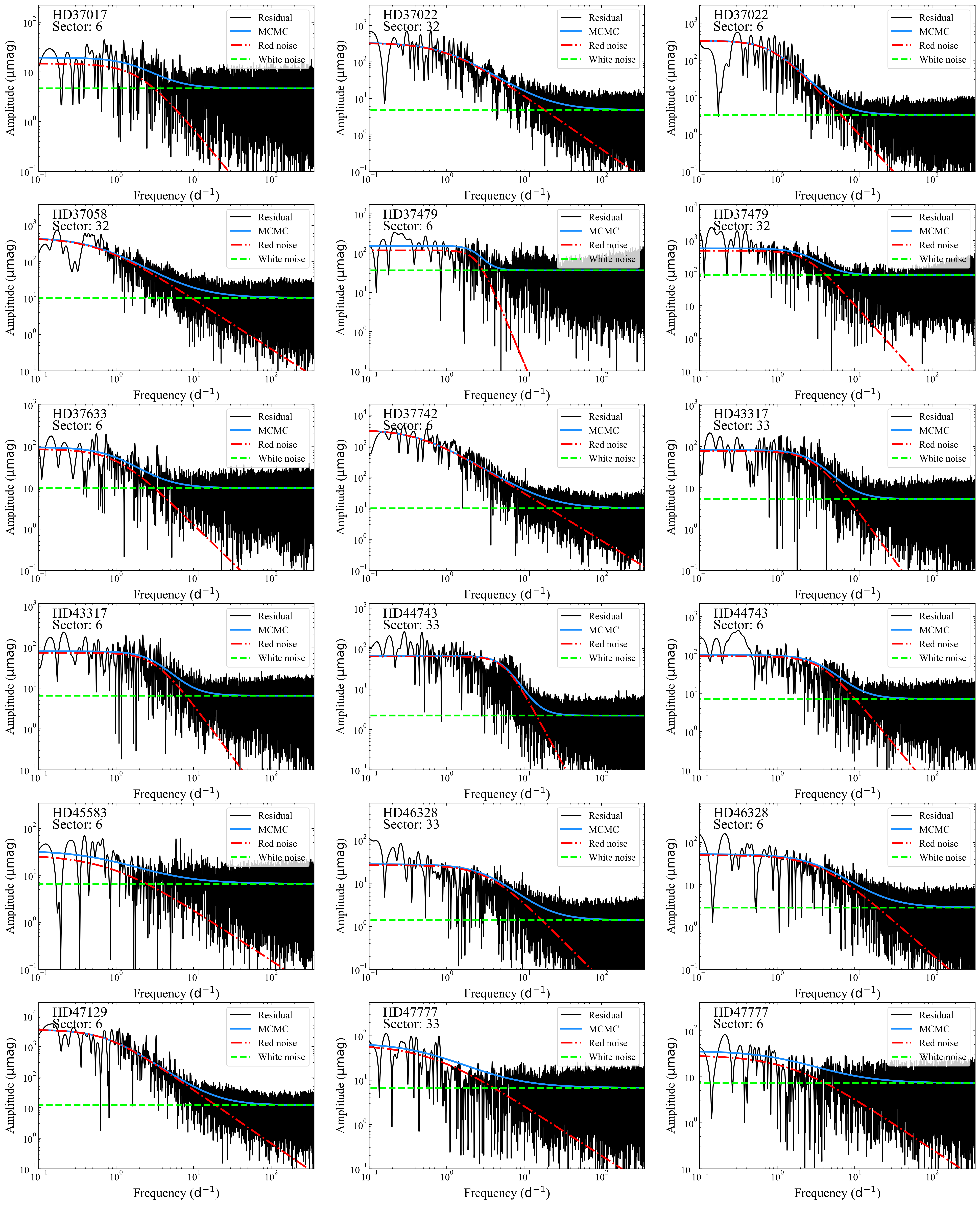}{1\textwidth}{}}
    \caption{continued.\label{Appendix C: Fig7}}
\end{figure}

\begin{figure}[ht]
    \gridline{\fig{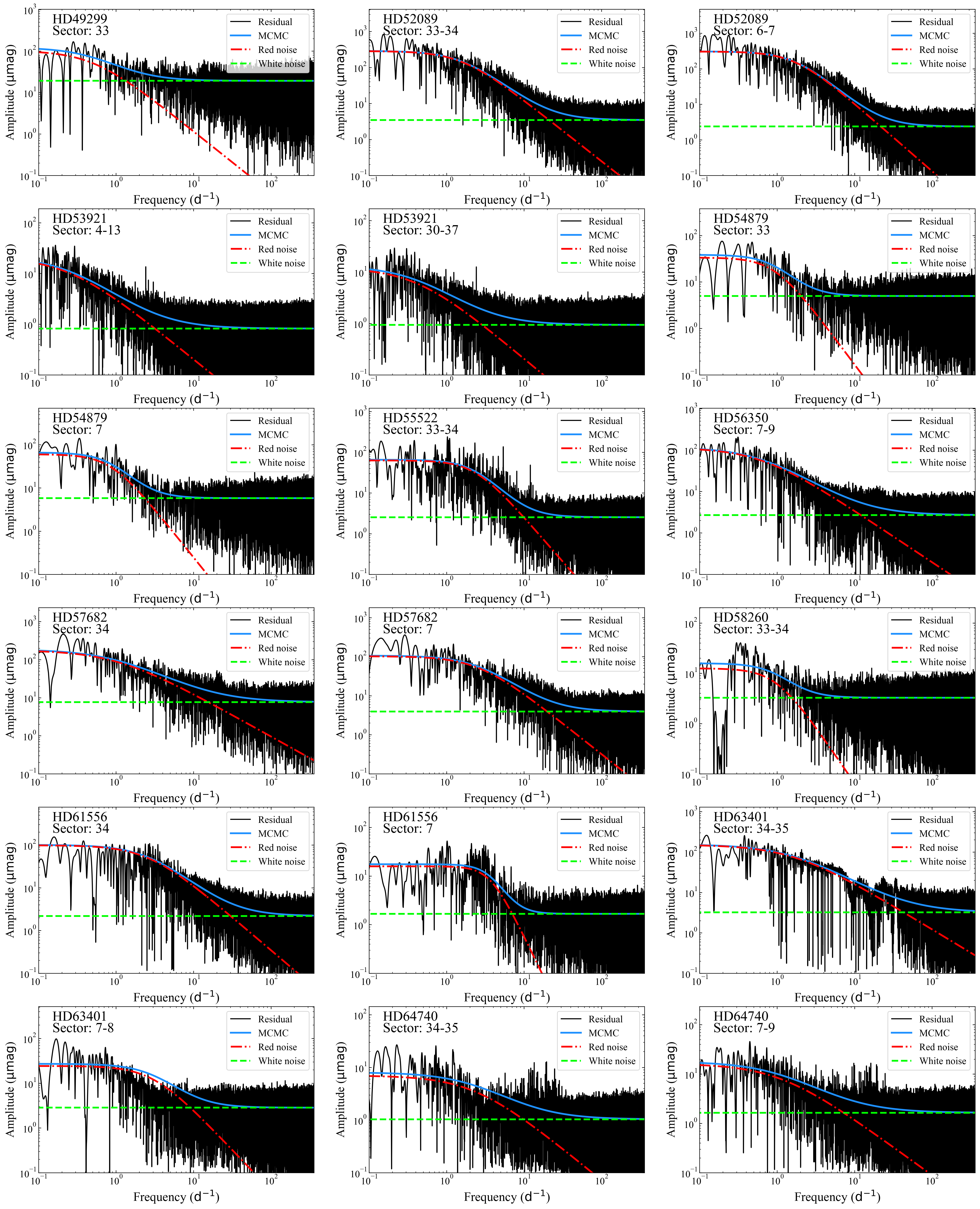}{1\textwidth}{}}
    \caption{continued.\label{Appendix C: Fig8}}
\end{figure}

\begin{figure}[ht]
    \gridline{\fig{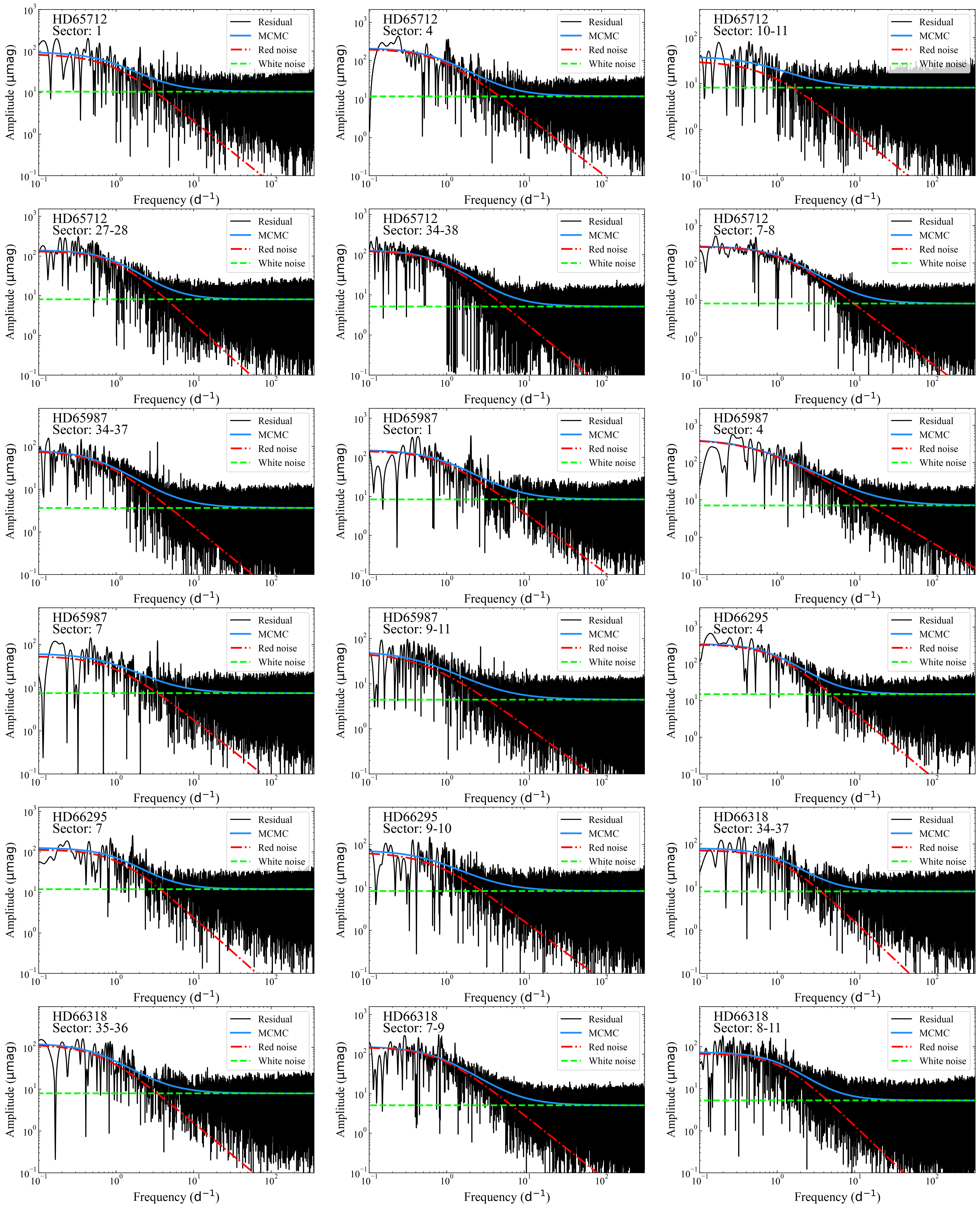}{1\textwidth}{}}
    \caption{continued.\label{Appendix C: Fig9}}
\end{figure}

\begin{figure}[ht]
    \gridline{\fig{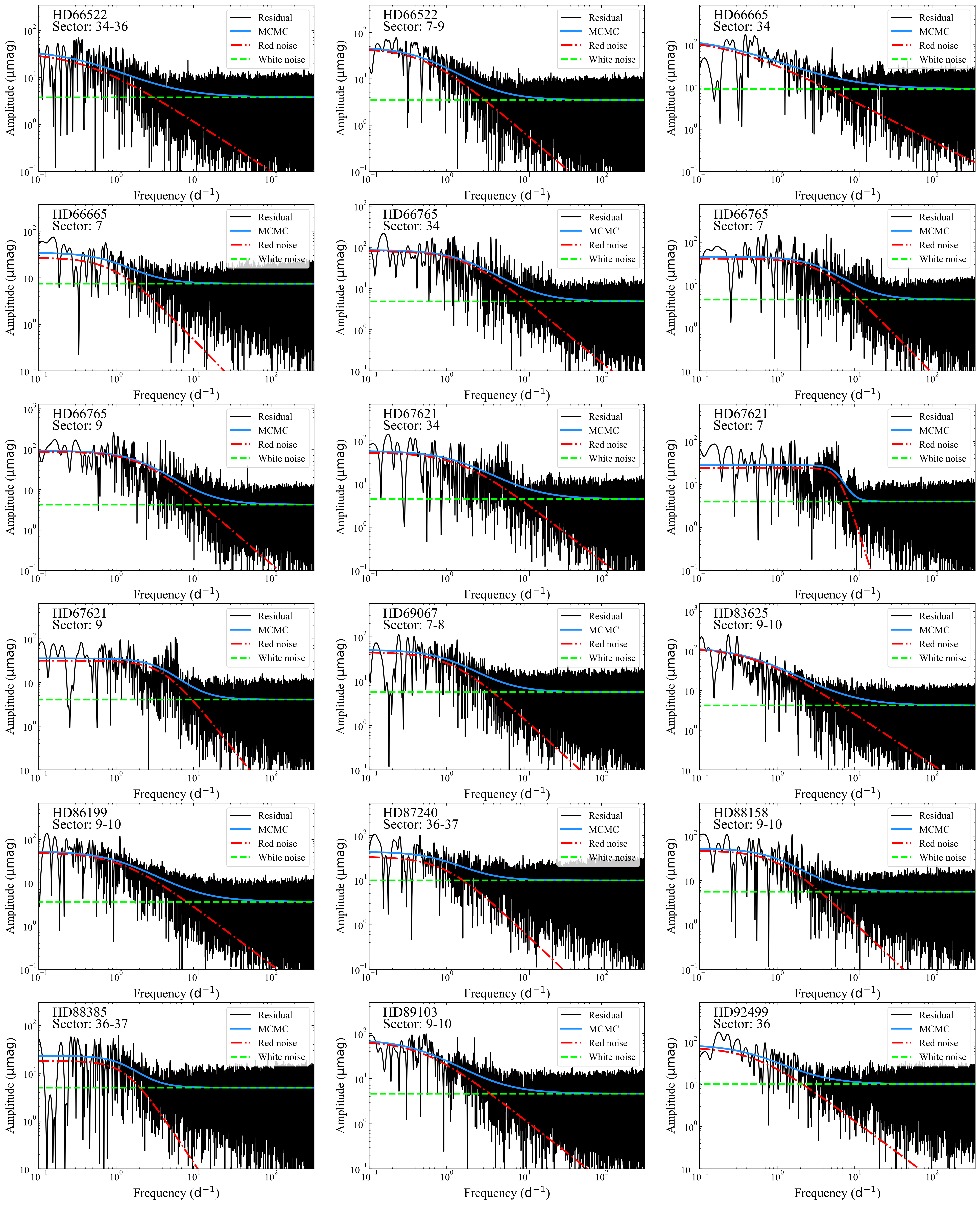}{1\textwidth}{}}
    \caption{continued.\label{Appendix C: Fig10}}
\end{figure}

\begin{figure}[ht]
    \gridline{\fig{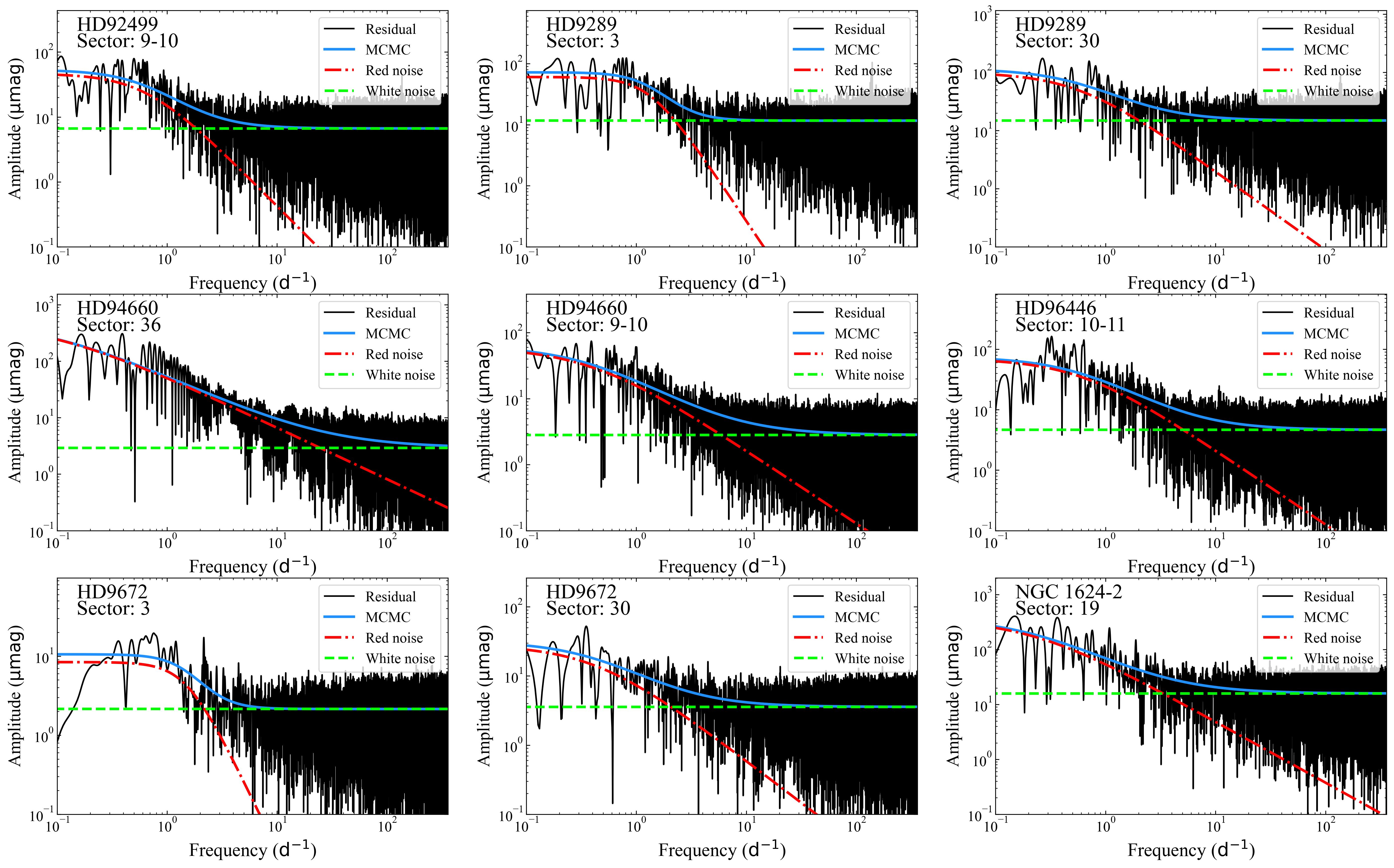}{1\textwidth}{}}
    \caption{continued.\label{Appendix C: Fig11}}
\end{figure}

\clearpage

\vspace{5mm}




\bibliography{sample631.bib}{}
\bibliographystyle{aasjournal}


\end{document}